\documentclass[letter,12pt,times,numbered,index,oneside]{Classes/PhDThesisPSnPDF}
\input{Preamble/preamble}

 \title{Radar Classification of Contiguous Activities of Daily Living} 		

\author{Ronny (Gerhard) Guendel}

\dept{Department of Electrical and Computer Engineering}

\university{Villanova University}
\crest{\includegraphics[width=0.5\textwidth]{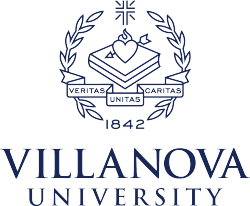}}

\advisor{Dr. Moeness G. Amin} 

\supervisorlinewidth{0.35\textwidth}

\advisorlinewidth{0.35\textwidth}

\renewcommand{\submissiontext}{In Partial Fulfillment\\ Of the Requirements for the Degree of\\
Master of Science in Electrical Engineering\\
Signal Processing and Communications (SPC)}

\degreedate{December 2019} 

\subject{LaTeX} \keywords{{Independent Study Thesis} {Engineering} {Villanova University}}

\ifdefineAbstract
\fi

\begin{document}

\frontmatter

\maketitle 

\thispagestyle{plain}
\setcounter{page}{2}

\begingroup
\centering
\null
\vfill
Copyright ~{\sffamily\textcopyright}~2019 by Ronny (Gerhard) Guendel
\\[0.5em]
All Rights Reserved
\vfill
\par
\endgroup
\clearpage


%
%
%
%
%
%
%
%
%
%
%
%
%


\begin{declaration}
	
I hereby declare that except where specific reference is made to the work of 
others, the contents of this \textit{thesis by publication} are original and have not been 
submitted in whole or in part for consideration for any other degree or 
qualification in this, or any other university. This master thesis is my own 
work and contains nothing which is the outcome of work done in collaboration 
with others, except as specified in the text and Acknowledgments. 


\end{declaration}


\begin{acknowledgements} 
This thesis is the result of my research at Villanova University. The project described in this research was supported in part by the Comcast Labs in Philadelphia. The contents are solely the responsibility of the author and do not necessarily represent the official views of the sponsors. First and foremost, I would like to acknowledge my indebtedness and render my warmest thanks to my advisor, Dr. Moeness G. Amin, who made this work possible. His guidance and expert advice have been invaluable throughout all stages of this work. I would like to thank my family and my friends, especially my parents and my sister, whose endless support has encouraged me through this project. Many people whom I cannot mention individually deserve thanks and appreciation for their support and help in the preparation of my thesis. I would like to express my gratitude to my thesis committee members, Dr. Bijan G. Mobasseri and Dr. Ahmad Hoorfar, for their invaluable guidance on my research. Also, I am grateful to Janice Moughan and the university staff and the International Student Office for their help in managing administrative tasks. Thanks to everyone who supported me.
	
\end{acknowledgements}

\begin{abstract}
We consider radar classifications of Activities of Daily Living (ADL) which can prove beneficial in fall detection, analysis of daily routines, and discerning physical and cognitive human conditions. We focus on contiguous motion classifications which follow and commensurate with the human ethogram of possible motion sequences. Contiguous motions can be closely connected with no clear time gap separations. In the proposed motion classification approach, we utilize the Radon transform applied to the radar range-map to detect the translation motion, whereas an energy detector is used to provide the onset and offset times of in-place motions, such as sitting down and standing up. It is shown that motion classifications give different results when performed forward and backward in time. The number of classes, thereby classification rates, considered by a classifier, is made variable depending on the current motion state and the possible transitioning activities in and out of the state. Motion examples are provided to delineate the performance of the proposed approach under typical sequences of human motions.
\end{abstract}



\tableofcontents
\listoftables
\listoffigures

\printnomenclature[3em]

\mainmatter

\chapter{Introduction} \label{chap:introduction}

\section{Motivation} \label{sec:motivation}
The combination of sensing technologies and data analytics is considered a powerful tool in efficient indoor monitoring of human activities 
\cite{1_aminSP, amin2017radar, 8_4801689, mobasseri2009time, 10_5404249, 9_7348882}. 
RF-based sensors offer fine controls for sensing configurations and work in all lighting and weather conditions. They promise non-intrusive, safe and reliable home automation and security.  Further, monitoring human indoor activities prove important for the vulnerable population, including the elderly and disable.  The principal task is to detect and classify motion abnormalities as well as anomalies in activity patterns that may be linked to deteriorating health conditions. In particular, falls are considered as an abnormal activity that should be accurately detected and classified with high sensitivity and specificity 
\cite{12_Hong2011, 13_Su2015, 14_Wu2015, LeKernec8746868}. 
Other daily activities can indicate, in their variants over times,  changes in routines and lifestyle as well as the state of physical, cognitive, and psychological health of the person. RF-based gesture recognition using hands and arms is also an important contactless technology for Man-Machine-Interface (MMI) 
\cite {kim2016hand, wang2016interacting, skaria2019hand, maminzz, zhang2016dynamic, GurbuzAmin:DLindoor}. 
Adding to the indoor applications, RF-based vital sign monitoring has vast medical use, as respiration and heart beats are essential diagnostic barometers for many health problems 
%
\cite{amin2017radar, 3_7842648, Seifert8613848, 5_7944373, 6_seifert2019RadarConf, shahFioranelli:ADL:8894733}. 
In this thesis, we focus on the classifications of ADL, dealing with motions as consecutive, and sometimes contiguous without clear time boundary and isolations. 

Human daily activities can be categorized into translation and in-place motions. Whereas the former mainly describes crawling and gait articulations, the latter is primarily associated with motions that do not exhibit considerable changes in range. In-place motions include sitting, standing, kneeling, and bending, each is performed without any stride. The accumulated range under each in-place is typically less than 1 meter. Fall can be considered a translation or in-place motion, depending on the motion corresponding range swath \cite{amin2019rf, aminGuendel2019radarConf, jokanovic2018fall}. 

In this work, we use a FMCW radar with range and Doppler resolution capabilities. We first separate translation and in-place motions using the Radon transform applied to the radar backscattering signals. The Radon transform which exploits the piece-wise linear behavior of range vs time corresponding to human walking.   This is revealed by computing the range-map that depicts the range vs slow-time of a moving target. Accordingly, horizontal lines correspond to in-place motions, whereas translation motions are manifested by lines with non-zero slopes. In the range-map, fast translation motions are associated with steep lines.  On the other hand, elderly gait typically leaves behind smaller signature slopes compared to those of young adults. Positive slopes describe range translation away from the radar, and negative slopes are a result of walking towards the radar. The Radon transform can reveal the transitions from translation to in-place motion and visa versa by capturing the time instants, or "breaking" points, of slope changes. Over the in-place motion time segments of ADL, characterized by horizontal lines in the range-map or small range variations, an  energy detector, in lieu of Radon transform, is applied to determine the onset and offset times of each motion. It is noted that we do not assume any underlying model guiding the received signal, which is viewed as as a deterministic signal with time-varying frequencies \cite{amin1992time, SetlurAmin:SPIE}. 

We use two-dimensional (2-D) PCA, a data-driven feature learning techniques followed by the Nearest Neighbors (NN) classifier. 
%
The 2-D PCA has shown to be very effective in motion classification.  It outperforms hand-crafted based motion classifications and offers competitive results to convolution and deep neural networks 
\cite{baris:DataCubeProcessing8691492, park2016microdoppler_bc, maminzz}. 
Both the target micro-Doppler signature, provided by the spectrograms and the target range-map are input to the 2-D PCA. It is important to note that the essence of this research contribution is not to devise new classifier but rather to address the contiguity issue of human motions and exploit the "ethogram" \cite{humanEthogram} of human activities which limit the possible contingent motions stemming from the present ones. In essence, we use 2-D PCA, NN classifier for demonstrating the concepts introduced.

We consider classifying consecutive motions, e.g., a stream of "sitting down", "standing up", "walking". Some of these consecutive motions are contiguous in the sense there is no clear time gap separation between two successive activities. We seek to exploit the logical sequence of human motion articulations. For example, we would not consider "sitting down" as a possible action following a previous "sitting down". This also applies to consecutive "standing up" actions. In the same practice, we would eliminate "walking" that immediately follows "bending down" without the occurrence of the intermediate motion of "rebounding up" from bending. More generally, we view motions as states, namely sitting, standing, walking, and laying states. The transition from one state to another is guided by the human ethogram that defines the possible sequences of motions.

The elimination of impossible motions in a sequence of motion leads to changing the candidate classes of motions to be classified at any given time. This, in turn, alters the size of the classifier confusion matrix, in lieu of using a fixed matrix dimension, which the case when dealing with all ADL as possibilities. We also utilize the fact that some actions and activities associated with some motion states are more accurately classified than those in other states. In this respect, situation can arise where classifications of current motions associated with one state encourage going backward in time and revisiting past states to gain more assertiveness on previously incurred motions. The work provides examples showing the above three main aspects of the proposed approach, namely, variable size confusion matrix and bidirectional classifications. These examples are based on 2-D PCA, NN classifier. Different arguments and conclusions using the same examples, but a different classifier, can be drawn; however, the two aspects of our approach remain applicable.

\section{Background} \label{sec:background}
\begin{figure}[tbph]
	\centering
	\includegraphics[width=0.99\linewidth]{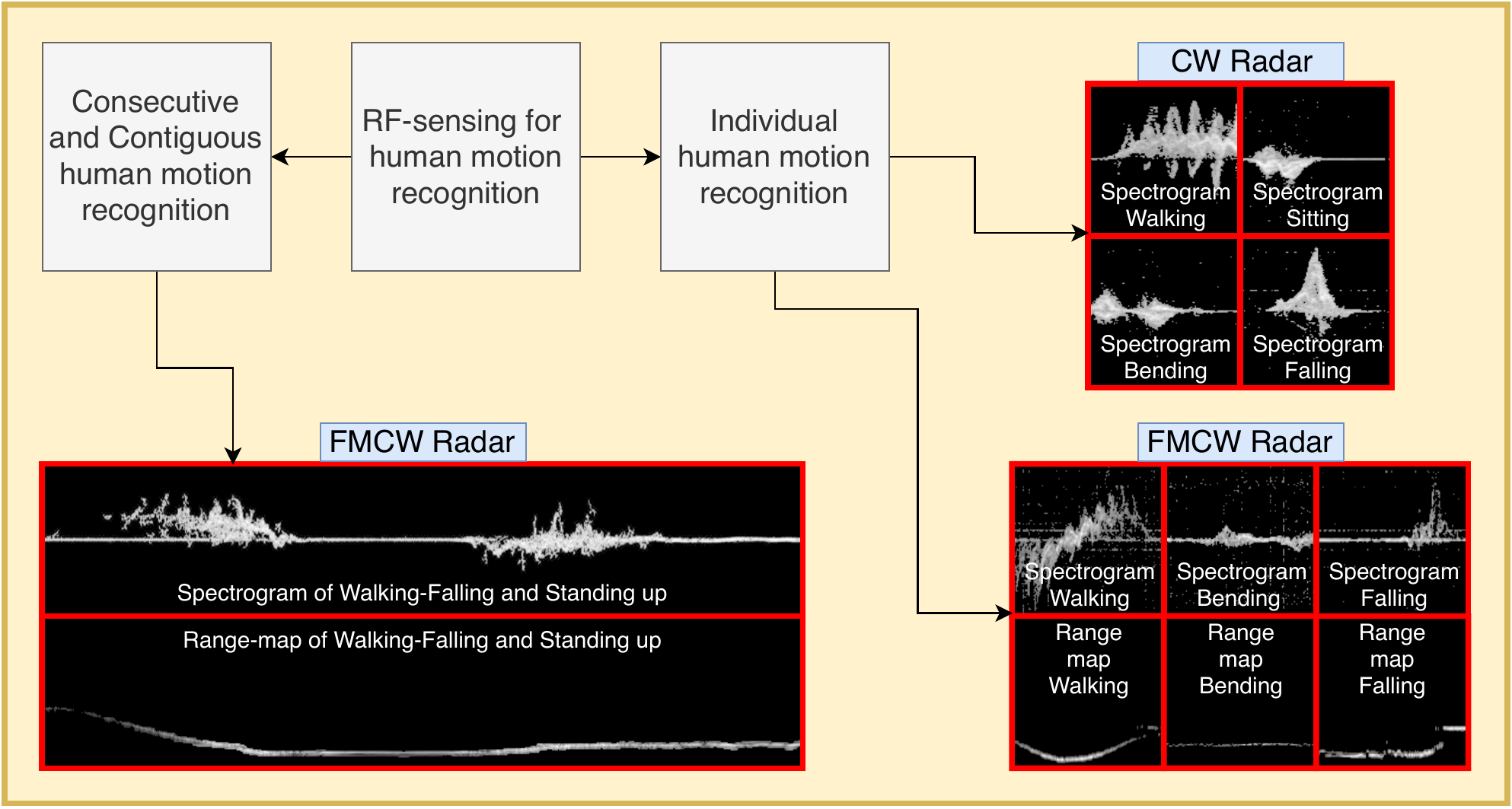}
	\caption{State-of-the-art methods (r.h.s.) and future approaches (l.h.s.) of ADL radar recognition.}
	\label{fig:stateoftheart}
\end{figure}

Machine learning combined with radar signal processing categorizing ADL human motion  has already become an area of high interest with different waveform modulation schemes, such as Continuous Waveform (CW) or Frequency-Modulated Continuous Waveform (FMCW) radars. The former only provides the Doppler signatures of the human motion, whereas the latter adds the range information. The Doppler or micro-Doppler signatures are revealed by the spectrogram. The spectrogram images and the range-map images (FMCW) are shown in Fig.~\ref{fig:stateoftheart} on the r.h.s. for few selected motions \cite{baris:GANbased:8835589}. 
Here, the data collection is basically performed on a predefined time window, which captures the individual motion. For such approaches, the determination of the onset and offset times of the activities are not performed. A variety of different feature extraction and classification methods based on spectrograms have been exploited and tested for their reliability and accuracy.  In general, the range-information has not even played a key role in ADL classification \cite{amin2016radar, amin2017radar}. 

In contrast, Fig.~\ref{fig:stateoftheart} shows on the l.h.s. a consecutive and contiguous motion sequence, collected by a FMCW radar. For multi motion sequences, possibly consisting of translation and in-place motions, some of which can be merged, existing motion classification approaches cannot be directly applied. It is noted that contiguous and consecutive motion sequences reflect a natural and realistic behavior of ADL. Hence, the work in this thesis begins with separating these consecutive motion actions by taking advantage of the information in the range-map. We incorporate effective human motion classification techniques, feature extraction methods, and detection of the onset and offset times  of each motion towards a final accurate and reliable automatic human motion rechgnition \cite{amin_Wu:Fall_PBC:7046290, baris:ASILOMAR:WidebandFall7869686, pcaKNN:8252120, rajaguru2017knn, 9_7348882}.

\section{Thesis organization}
The thesis is organized as follows. In Chapter~\ref{chap:radarSystem}, the system model and the pre-processing steps are discussed, along with the 2-D PCA feature extraction method, and the fusion classifier that combines spectrogram and range-map features. In Chapter~\ref{chap:radon_PBC}, the motion capturing method, supported by the energy detector and range-map processing based on the Radon transform is presented. Chapter~\ref{chap:State_expResults} describes motions as states and discusses transient activities forward and backward in time, along with the experimental setup, with key examples showing the essence of our approach. Finally, conclusions are provided in Chapter \ref{chap:Conclusion}.

\chapter{\label{chap:radarSystem}Radar system, pre-processing and classification}

\section{\label{sec:RadarModel} Radar system and data analysis}

\subsection{\label{subsec:RadarModel}Radar model}
\begin{figure}
	\centering
	\includegraphics[width=0.7\linewidth]{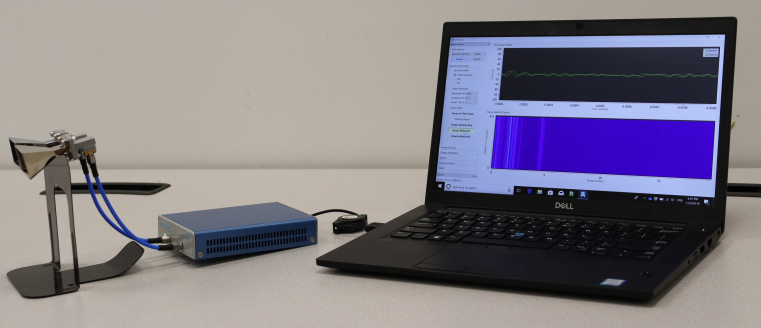}
	\caption[K-band radar Ancortek SDR-KIT 2500B.]{K-band radar Ancortek SDR-KIT 2500B~\cite{website2500B}.}
	\label{fig:ancortek2500bimg0096}
\end{figure}

The data collection was performed using SDR-KIT 2500B \cite{website2500B} (Fig.~\ref{fig:ancortek2500bimg0096}), which is developed by Ancortek, Inc. 
The FMCW radar operates with a center frequency $25~GHz$ and bandwidth $2~GHz$. 
\begin{figure}[!htb] 
	\centering
	\includegraphics[scale=.7]{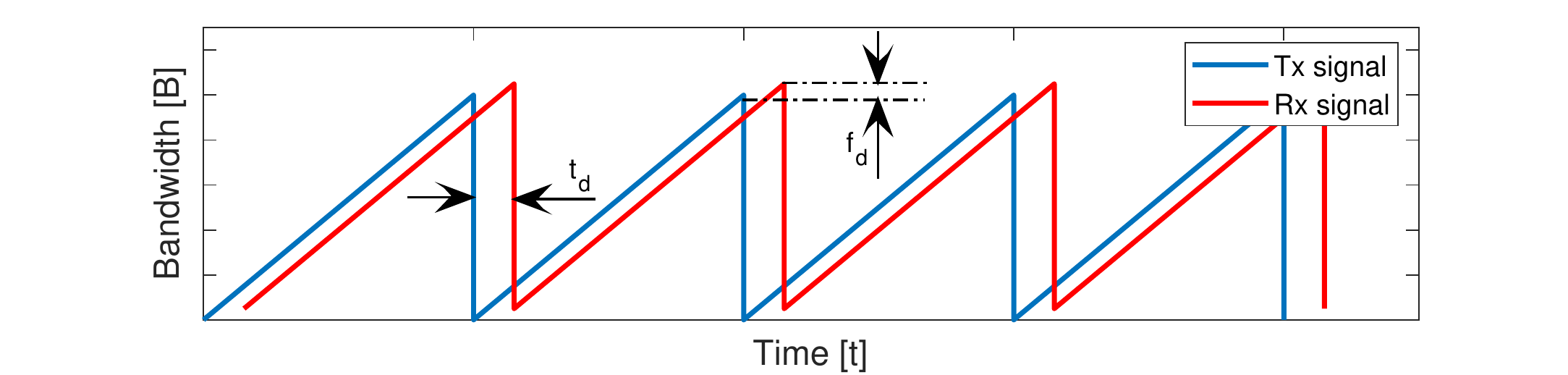}
	\caption{Illustration of a FMCW Sawtooth transmitted and received signal model.}
	\label{fig:FMCWSawtooth}
\end{figure}
The Fig.~\ref{fig:FMCWSawtooth} illustrates the transmitted (Tx) signal and the received (Rx) signal for four pulse repetition intervals (PRIs), schematically, for the used radar system. The  PRI is  $1~ms$, and the range resolution ($ RR $) is ~$7.5~cm$, which is computed as, 
\begin{equation}\label{eq:rr}
RR = \frac{c_0}{2B}
\end{equation}
with $ c_0 $ as the speed of light and $B$ indicates the bandwidth. The transmitted signal is,
\begin{equation}
S_{Tx}(t) = A_T \cdot cos[2\pi (f_c t +\frac{1}{2}\alpha t^2)]
\end{equation}
\noindent where  $\alpha$ is  the chirp rate given by $B/T$. The received signal is,
\begin{equation}
S_{Rx}(t) = A_R \cdot  cos[2\pi (f_c (t-\tau) +\alpha (\frac{1}{2} t^2 -\tau\cdot t) + f_D \cdot t)]
\end{equation}
\noindent where $\tau$ is the two-way travel time, $f_D$ is the Doppler shift assuming constant velocity target. $A_R$, the signal amplitude, is given by, 
\begin{equation}
A_R = \frac{G \lambda \sqrt{P \sigma}}{\sqrt{(4 \pi)^3}R^2 \sqrt{L_s} \sqrt{L_a}}
\end{equation}
\noindent with the  antenna gain,  $G$, and the transmitted power , $P$. $\sigma$ is the radar cross-section (RCS), $L_a$ and $L_s$ indicate the atmospheric and system loss, respectively. The complex baseband signal is expressed in terms of the in-phase and quadrature components as, 
\begin{equation}
s(t) = I(t) + jQ(t) = Ae^{\psi (t)}
\end{equation}
\noindent where ($\psi(t)$) is the signal phase \cite{baris:DataCubeProcessing8691492, art:fmcwRadarModel}. This signal is used in all follow-on analysis.

\subsection{Range-map computation \label{subsec:rangemap}}
For the computation of the target range profile, the matched filtered of the radar signal return is first represented by a two-dimensional matrix, $s(n,m)$. The Discrete Fourier Transform (DFT) applied to each column, corresponding to one $PRI$, provides the target range information. The range-map, $R(p,m)$, is generated by incorporating the consecutive $PRIs$, and is given by, 
\begin{equation} \label{eq:STFT:RM}
R(p,m) = \frac{1}{N} \sum_{n=0}^{N-1} s(n,m) exp(-j 2 \pi \frac{p n}{N})
\end{equation}
\noindent where $p = 0, . . . ,N - 1$,  $N$ is the number of samples, or range bins, in one $PRI$,  and $m = 0,...,M-1$, where $M$ represents the total number of $PRIs$ considered. In the data collection experiments, we set N and M to be $512$ and $12,000$ (e.g. for twelve seconds of data), respectively. 

\subsection{Micro-Doppler signature}
To obtain the target micro-Doppler signature, we first sum the data over the range bins of interest as, 
\begin{equation} \label{eq:rangebinsum}
V(m) = \sum_{r =r_1}^{r_2} R(r,m)
\end{equation}
\noindent where $r_1$ and $r_2$ are the minimum and maximum range bins considered, set to  $10$  and $128$, respectively. This corresponds to a range swath from $0.75~m$ to $9.6~m$ \cite{fmcwRadar_8443507}. The short-time Fourier transform (STFT) is then applied to $V(m)$, and its magnitude square, i.e., spectrogram, is computed to yield the micro-Doppler signature, $MD(n,k)$,
\begin{equation} \label{eq:MDequation}
MD(n,k) =  \left|	\sum_{m=0}^{L-1}w(m)V(n-m)exp(-j 2 \pi \frac{m k}{L}) \right| ^2
\end{equation}
\noindent A Hanning window $w(m)$ of size $L = 128$ is applied to reduce the sidelobes \cite{art:hanningWindow467238}. 
Eq.~\ref{eq:MDequation} is computed at several windows positions separated by $8$~samples, which corresponds to $94~\%$ window overlapping. The spectrogram is resized with $128$~samples for Doppler scaling and $32~samples$ ($1~sec$) in slow-time. The same resizing process is applied for range-map images \cite{aminGuendel:IET}.

\section{Pre-processing \label{sec:preprocessing}}
This section provides the required pre-processing steps for the range-map and the micro-Doppler spectrogram. Neither the applied Radon transform nor the PBC gives reliable results on the initially computed range-map or the micro-Doppler spectrogram, respectively. 
     
\subsection{Range-map pre-processing\label{subsec:rangemappreproc}}

\begin{figure}[ht]
	\centering
	\begin{subfigure}[b]{0.49\textwidth}
		\includegraphics[width=\linewidth, frame]{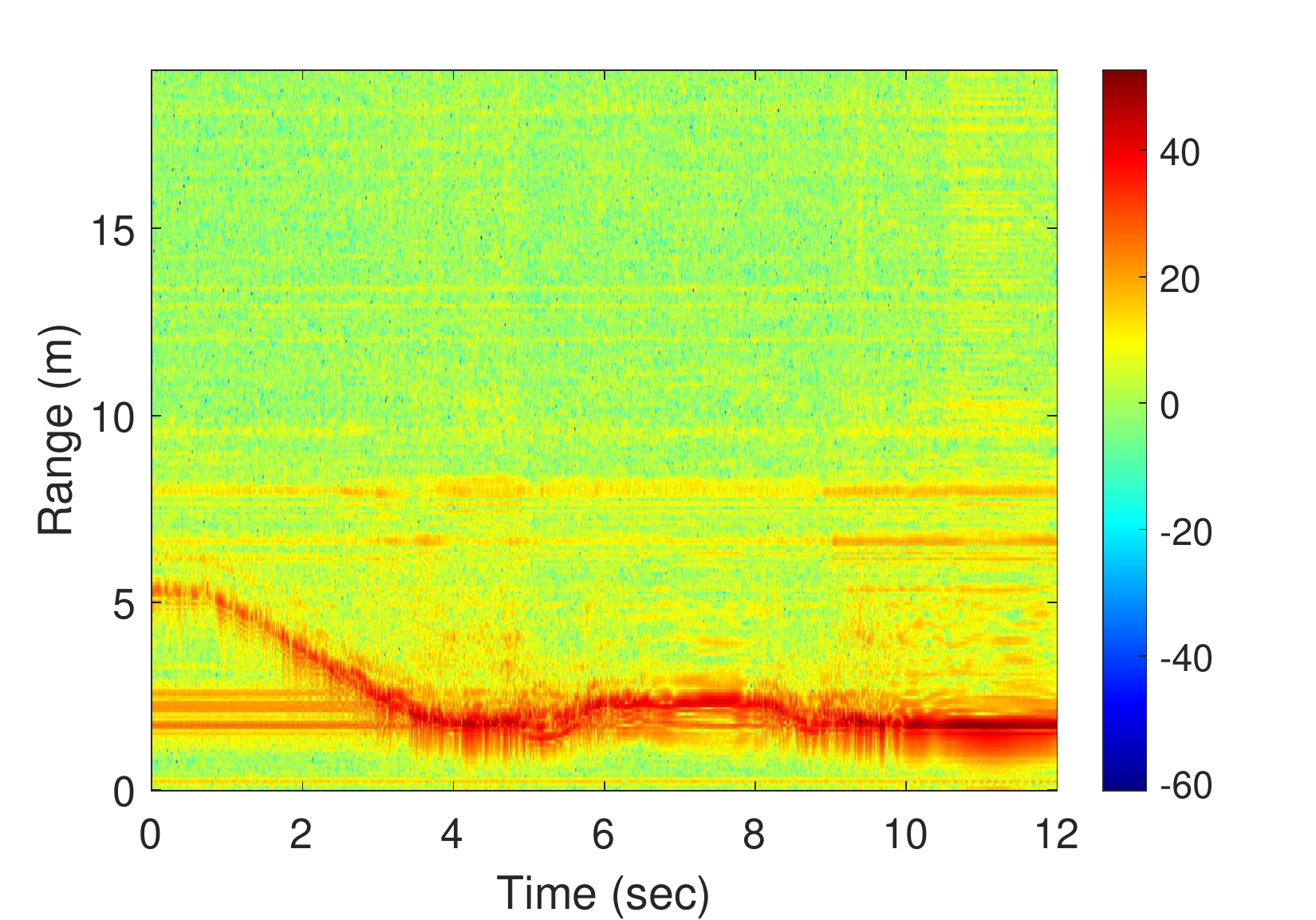}
		\caption{Range-map profile}
		\label{subfig:RM_1_Raw}
	\end{subfigure}
	~ 
	\begin{subfigure}[b]{0.49\textwidth}
		\includegraphics[width=\linewidth, frame]{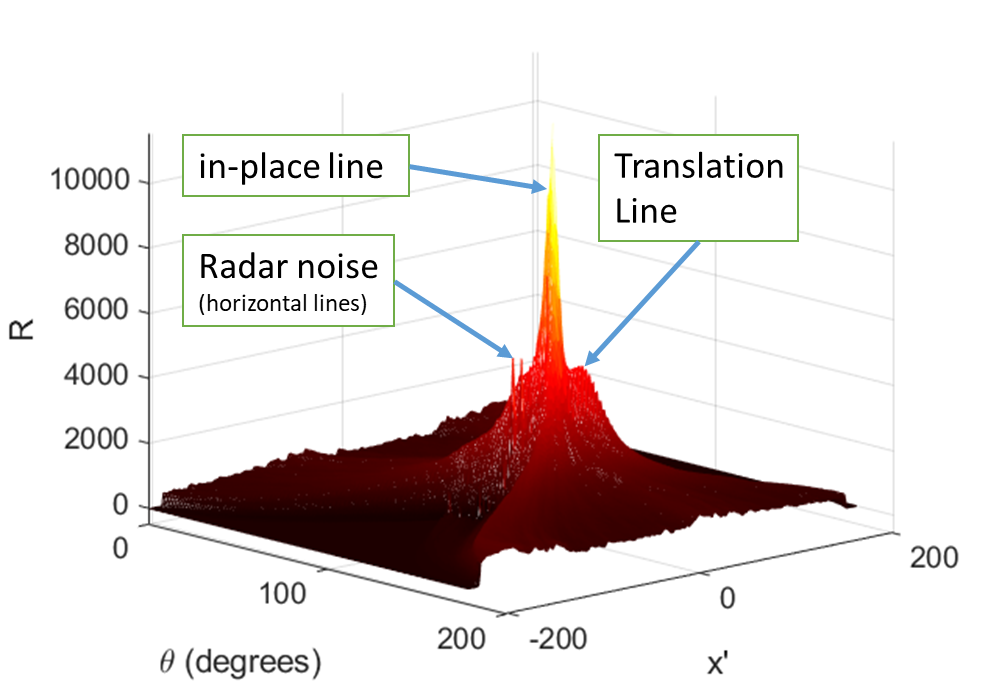}
		\caption{Radon transform}
		\label{subfig:RM_1_Raw_Radontrans}
	\end{subfigure}
	\caption{Range map for walking followed by sitting down and standing up without pre-processing steps and the performed Radon transform.\label{fig:RM_1_Raw}}
\end{figure}

The Eq.~\ref{eq:STFT:RM} is applied on the radar received raw data $ s(n,m) $ to compute the complex range-map matrix. The range map figure is then computed as, 
\begin{equation}\label{eq:}
RM = 10 \cdot log_{10} \left(\sqrt{real^2(R) + imag^2(R)}\right)
\end{equation} 
with $ R $ as the complex range-map matrix. The range-map ($ RM $) is converted to a logarithmic magnitude scale, shown in Fig.~\ref{subfig:RM_1_Raw}. 
After down-sampling the range-map ($ RM $) to an image of $M \times N =  128 \times 384$ samples (for $ 12s $ of data), the Radon transform is applied and shown in Fig.~\ref{subfig:RM_1_Raw_Radontrans}.  Noisy horizontal lines may cause no clear peaks in the Radon graph to be assigned to a persons translation and in-place profile. Furthermore, the \textit{Tranlation Line} in Fig.~\ref{subfig:RM_1_Raw_Radontrans} is blurry and unclear. To mitigate the problem, the range-map should be cleaned out by using multiple steps explained in the follow-on sections, since all lines except the true target line, compromise the Radon transform.

\subsubsection{Columnwise normalization of the the Range-map} 
\begin{figure}[ht]
	\centering
	\begin{subfigure}[b]{0.49\textwidth}
		\includegraphics[width=\linewidth, frame]{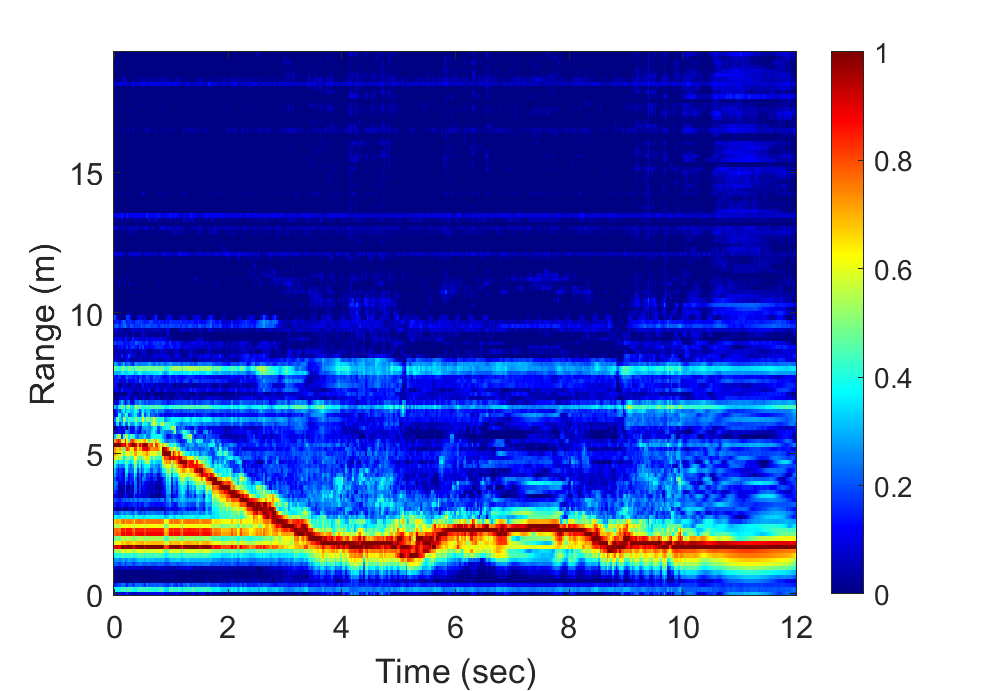}
		\caption{Range-map profile}
		\label{subfig:RM_2_norm}
	\end{subfigure}
	~ 
	\begin{subfigure}[b]{0.49\textwidth}
		\includegraphics[width=\linewidth, frame]{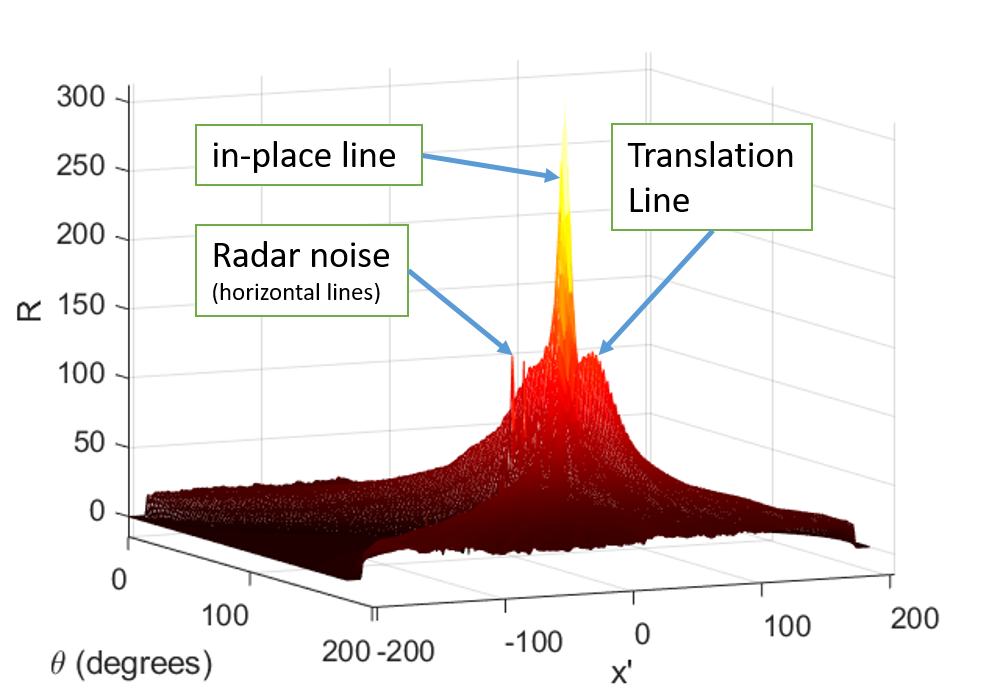}
		\caption{Radon transform}
		\label{subfig:RM_2_norm_Radontrans}
	\end{subfigure}
	\caption{Columnwise normalized range map and the performed Radon transform.\label{fig:RM_2_norm}}
\end{figure}
The normalization of the columns in the range-profile has shown good results to strip the signature from dependence on target distance. We divide each sample of the range-map, $ RM $, by their column maximum value. The elementwise operation is known as Hadamard division and is computed as, 
\begin{equation}\label{eq:hadamarddivison}
\begin{split}
\begin{bmatrix}
	RM_{1,1}       &   RM_{1,2}       & \dots     &   RM_{1,N}       \\
	RM_{2,1}       &   RM_{2,2}       & \dots     &   RM_{2,N}       \\
	\vdots  &  \vdots   &   \vdots  &   \vdots  \\   
	RM_{M,1}       &   RM_{M,2}       & \dots     &   RM_{M,N}       \\
	\end{bmatrix}
	\oslash
\begin{bmatrix}
max_1(RM_{m})       &   max_2(RM_{m})       & \dots     &   max_N(RM_{m})       \\
max_1(RM_{m})       &   max_2(RM_{m})       & \dots     &   max_N(RM_{m})       \\
\vdots  &  \vdots   &   \vdots  &   \vdots  \\   
max_1(RM_{m})       &   max_2(RM_{m})       & \dots     &   max_N(RM_{m})       \\
\end{bmatrix}\\
=
\begin{bmatrix}
\widetilde{RM}_{1,1}       &   \widetilde{RM}_{1,2}       & \dots     &   \widetilde{RM}_{1,N}       \\
\widetilde{RM}_{2,1}       &   \widetilde{RM}_{2,2}       & \dots     &   \widetilde{RM}_{2,N}       \\
\vdots  &  \vdots   &   \vdots  &   \vdots  \\   
\widetilde{RM}_{M,1}       &   \widetilde{RM}_{M,2}       & \dots     &   \widetilde{RM}_{M,N}       \\ 
\end{bmatrix}
\end{split}	
\end{equation}
with $max_n(RM_{m})$ for $ n  = 1,2,\dots,N$ as the maximal individual column value. $ \widetilde{RM} $ is the normalized range-map matrix shown in Fig.~\ref{subfig:RM_2_norm}. 
Further, it can be seen in the Radon transform (Fig.~\ref{subfig:RM_2_norm_Radontrans}) that the horizontal lines are very much maintained, whereas the \textit{Tranlation Line} becomes slightly stronger due to distance compensation. 

\subsubsection{eCLEAN\label{subsubsec:eclean_RM}}
\begin{figure}[ht]
	\centering
	\includegraphics[width=0.49\linewidth, frame]{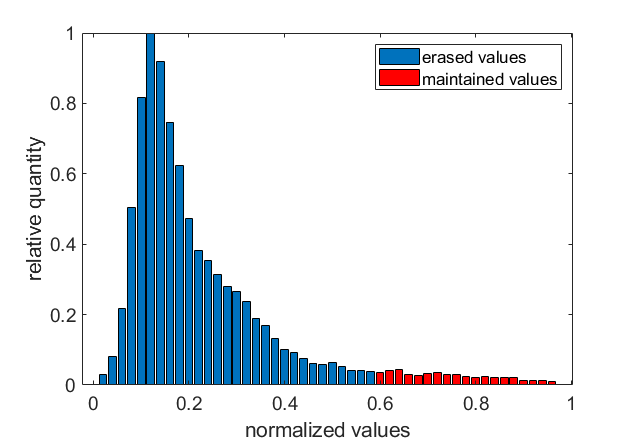}
	\caption{Histogram of the \textit{eCLEAN} algorithm for the range-map.\label{fig:histogr}}
\end{figure}
\begin{figure}[ht]
	\centering
	\begin{subfigure}[b]{0.49\textwidth}
		\includegraphics[width=\linewidth, frame]{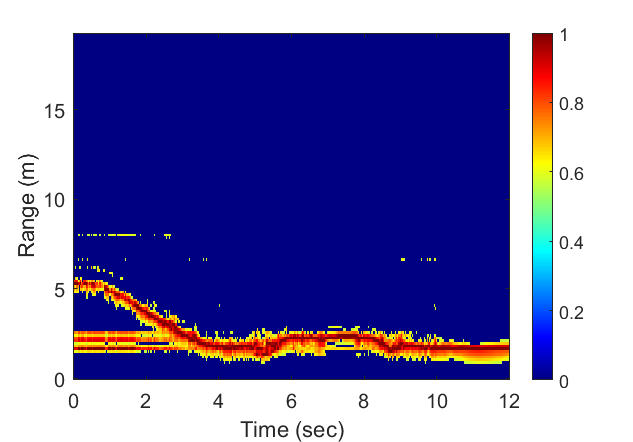}
		\caption{Range-map profile}
		\label{subfig:RM_3_eCLEAN}
	\end{subfigure}
	~ 
	\begin{subfigure}[b]{0.49\textwidth}
		\includegraphics[width=\linewidth, frame]{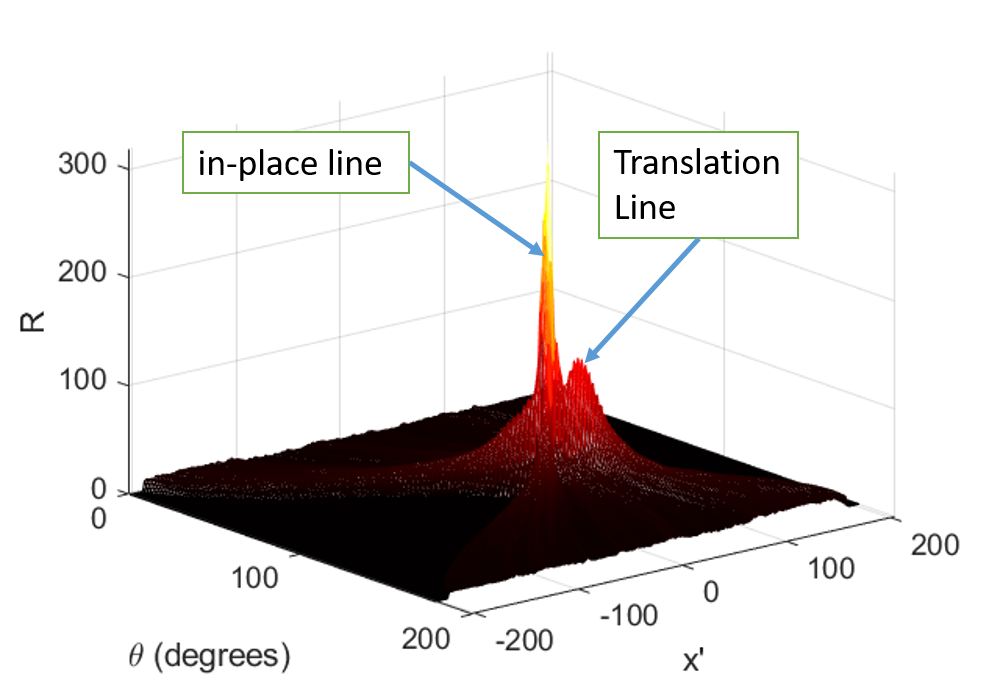}
		\caption{Radon transform}
		\label{subfig:RM_3_eCLEAN_Radontrans}
	\end{subfigure}
	\caption{Columnwise normalized range map and the performed Radon transform.\label{fig:RM_3_eCLEAN}}
\end{figure}
\noindent
To reduce the the noise enhancement in the range-map, we apply the \textit{eCLEAN} algorithm \cite{eclean:Kulpa, baris:DataCubeProcessing8691492, baris:GANbased:8835589}. The algorithm computes a histogram from the matrix $ \widetilde{RM} $, as shown in Fig.~\ref{fig:histogr}. The values belonging to the $ 20 $ highest histogram bins are maintained (shown in red), whereas the values belonging to the lower histogram bins are set to $ 0 $.
The result is shown in Fig.~\ref{subfig:RM_3_eCLEAN}, where most of the noise is eliminated. Accordingly, the Radon transform (Fig.~\ref{subfig:RM_3_eCLEAN_Radontrans}) does not show the extraneous horizontal lines due to the radar itself. Further, the peak of the \textit{Translation Line} becomes more dominant than before.  

\subsubsection{Outlier removal\label{subsubsec:outlier_RM}} 
\begin{figure}[ht]
	\centering
	\begin{subfigure}[b]{0.49\textwidth}
		\includegraphics[width=\linewidth, frame]{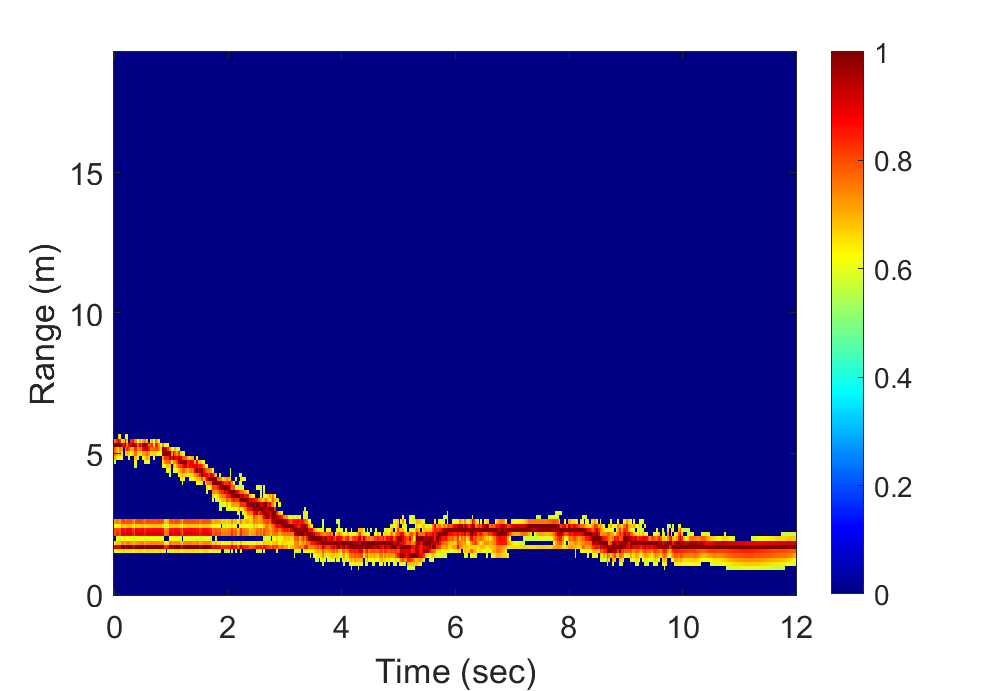}
		\caption{Range-map; outlier removed}
		\label{subfig:RM_4_OLR}
	\end{subfigure}
	~ 
	\begin{subfigure}[b]{0.49\textwidth}
		\includegraphics[width=\linewidth, frame]{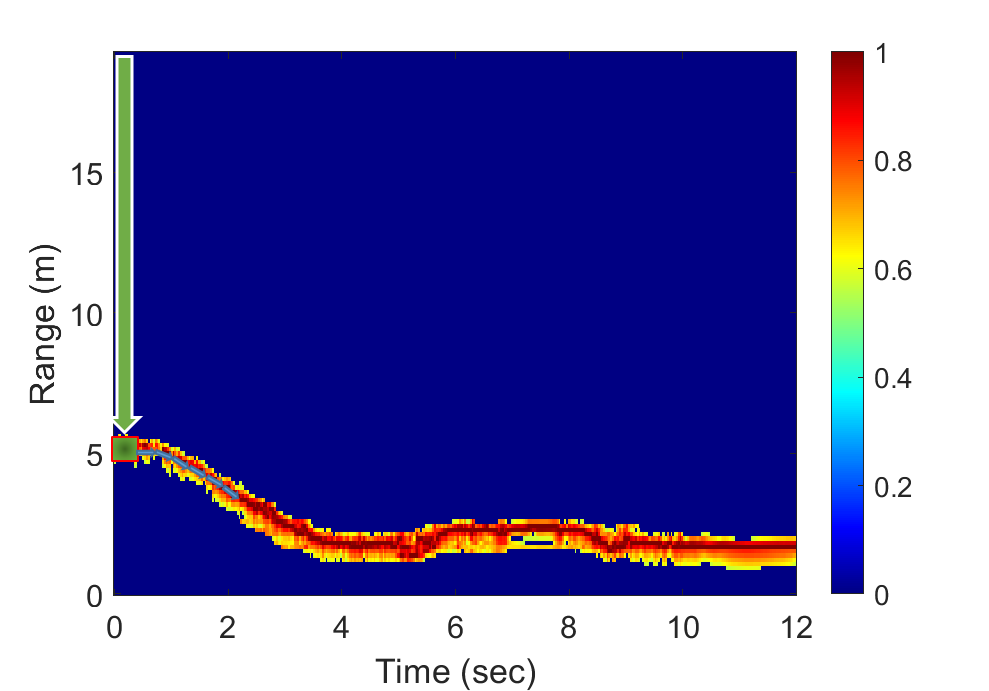}
		\caption{Range-map; kernel cleaning}
		\label{subfig:RM_5_kernel_6}
	\end{subfigure}
	\\
	\centering
	\begin{subfigure}[b]{0.49\textwidth}
		\includegraphics[width=\linewidth, frame]{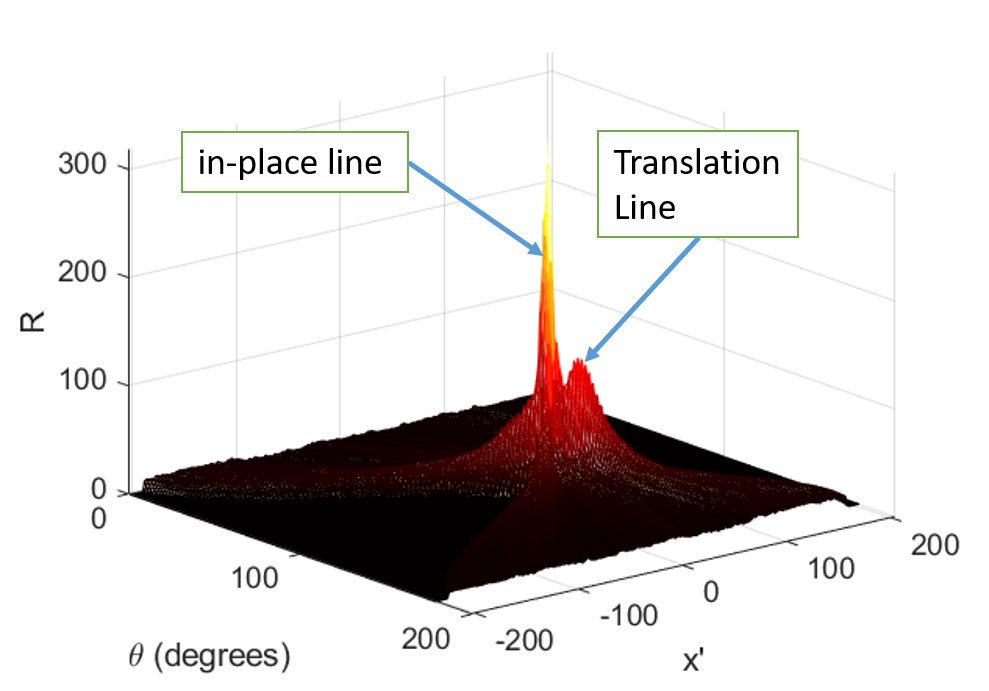}
		\caption{Radon transform on (a)}
		\label{subfig:Radon_4_OLR_2}
	\end{subfigure}
	~ 
	\begin{subfigure}[b]{0.49\textwidth}
		\includegraphics[width=\linewidth, frame]{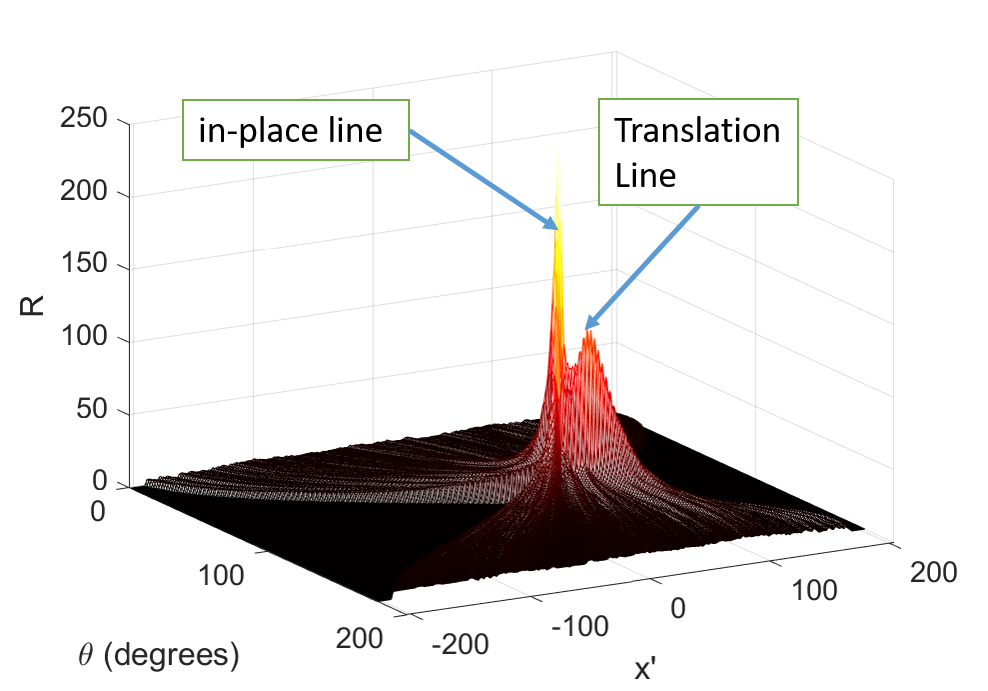}
		\caption{Radon transform on (b)}
		\label{subfig:Radon_5_kernel_6_2}
	\end{subfigure}
	\caption[The figure set shows the range-maps after removing the outlier points and after the kernel cleaning the Radon transform.]{The figure set shows (a) the range-map after removing the outlier points, and (b) shows the range-map after kernel cleaning. (c) and (d) display the Radon transform on (a) and (b). \label{fig:RM_5_4_OLR_kernel}}
\end{figure}

Below, we explain the processing step to remove outlier points. Outlier points (Pixel) may confuse the Radon transform. Furthermore, they can lower the classification outcome, since these points contain inappropriate features. The outlier points are removed by a using the \textsc{Matlab} function {\lstinline!bwareaopen(RM-eClean,50)!} with a default value of $ 50 $ pixel for a matrix size of $ 128x384 $ for range map processing. The function finds independent isolated pixel clusters of fewer than the default value. The isolated pixel clusters are erased from the image.  The result is shown in Fig.~\ref{subfig:RM_4_OLR}.

\subsubsection{Kernel cleaning\label{subsubsec:kernelCleaning}}
The kernel cleaning algorithm is a necessary computational step to improve the performance of the Radon transform. Furthermore, the algorithm limits the width (spread) of the target line to the window size of $ 6 $ pixel (according to Algorithm~\ref{code:Pseudocode}) such that no biased range profile depending on targets line width occurs. 

Fig.~\ref{subfig:RM_4_OLR} shows the input image. The algorithm begins by finding the farthest target distance at slow-time bin $ t=0s $, as shown with the green arrow in Fig.~\ref{subfig:RM_5_kernel_6}. Here, a kernel of the window size (6 pixel) is placed over the target line, as shown in Fig.~\ref{subfig:RM_5_kernel_6} with the red/green square. The summed power within the placed kernel window is indicated as {\lstinline!pwr2!}. {\lstinline!pwr1!} and {\lstinline!pwr3!} are the summed power of the window shifted one pixel away or towards the radar, respectively, and listed in the Algorithm~\ref{code:Pseudocode}. The highest power indicates the window of which the $ 6x6 $ pixels kernel is mirrored in the empty matrix {\lstinline!RMC!}. Then, the algorithm shifts one sample in slow-time ({\lstinline!n = n + 1!}) and repeats the routine until ({\lstinline!n == 384-Win!}) the end of the last six slow-time bins are reached. It is noticed that the algorithm can just go forward in slow-time; therefore, it cannot follow the noisy horizontal line at distance $ 2~m $ between $ 0~s $ and $ 3~s $ in slow-time. 

For comparison, Fig.~\ref{subfig:Radon_4_OLR_2} and Fig.~\ref{subfig:Radon_5_kernel_6_2} show the result of the Radon transform before and after the kernel cleaning algorithm. Here, the magnitude of the peaks for the \textit{in-place Line} and the \textit{Translation Line} are more leveled out. 
\begin{algorithm}[htbp]
	{\footnotesize \KwData{range-map: RM}
	\KwResult{Cleaned range-map: RMC}
	\textbf{Initialization:} RMC = zeros(128,384); pwr = 0; pwr1 = 0; pwr2 = 0; pwr3 = 0; m = 128; n = 1; Win = 6\;
	\While{pwr == 0 \&\& n ==1\tcc*[r]{While loop counts down to 5.5m until the target line is detected}}{
		pwr = sum(RM[m,n:n+Win])\tcc*[r]{Computing the power at range-bin m with the width of 6 pixel}\
		m = m-1\tcc*[r]{Counting towards the radar}\
		\eIf{pwr != 0\tcc*[r]{It runs into the "if" condition, when the window line of 6 pix hits the target line at 5.5m}}{
			\While{n != (384-Win)}{
			pwr1 = sum(RM[m-Win+1:m+1,n:n+Win]) \tcc*[r]{Comp. pwr1; shifted 1 range-bin away}\
			pwr2 = sum(RM[m-Win:m,n:n+Win]) \tcc*[r]{Comp. pwr2 as for the previous range bin}\
			pwr3 = sum(RM[m-Win-1:m-1,n:n+Win]) \tcc*[r]{Compu. pwr3; shifted 1 range-bin towards}\
			\uIf{pwr1 >= (pwr2||pwr3)\tcc*[r]{pwr1 is the highest}}{
				RMC[m-Win+1:m+1,n:n+Win] = RM[m-Win+1:m+1,n:n+Win]\tcc*[r]{Copying the pixel into the zero matrix RMC}\
				m = m +1\tcc*[r]{Counting away the radar}\ 
			}
			\uElseIf{pwr2 >= (pwr1||pwr3)\tcc*[r]{pwr2 is the highest}}{
				RMC([m-Win+1:m+1,n:n+Win]) = RM([m-Win:m,n:n+Win]\tcc*[r]{Copying the pixel into the zero matrix RMC}\ \tcc*[r]{distance to the radar m maintained}\ 
				
			}
			\Else{\tcc*[r]{pwr3 is the highest}
				RMC[m-Win-1:m-1,n:n+Win] = RM[m-Win-1:m-1,n:n+Win]\tcc*[r]{Copying the pixel into the zero matrix RMC}\
				m = m -1\tcc*[r]{Counting towards the radar}\ 
			}
		n = n + 1\tcc*[r]{counting one slow-time bin to the right}\
		}
		}{
			return\;
		}
	}}
	\caption{Kernel clean algorithm for range-map processing.\label{code:Pseudocode}}
\end{algorithm}

\begin{figure}[ht]
	\centering
	\begin{subfigure}[b]{0.49\textwidth}
		\includegraphics[width=\linewidth, frame]{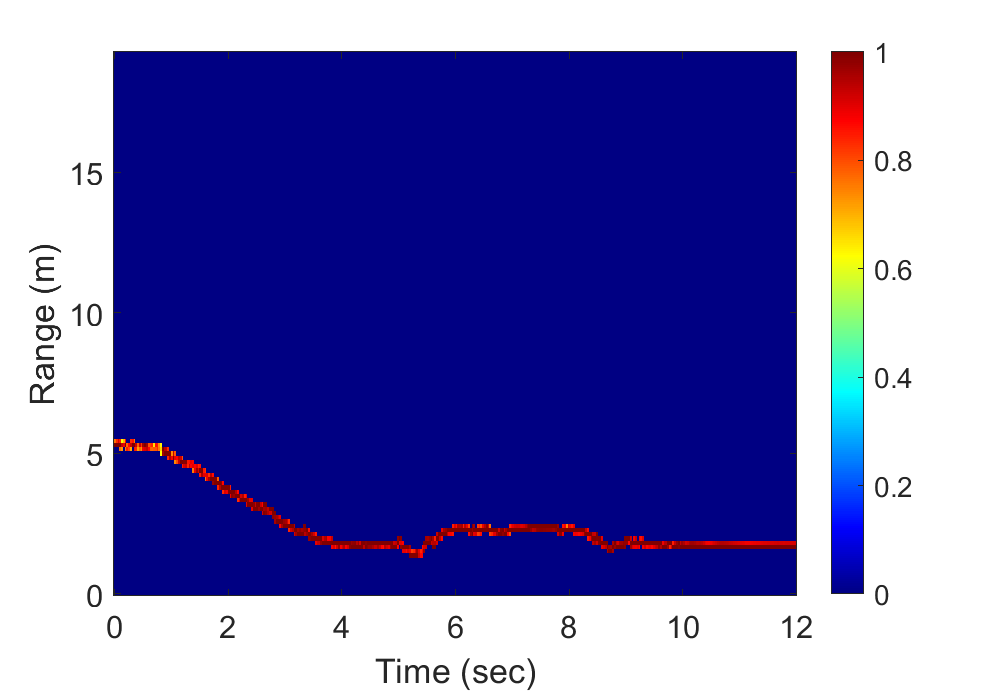}
		\caption{Range-map cleaning with kernel $ 1 $}
		\label{subfig:RM_5_kernel_1}
	\end{subfigure}
	~ 
	\begin{subfigure}[b]{0.49\textwidth}
		\includegraphics[width=\linewidth, frame]{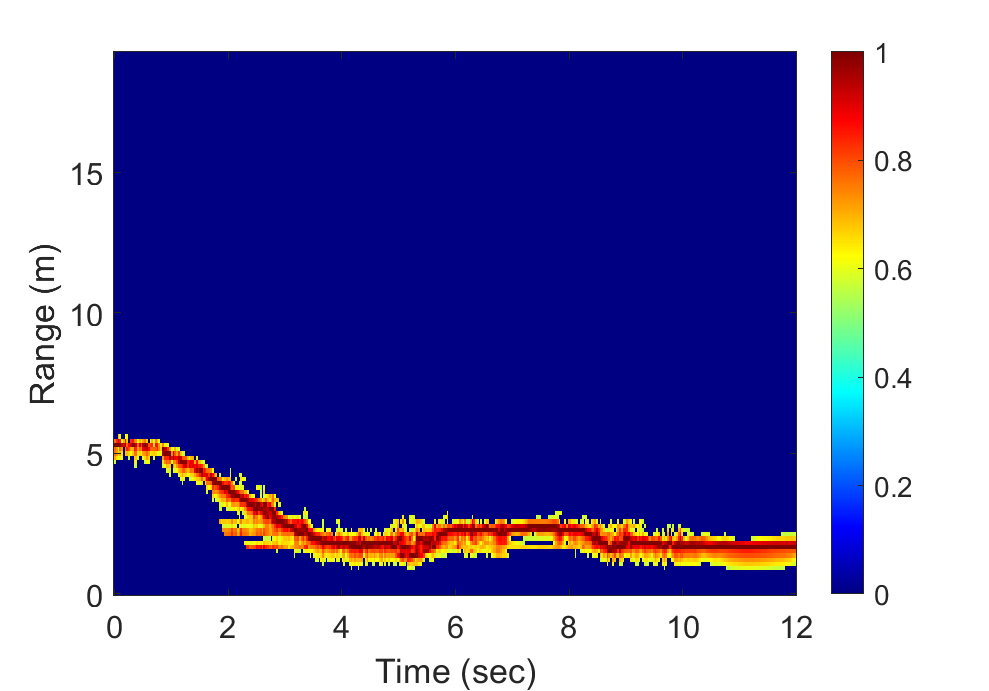}
		\caption{Range-map cleaning with kernel $ 12x12 $}
		\label{subfig:RM_5_kernel_12}
	\end{subfigure}
	\\
	\centering
	\begin{subfigure}[b]{0.49\textwidth}
		\includegraphics[width=\linewidth, frame]{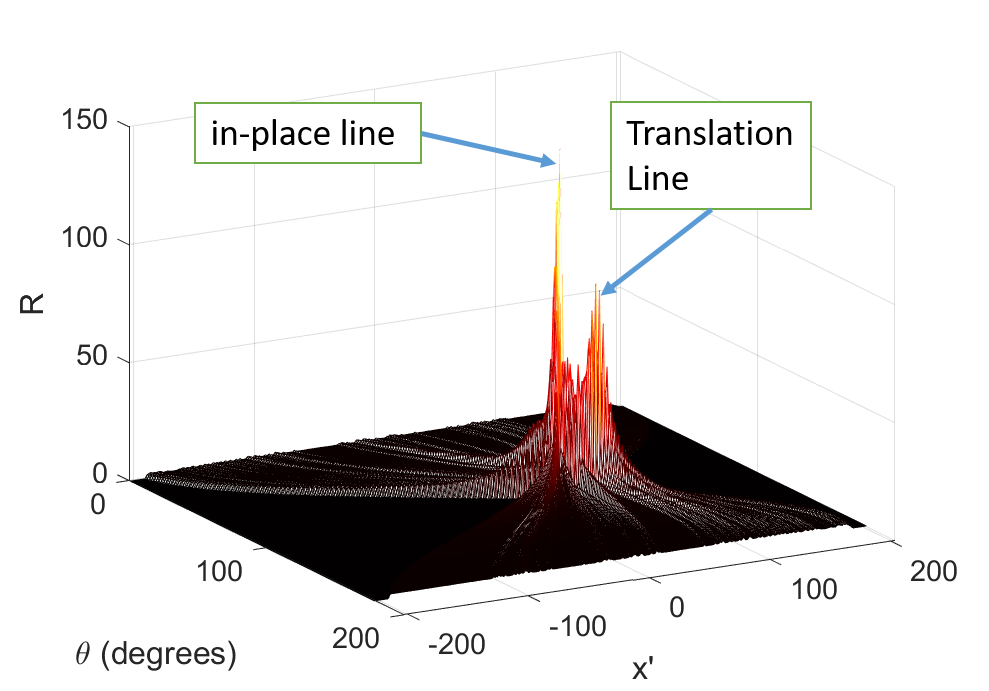}
		\caption{Radon transform on (a)}
		\label{subfig:Radon_5_kernel_1_2}
	\end{subfigure}
	~ 
	\begin{subfigure}[b]{0.49\textwidth}
		\includegraphics[width=\linewidth, frame]{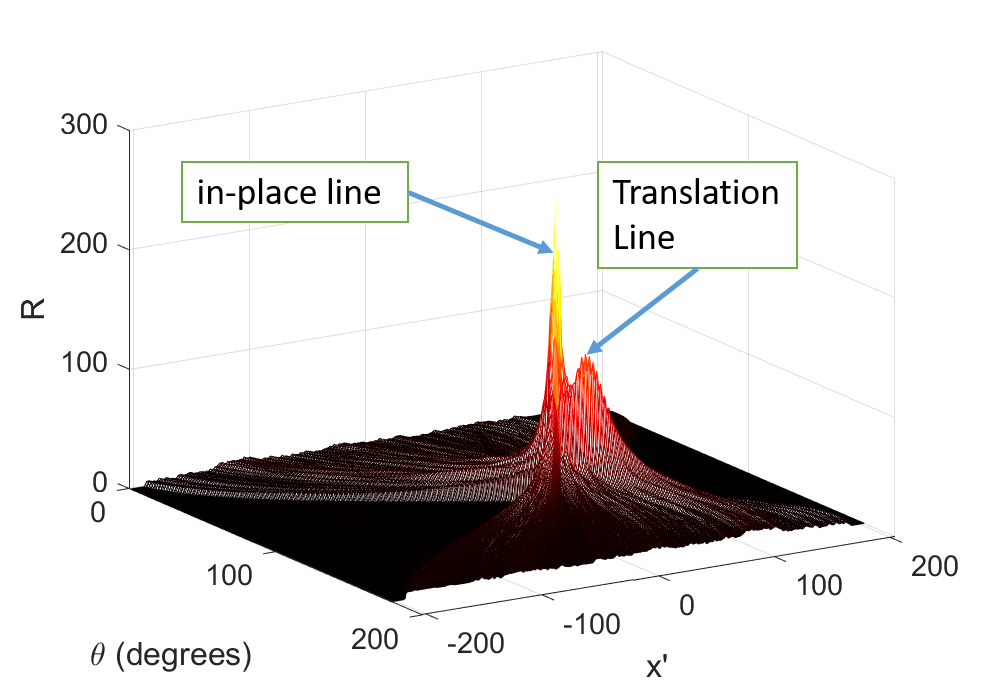}
		\caption{Radon transform on (b)}
		\label{subfig:Radon_5_kernel_12_2}
	\end{subfigure}
	\caption[The figure set shows the cleaned range-map by using a different kernel size with the Radon transform.]{The figure set shows (a) the cleaned range-map by using kernel size $ 1 $ and (b) shows the cleaned range-map with kernel size $ 12x12 $. (c) and (d) display the Radon transform on (a) and (b). \label{fig:RM_5_kernel_1_12}}
\end{figure}

The kernel size of $ 6x6 $ gives the best results and has been found iteratively. An example of choosing the kernel size to $ 1 $ and $ 12 $ are shown in Fig.~\ref{fig:RM_5_kernel_1_12}. According to the kernel size $ 1 $, a small kernel yields to sharp peaks in the Radon transform, whereas multiple peaks between the \textit{in-place Line} and the \textit{translation line} can cause confusion in finding the true peaks of the two lines. On the other hand, a too widely chosen kernel leads to a blurry and imprecise peak in the Radon transform.   

 
\subsection{Micro-Doppler spectrogram pre-processing\label{subsec:rangemappreproc}}
\begin{figure}[ht]
	\centering
	\begin{subfigure}[b]{0.49\textwidth}
		\includegraphics[width=\linewidth, frame]{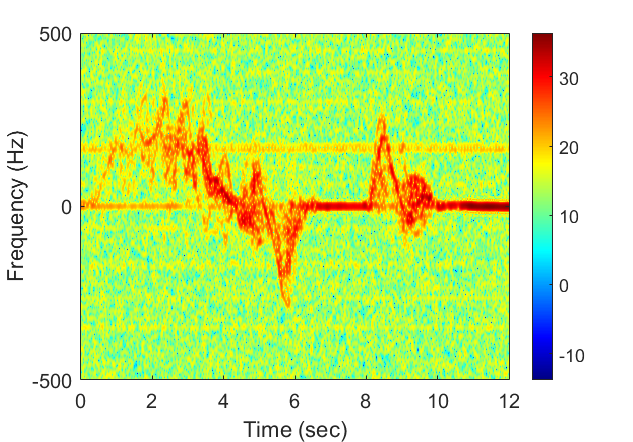}
		\caption{Raw micro-Doppler spectrogram}
		\label{subfig:MD_1_Raw}
	\end{subfigure}
	~ 
	\begin{subfigure}[b]{0.49\textwidth}
		\includegraphics[width=\linewidth, frame]{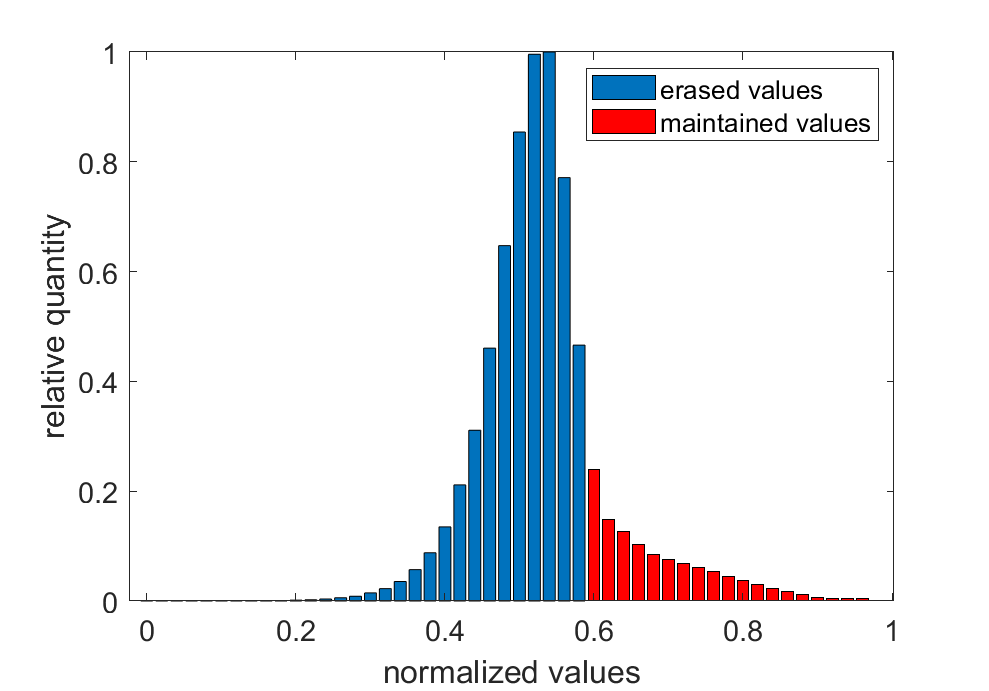}
		\caption{Histogram of the eCLEAN algorithm for the spectrogram}
		\label{subfig:eCLEAN_2_MD_histogram}
	\end{subfigure}
	 \\
	\centering
	\begin{subfigure}[b]{0.49\textwidth}
		\includegraphics[width=\linewidth, frame]{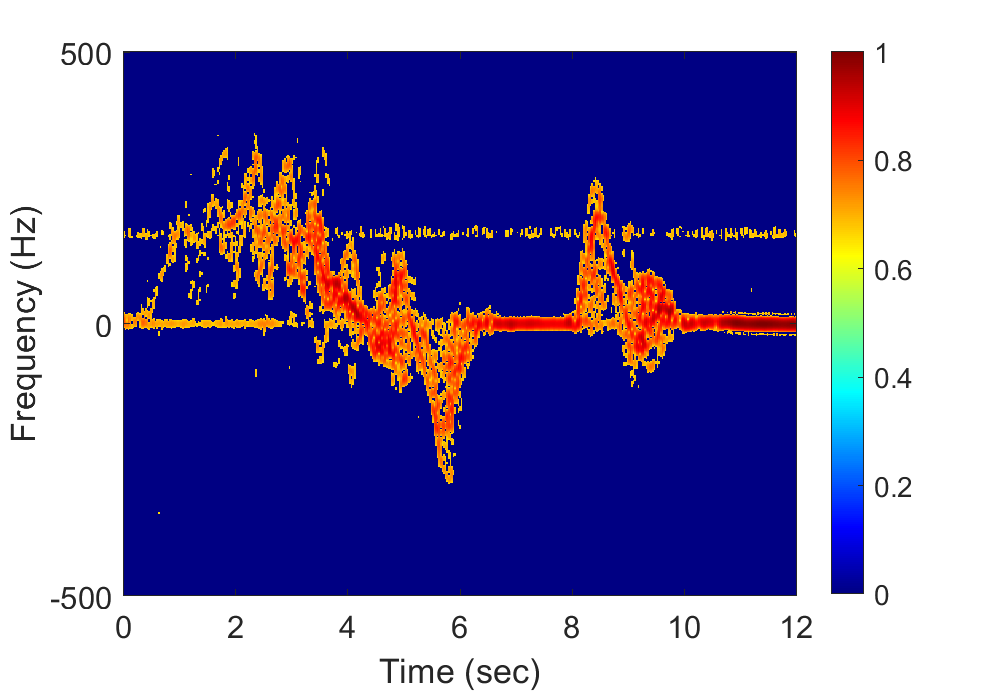}
		\caption{Micro-Doppler spectrogram after eCLEAN}
		\label{subfig:MD_2_eC}
	\end{subfigure}
	~ 
	\begin{subfigure}[b]{0.49\textwidth}
`		\includegraphics[width=\linewidth, frame]{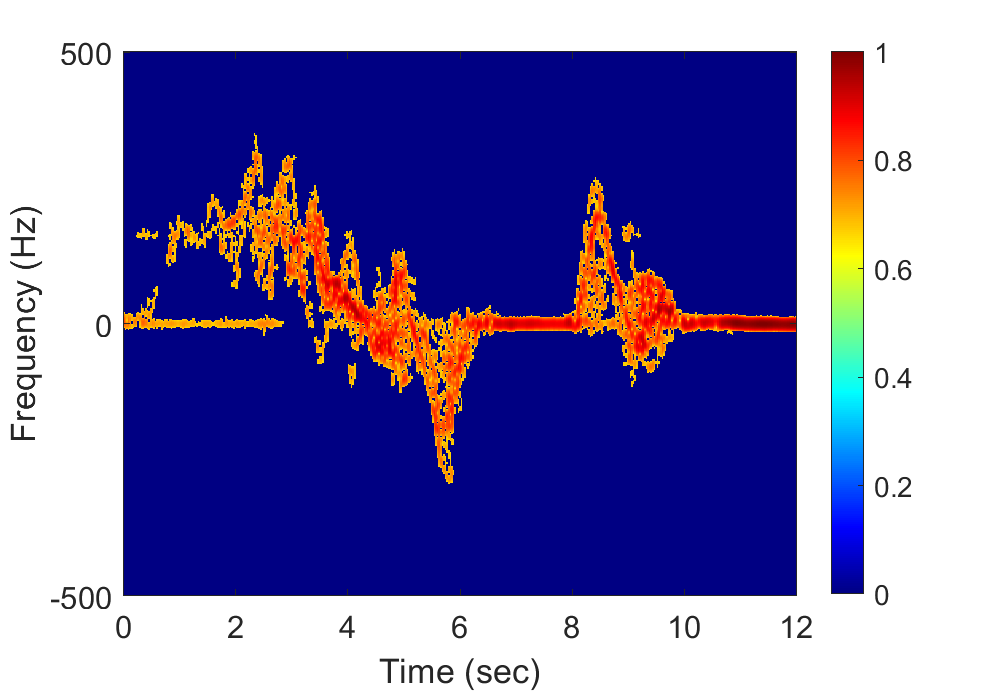}
		\caption{Micro-Doppler spectrogram after outlier removal}
		\label{subfig:MD_3_OLR}
	\end{subfigure}
	\caption[The figure set shows initial micro-Doppler spectrogram, the histogram of the \textit{eCLEAN} algorithm, spectrogram after the \textit{eCLEAN} algorithm, and after removing the outlier pixels.]{The figure set shows (a) initial micro-Doppler spectrogram, (b) shows the histogram values of the \textit{eCLEAN} algorithm. The micro-Doppler spectrogram after the \textit{eCLEAN} algorithm and after removing outlier pixels are shown in (c) and (d). \label{fig:MD_cleaning}}
\end{figure}

The outcome of the Eq.~\ref{eq:MDequation} provides the micro-Doppler spectrogram, $ MD $, by using the range-bins between $ 10 $ and $ 128 $. The spectrogram (Fig.~\ref{subfig:MD_1_Raw}) is scaled to a logarithmic scale as for the range-map and provides the initial image for the next processing step, namely, the \textit{eCLEAN} algorithm.

\subsubsection{eCLEAN\label{subsubsec:eclean_MD}} 
The \textit{eCLEAN} algorithm is also applied used for eliminating the majority of the noise in the spectrogram. The upper $ 60\% $ of the pixel values, according to Fig.~\ref{subfig:eCLEAN_2_MD_histogram}, are maintained to generate the spectrogram (Fig.~\ref{subfig:MD_2_eC}). It can be seen that some pixel clouds exist in the micro-Doppler spectrogram. Additionally, there occurs a horizontal line at approximately $ 200Hz $ Doppler frequency.

\subsubsection{Outlier removal\label{subsubsec:outlier_MD}}
The micro-Doppler spectrogram shown in Fig.~\ref{subfig:MD_2_eC} provides a clear motion sequence, whereas the horizontal line and the outlier points can adversely affect the classifier and the energy detector (PDC) of finding the onset and the offset times of a motion. Therefore, the outlier removal function with a default value of $ 150 $ pixel for the micro-Doppler spectrogram is applied. The result is shown in Fig.~\ref{subfig:MD_3_OLR}.

\subsection{Conclusion to pre-processing}
\begin{figure}[ht]
	\centering
	\includegraphics[width=\linewidth]{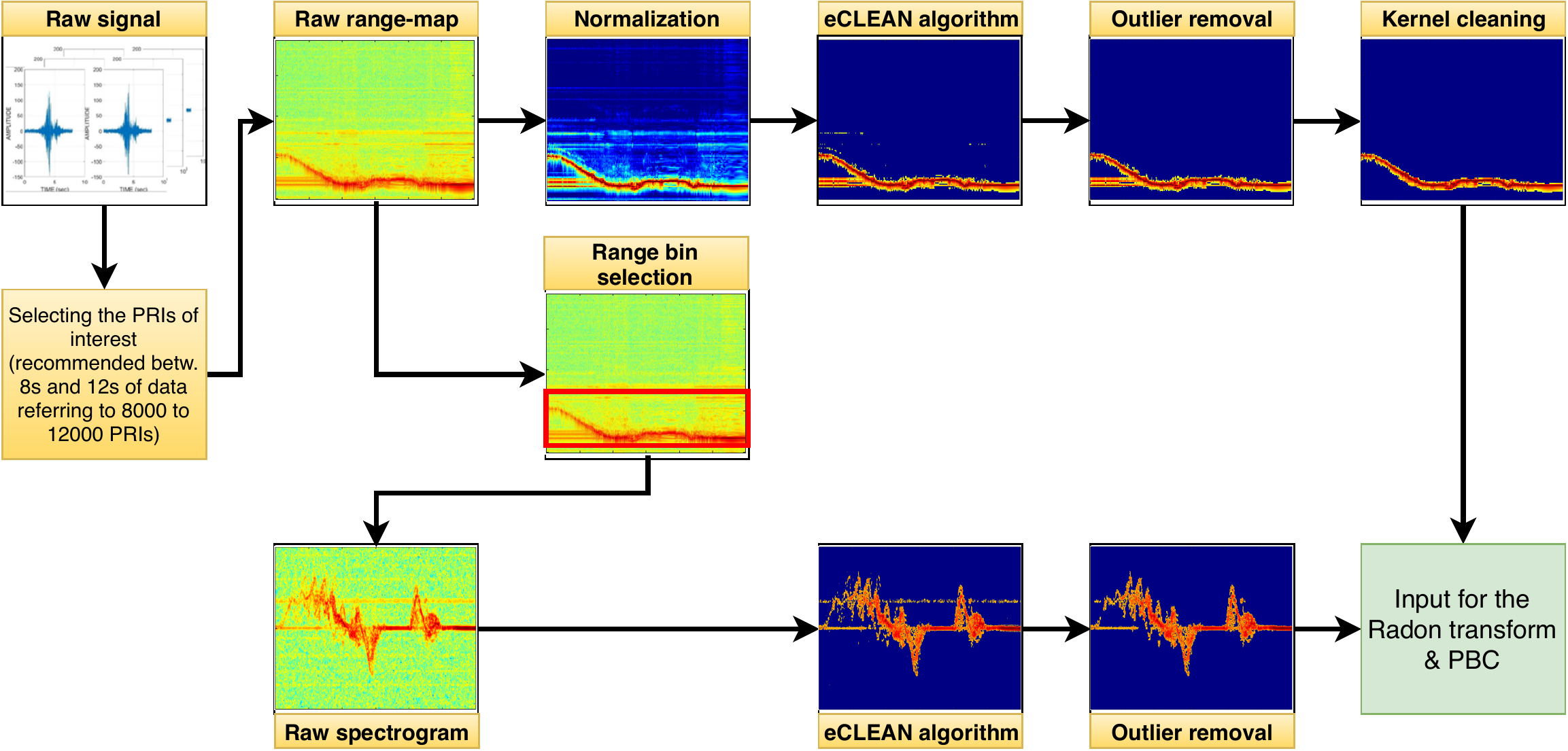}
	\caption{The flow diagram shows the pre-processing steps for the range-map and the micro-Doppler spectrogram. \label{fig:ProcessingSteps}}
\end{figure}
%
\noindent
In the previous sections, the individual pre-processing steps are explained. This section visualizes the individually dependent steps for the range-map and the spectrogram, with a flow chart shown in  Fig.~\ref{fig:ProcessingSteps}. The Ancortek Radar system provides the \textit{Raw signal}, which is used after selection of the PRIs of interest, to compute the \textit{Raw range-map}. The \textit{Raw range-map} is further processed by the \textit{Normalization}. 

From the \textit{Raw range-map}, the  range bins of interest are selected to process the \textit{Raw spectrogram}. The noise elimination by using the \textit{eCLEAN}  algorithm is performed on both the range-map and the spectrogram. The same applies for the \textit{Outlier removal} function. The \textit{Kernel cleaning} is only applied on the range-map. The final range-map and spectrogram are used for further motion separation techniques through the Radon transform and the PBC, explained in Chapter~\ref{chap:radon_PBC}.

\section{Feature extraction and classification\label{sec:featureExtr}}
This section introduces the reader to the used feature extraction method of the Two-Dimensional Principal Component Analysis (2-D~PCA). The extracted features are classified with the Nearest Neighbor classifier, while we illustrate the Fusion method of classifying spectrogram and range features, simultaneously. Finally in this section, reconstructed micro-Doppler and range-map images are visualizing the importance of using just few principal component vectors. 

\subsection{Two-dimensional principal component analysis and nearest neighbor classification}
The Two-Dimensional Principal Component Analysis (2-D~PCA) is used for feature extraction, followed by the Nearest Neighbour (NN) classifier. The covariance matrix $ H $ is computed as, 
\begin{equation}\label{ImageCovMatr}
H = \frac{1}{I} \displaystyle\sum_{i=1}^{I} (X^{(i)}-\bar{X})^T\cdot(X^{(i)} -\bar{X})
\end{equation}
\noindent
where $\bar{X} \in \mathbb{R}^{\eta \times \eta}$  is the mean image, as $\bar{X}=\frac{1}{I}\sum_{i=1}^I X^{(i)}$. $I$ is the total number of images in the training data. In the above equitation, $X^{(i)} \in \mathbb{R}^{\eta \times \eta}$ is the $i$-th micro-Doppler or range-map image, computed form $MD(n,k)$ or $R(p,m)$, respectively. From the eigendecomposition of $ H $, the eigenvalues ($\lambda_i$) and eigenvectors ($\nu_i$)  are extracted, such that $ J(\Phi)=\Phi^T H \Phi $. The eigenvectors corresponding to the  $d$ largest eigenvalues form the matrix $\Phi = [\nu_1, \nu_2, ..., \nu_d$]. The default setting was $d_{MD}= 14$ and $d_{RM}=4$ for micro-Doppler and range-map, respectively. The individual training images $X^{(i)}$ are projected onto  the $d$-dimensional subspace matrix to compute the principal component matrix as,
\begin{equation}\label{eq:2dPCA}
Y = X\Phi
\end{equation} 
The micro-Doppler image, $Y_{MD}$,  is of  dimension $\mathbb{R}^{\eta \times d_{MD}}$, and the range-map image, $Y_{RM}$, is of dimension $\mathbb{R}^{\eta \times d_{RM}}$. 

\begin{figure}[ht]
	\centering
	\includegraphics[width=\linewidth]{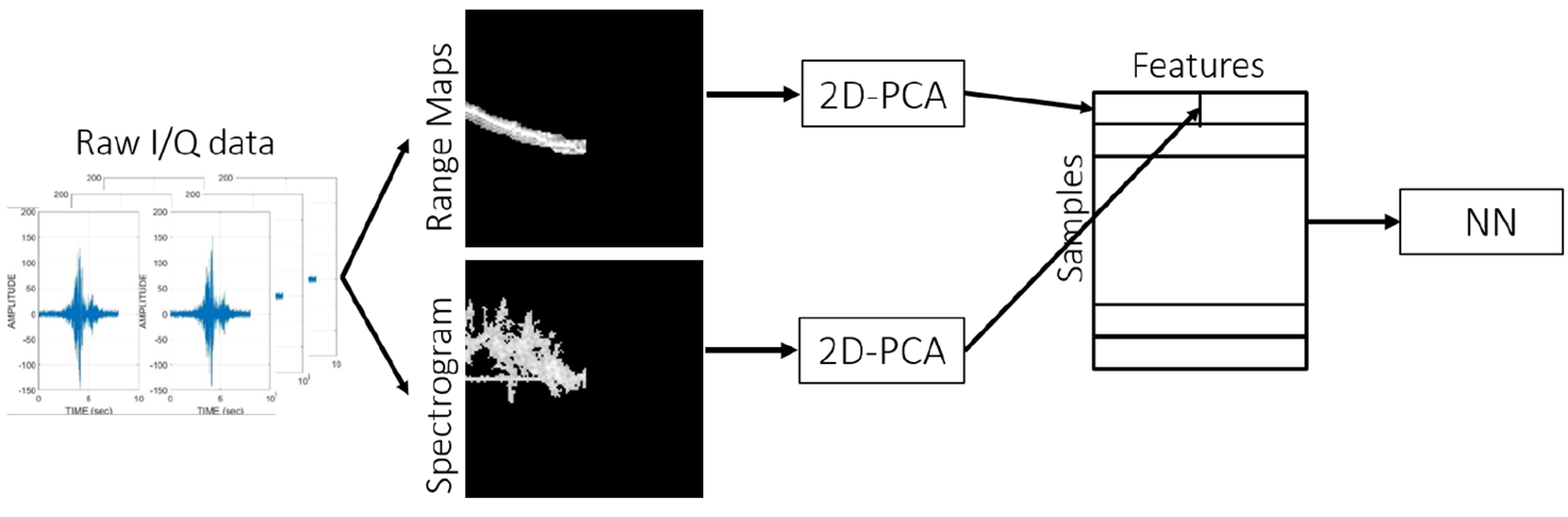}
	\caption[The vectorized principal component matrices of the micro-Doppler and range-map image form the  principal component vectors for fusion classification]{The vectorized principal component matrices of the micro-Doppler and range-map image form the  principal component vectors for fusion classification~ \cite{baris:FusionPCA8835840}.} \label{fig:fusionGraph}
\end{figure}

The individual test images are projected using the same procedure to provide the feature matrix $Y_{MD}(Test)$ and $Y_{RM}(Test)$. 
The NN classifier operates on the fused vectorized and concatenated micro-Doppler and range-map feature vectors, as shown in Fig.~\ref{fig:fusionGraph}.

\noindent
According to the run-time and classification rate, the 2-D~PCA combined with a NN classifier is comparable to other state-of-the-art classification methods 
\cite{baris:FusionPCA8835840, baris:GANbased:8835589}. 
As stated in the introduction, this thesis does not seek to provide new classification methodology, but rather defines a framework to exploit logical motion sequences \cite{aminGuendel:IET}. It is important to note that some work on PCA only uses a specific principal component that best characterize the class distinctions \cite{Hoorfar:4815956}. We did not sort the principal components according to their relevance, and used all of them.  
%

\subsection{Visualization of reconstructing images\label{subsec:visualPCA}}
\begin{figure}[tbph]
	\centering
	\includegraphics[width=0.6\linewidth]{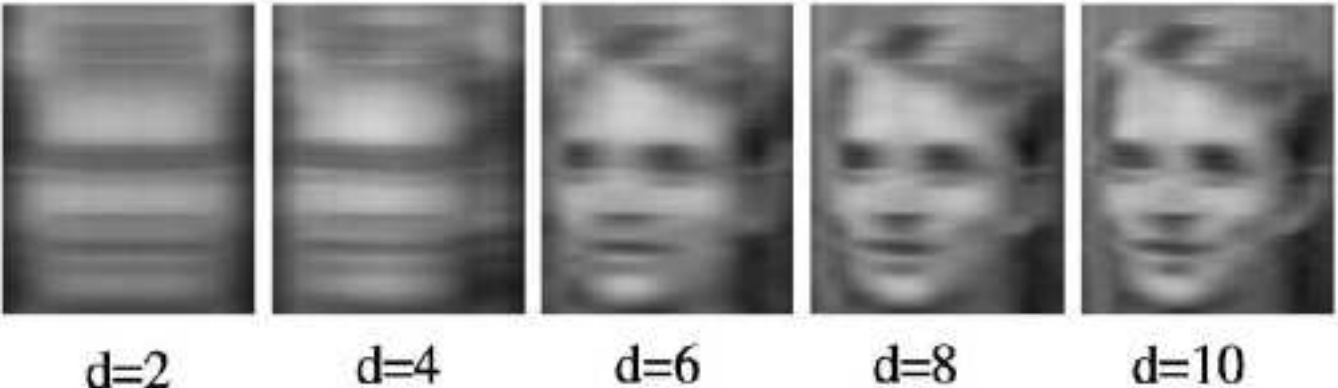}
	\caption[Shows the effect by using different numbers of the first eigenvectors]{Shows the effect by using different numbers of the first eigenvectors $ d $~\cite{chihaoui2016survey}.}
	\label{fig:2dpcafaces}
\end{figure}
Principal analysis is a powerful tool for feature extraction. However, it is challenging to determine the proper number of principal values. Therefore, paper \cite{chihaoui2016survey} has investigated reconstructing faces for face detection by using, e.g., 2D-PCA. It is shown in Fig.~\ref{fig:2dpcafaces} that the reconstructed faces provide acceptable results after $ d=6 $ principal values. 
%
\begin{figure}[htb]
	\begin{center}
		\begin{minipage}[t]{.13\linewidth}
			\centering
			{\footnotesize ~~~~~~Orig. Im.}\\
			\rotatebox[origin=l]{90}{{\small ~Freq. (Hz)}}\includegraphics[width=\linewidth]{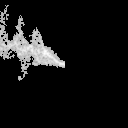}
			\\{\small~~~~~Time (sec)\\~~~~~~~~(a)}\label{subfig:ImV_MD_orig}
		\end{minipage}
		~~~
		\begin{minipage}[t]{.13\linewidth}
			\centering
			{\footnotesize ~~~~~~~~$d_{MD}$=1}\\
			\rotatebox[origin=l]{90}{{\small ~Freq. (Hz)}}\includegraphics[width=\linewidth]{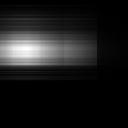}
			\\{\small~~~~~Time (sec)\\~~~~~~~~(b)}\label{subfig:ImV_MD_1}
		\end{minipage}
		~~~
		\begin{minipage}[t]{.13\linewidth}
			\centering
			{\footnotesize ~~~~~~~~$d_{MD}$=2}\\
			\rotatebox[origin=l]{90}{{\small ~Freq. (Hz)}}\includegraphics[width=\linewidth]{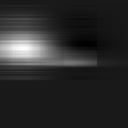}
			\\{\small~~~~~Time (sec)\\~~~~~~~~(c)}\label{subfig:ImV_MD_2}
		\end{minipage}
		~~~
		\begin{minipage}[t]{.13\linewidth}
			\centering
			{\footnotesize ~~~~~~~~$d_{MD}$=3}\\
			\rotatebox[origin=l]{90}{{\small ~Freq. (Hz)}}\includegraphics[width=\linewidth]{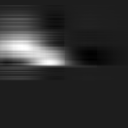}
			\\{\small~~~~~Time (sec)\\~~~~~~~~(d)}\label{subfig:ImV_MD_3}
		\end{minipage}
		~~~
		\begin{minipage}[t]{.13\linewidth}
			\centering
			{\footnotesize ~~~~~~~~$d_{MD}$=4}\\
			\rotatebox[origin=l]{90}{{\small ~Freq. (Hz)}}\includegraphics[width=\linewidth]{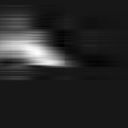}
			\\{\small~~~~~Time (sec)\\~~~~~~~~(e)}\label{subfig:ImV_MD_4}
		\end{minipage}
		~~~
		\begin{minipage}[t]{.13\linewidth}
			\centering
			{\footnotesize ~~~~~~~~$d_{MD}$=6}\\
			\rotatebox[origin=l]{90}{{\small ~Freq. (Hz)}}\includegraphics[width=\linewidth]{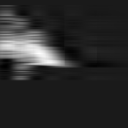}
			\\{\small~~~~~Time (sec)\\~~~~~~~~(f)}\label{subfig:ImV_MD_6}
		\end{minipage}
		\\
		\vspace{.3cm}	
		\begin{minipage}[t]{.13\linewidth}
			\centering
			{\footnotesize ~~~~~~~~$d_{MD}$=10}\\
			\rotatebox[origin=l]{90}{{\small ~Freq. (Hz)}}\includegraphics[width=\linewidth]{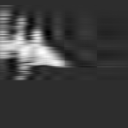}
			\\{\small~~~~~Time (sec)\\~~~~~~~~(g)}\label{subfig:ImV_MD_10}
		\end{minipage}
		~~~
		\begin{minipage}[t]{.13\linewidth}
			\centering
			{\footnotesize ~~~~~~~~$d_{MD}$=14}\\
			\rotatebox[origin=l]{90}{{\small ~Freq. (Hz)}}\includegraphics[width=\linewidth]{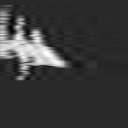}
			\\{\small~~~~~Time (sec)\\~~~~~~~~(h)}\label{subfig:ImV_MD_14}
		\end{minipage}
		~~~
		\begin{minipage}[t]{.13\linewidth}
			\centering
			{\footnotesize ~~~~~~~~$d_{MD}$=18}\\
			\rotatebox[origin=l]{90}{{\small ~Freq. (Hz)}}\includegraphics[width=\linewidth]{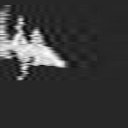}
			\\{\small~~~~~Time (sec)\\~~~~~~~~(i)}\label{subfig:ImV_MD_18}
		\end{minipage}
		~~~
		\begin{minipage}[t]{.13\linewidth}
			\centering
			{\footnotesize ~~~~~~~~$d_{MD}$=22}\\
			\rotatebox[origin=l]{90}{{\small ~Freq. (Hz)}}\includegraphics[width=\linewidth]{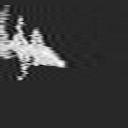}
			\\{\small~~~~~Time (sec)\\~~~~~~~~(j)}\label{subfig:ImV_MD_22}
		\end{minipage}
		~~~
		\begin{minipage}[t]{.13\linewidth}
			\centering
			{\footnotesize ~~~~~~~~$d_{MD}$=50}\\
			\rotatebox[origin=l]{90}{{\small ~Freq. (Hz)}}\includegraphics[width=\linewidth]{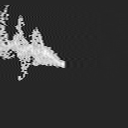}
			\\{\small~~~~~Time (sec)\\~~~~~~~~(k)}\label{subfig:ImV_MD_50}
		\end{minipage}
		~~~
		\begin{minipage}[t]{.13\linewidth}
			\centering
			{\footnotesize ~~~~~~~~$d_{MD}$=128}\\
			\rotatebox[origin=l]{90}{{\small ~Freq. (Hz)}}\includegraphics[width=\linewidth]{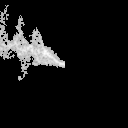}
			\\{\small~~~~~Time (sec)\\~~~~~~~~(l)}\label{subfig:ImV_MD_128}
		\end{minipage}
	\end{center}
	\caption[The images show the reconstructed micro-Doppler images by using different numbers of eigenvectors.]{The images show the reconstructed micro-Doppler images by using different numbers of eigenvectors. (a) shows the original image for comparison.\label{fig:ImV_MD}}
\end{figure} 

\begin{figure}[htb]
	\begin{center}
		\begin{minipage}[t]{.13\linewidth}
			\centering
			{\footnotesize ~~~~~~Orig. Im.}\\
			\rotatebox[origin=l]{90}{{\small ~Range (m)}}\includegraphics[width=\linewidth]{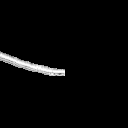}
			\\{\small~~~~~Time (sec)\\~~~~~~~~(a)}\label{subfig:ImV_RM_orig}
		\end{minipage}
		~~~
		\begin{minipage}[t]{.13\linewidth}
			\centering
			{\footnotesize ~~~~~~~~$d_{RM}$=1}\\
			\rotatebox[origin=l]{90}{{\small ~Range (m)}}\includegraphics[width=\linewidth]{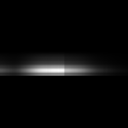}
			\\{\small~~~~~Time (sec)\\~~~~~~~~(b)}\label{subfig:ImV_RM_1}
		\end{minipage}
		~~~
		\begin{minipage}[t]{.13\linewidth}
			\centering
			{\footnotesize ~~~~~~~~$d_{RM}$=2}\\
			\rotatebox[origin=l]{90}{{\small ~Range (m)}}\includegraphics[width=\linewidth]{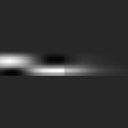}
			\\{\small~~~~~Time (sec)\\~~~~~~~~(c)}\label{subfig:ImV_RM_2}
		\end{minipage}
		~~~
		\begin{minipage}[t]{.13\linewidth}
			\centering
			{\footnotesize ~~~~~~~~$d_{RM}$=3}\\
			\rotatebox[origin=l]{90}{{\small ~Range (m)}}\includegraphics[width=\linewidth]{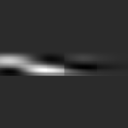}
			\\{\small~~~~~Time (sec)\\~~~~~~~~(d)}\label{subfig:ImV_RM_3}
		\end{minipage}
		~~~
		\begin{minipage}[t]{.13\linewidth}
			\centering
			{\footnotesize ~~~~~~~~$d_{RM}$=4}\\
			\rotatebox[origin=l]{90}{{\small ~Range (m)}}\includegraphics[width=\linewidth]{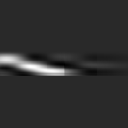}
			\\{\small~~~~~Time (sec)\\~~~~~~~~(e)}\label{subfig:ImV_RM_4}
		\end{minipage}
		~~~
		\begin{minipage}[t]{.13\linewidth}
			\centering
			{\footnotesize ~~~~~~~~$d_{RM}$=6}\\
			\rotatebox[origin=l]{90}{{\small ~Range (m)}}\includegraphics[width=\linewidth]{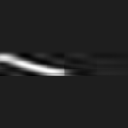}
			\\{\small~~~~~Time (sec)\\~~~~~~~~(f)}\label{subfig:ImV_RM_6}
		\end{minipage}
		\\
		\vspace{.3cm}	
		\begin{minipage}[t]{.13\linewidth}
			\centering
			{\footnotesize ~~~~~~~~$d_{RM}$=10}\\
			\rotatebox[origin=l]{90}{{\small ~Range (m)}}\includegraphics[width=\linewidth]{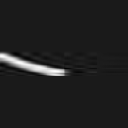}
			\\{\small~~~~~Time (sec)\\~~~~~~~~(g)}\label{subfig:ImV_RM_10}
		\end{minipage}
		~~~
		\begin{minipage}[t]{.13\linewidth}
			\centering
			{\footnotesize ~~~~~~~~$d_{RM}$=14}\\
			\rotatebox[origin=l]{90}{{\small ~Range (m)}}\includegraphics[width=\linewidth]{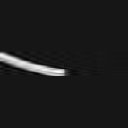}
			\\{\small~~~~~Time (sec)\\~~~~~~~~(h)}\label{subfig:ImV_RM_14}
		\end{minipage}
		~~~
		\begin{minipage}[t]{.13\linewidth}
			\centering
			{\footnotesize ~~~~~~~~$d_{RM}$=18}\\
			\rotatebox[origin=l]{90}{{\small ~Range (m)}}\includegraphics[width=\linewidth]{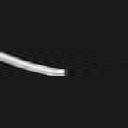}
			\\{\small~~~~~Time (sec)\\~~~~~~~~(i)}\label{subfig:ImV_RM_18}
		\end{minipage}
		~~~
		\begin{minipage}[t]{.13\linewidth}
			\centering
			{\footnotesize ~~~~~~~~$d_{RM}$=22}\\
			\rotatebox[origin=l]{90}{{\small ~Range (m)}}\includegraphics[width=\linewidth]{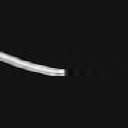}
			\\{\small~~~~~Time (sec)\\~~~~~~~~(j)}\label{subfig:ImV_RM_22}
		\end{minipage}
		~~~
		\begin{minipage}[t]{.13\linewidth}
			\centering
			{\footnotesize ~~~~~~~~$d_{RM}$=50}\\
			\rotatebox[origin=l]{90}{{\small ~Range (m)}}\includegraphics[width=\linewidth]{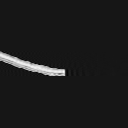}
			\\{\small~~~~~Time (sec)\\~~~~~~~~(k)}\label{subfig:ImV_RM_50}
		\end{minipage}
		~~~
		\begin{minipage}[t]{.13\linewidth}
			\centering
			{\footnotesize ~~~~~~~~$d_{RM}$=128}\\
			\rotatebox[origin=l]{90}{{\small ~Range (m)}}\includegraphics[width=\linewidth]{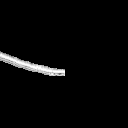}
			\\{\small~~~~~Time (sec)\\~~~~~~~~(l)}\label{subfig:ImV_RM_128}
		\end{minipage}
	\end{center}
	\caption[The images show the reconstructed range-maps by using different numbers of eigenvectors.]{The images show the reconstructed range-maps by using different numbers of eigenvectors. (a) shows the original image for comparison.\label{fig:ImV_RM}}
\end{figure} 
%
The same reconstructing process has been applied for a walking motion for a micro-Doppler image and the corresponding range-map image. The original images are shown in Fig.~\ref{fig:ImV_MD}{\color{blue}.a} and Fig.~\ref{fig:ImV_RM}{\color{blue}.a}.  
Typically, we compute the principal component matrix by using Eq.~\ref{eq:2dPCA} ($ Y = X\Phi $), where $ X $ is the image and $ \Phi $ is the projection matrix formed from the eigenvectors corresponding to the $ d $ largest eigenvalues. The same equation is used to reconstruct image $ \hat{X} $ as, 
\begin{subequations}
	\begin{align} \label{eq:pcaReconstr_1}
		\hat{X} 	&=   Y\Phi^+	\\[.15in]
		\label{eq:pcaReconstr_2}
		\Phi^+ 		&= 	(\Phi^T \Phi)^{-1} \Phi^T \\[.15in]
		\label{eq:pcaReconstr_3}
		\hat{X}  	&=  Y (\Phi^T \Phi)^{-1} \Phi^T
	\end{align}
\end{subequations}
where $ \Phi^+ $ is is the pseudo inverse matrix of $ \Phi $. Eq.~\ref{eq:pcaReconstr_3} is used to recompute the images in Fig.~\ref{fig:ImV_MD} and Fig.~\ref{fig:ImV_RM}.

The micro-Doppler images in Fig.~\ref{fig:ImV_MD} show an acceptable reconstructed micro-Doppler image  by using $ d_{MD} = 10 $, where the spikes from the arms and the legs become visible. This confirms also the choice default value of $ d_{MD} = 14 $, which have been found determined based on the highest classification rate by using the k-NN. The same applies for the range-map images in Fig.~\ref{fig:ImV_RM}, where the default value is $ d_{RM} = 4 $. By using $ d_{RM} = 6 $, a nearly full reconstructed range-map can be observed.

\chapter{Motion capturing by the Radon transform and the energy detection\label{chap:radon_PBC}}

%
The Radon transform is considered an effective tool in detecting dominant contour schemes in images, especially in medical image processing, whereas the inverse Radon transform is typically used for reconstructing images from medical CT scans \cite{29_wininger2013basis}. In the underlying problem, we apply the Radon transform to the range-map of the radar signal returns to detect pertinent line structures.
The Radon transform, instead of a motion tracker \cite{book:principleOfModernRadar}, can simply reveal the transitions from translation to in-place motion and visa versa by capturing the "breaking" points, or time instants of changing slopes.

In employing the Radon transform for human motion recognition, we recognize that horizontal lines in the range-map correspond to in-place motions with no noticeable range extent, whereas lines with non-zero inclinations represent continuous changes in range gates stemming from motion translations, such as walking. It may be expected that acceleration or deceleration gives rise to curvy signatures in the range-map which warrant the use of the Hough  transform \cite{30_4599159}. However, these actions cause very minor changes in the values of the slopes or in the intersection points. Examples of in-place motions are sitting down, standing up from sitting, or bending while standing as well as from a sitting position. 
Successful separations between translation and in-place motion categories helps in identifying the human "actions" vs "resting" modes at home. Information in the range-map alone is sufficient to quantify the human activity as one of the two aforementioned categories. In this respect, any follow-on classification of human motions using spectrograms for micro-Doppler signature estimation can proceed without confusing one category with another. For example, classifiers applied to time intervals of in-place motions will not have "walking" as a possible outcome \cite{aminGuendel:IET}. 

\section{Application of the Radon transform}
\label{subsec:radonTrans}
\begin{figure}[ht]
	\centering
	\includegraphics[width=0.9\linewidth]{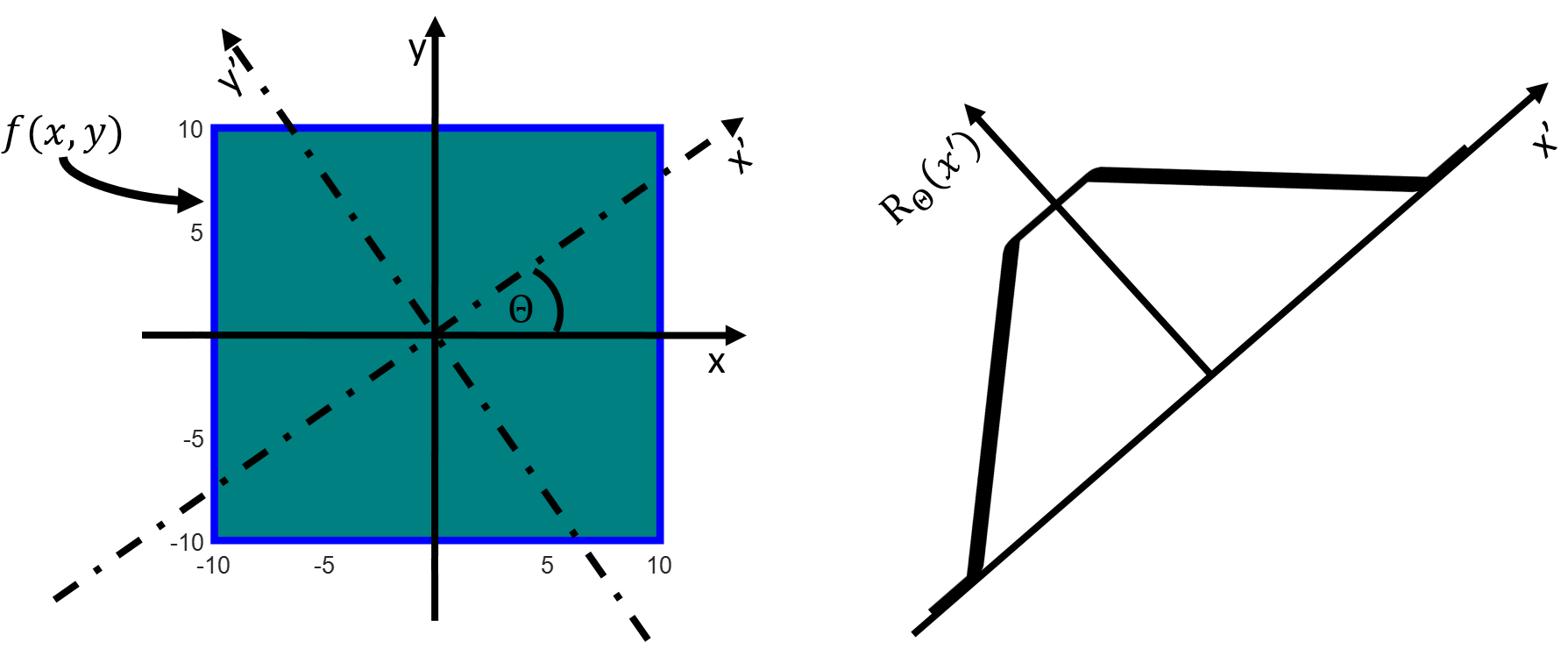}
	\vspace{.5em}
	\caption{Radon transform operation. \label{fig:radontransExampl}}
\end{figure}

The Radon transform converts a Cartesian coordinate system $ (x,y) $ to an angle and a distance from a center point representation $ (\theta, x') $. It integrates the pixel values over  $ y' $, with each pixel projected onto the new subspace abscissa $ x' $. The projection is described as, $ x' = x\cdot cos\theta + y\cdot sin\theta $. The Radon transform is demonstrated in Fig.~\ref{fig:radontransExampl}. The computation of $R_\theta$ over the coordinate $y'$ is given by,

\begin{equation}
\label{eq:RadonEquation4}
{R_\theta }(x') = \int\limits_{ - \infty }^\infty  {B(x'\cos \theta  - y'\sin \theta ,x'\sin \theta  + y'\cos \theta )dy'} 
\end{equation}

\begin{figure}[ht]
	\includegraphics[width=0.99\linewidth]{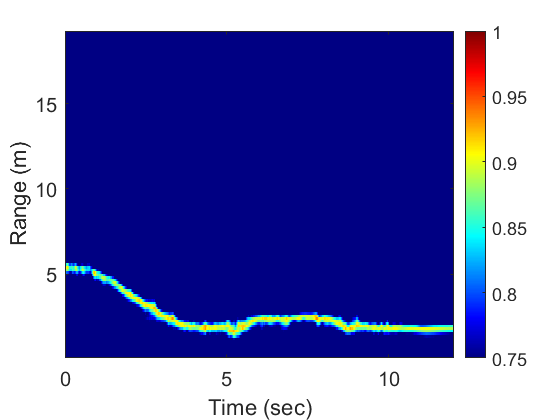}
	\caption{Range-map for the motion sequence.\label{subfig:RM4}}
\end{figure}

\noindent
where we consider the range-map as an image with each sample converted to a decibel absolute value. This image, provided by Eq.~\ref{eq:STFT:RM}, is of size $ M \times N = 256 \times 12,000 $ (for $12sec$), where $ M $ and $ N $ represent the number of range bins (rows) and slow-times (columns), respectively. Image resizing is performed by uniform sub-sampling over range and slow-time to produce a compressed image size, referred to as $ B(m,n)$,  of dimension $ 128 \times 384 $. To improve the end result, the down-sampled image $ B(m,n)$ is filtered with a ($ 3\times 3 $) smoothing kernel of unit value coefficients. Fig.~\ref{subfig:RM4} shows an example of the resized, filtered, and thresholded range-map where the person walks, then assumes two consecutive in-place motions, namely, sitting down and standing up. 

\begin{figure}[ht] 
	\includegraphics[width=\linewidth]{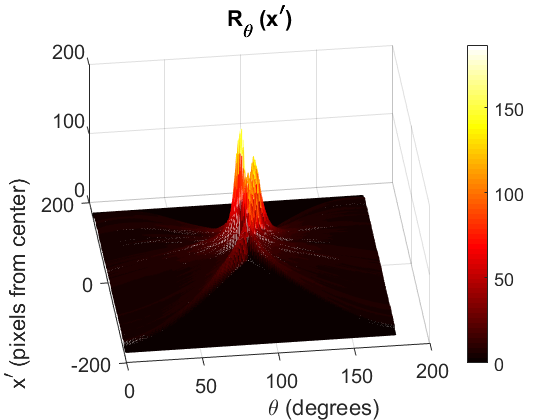}
	\caption{Radon transform matrix $R_\theta(x')$, 3D representation.}
	\label{subfig:RT1}
\end{figure}

\begin{figure}[ht]
	\includegraphics[width=0.99\linewidth]{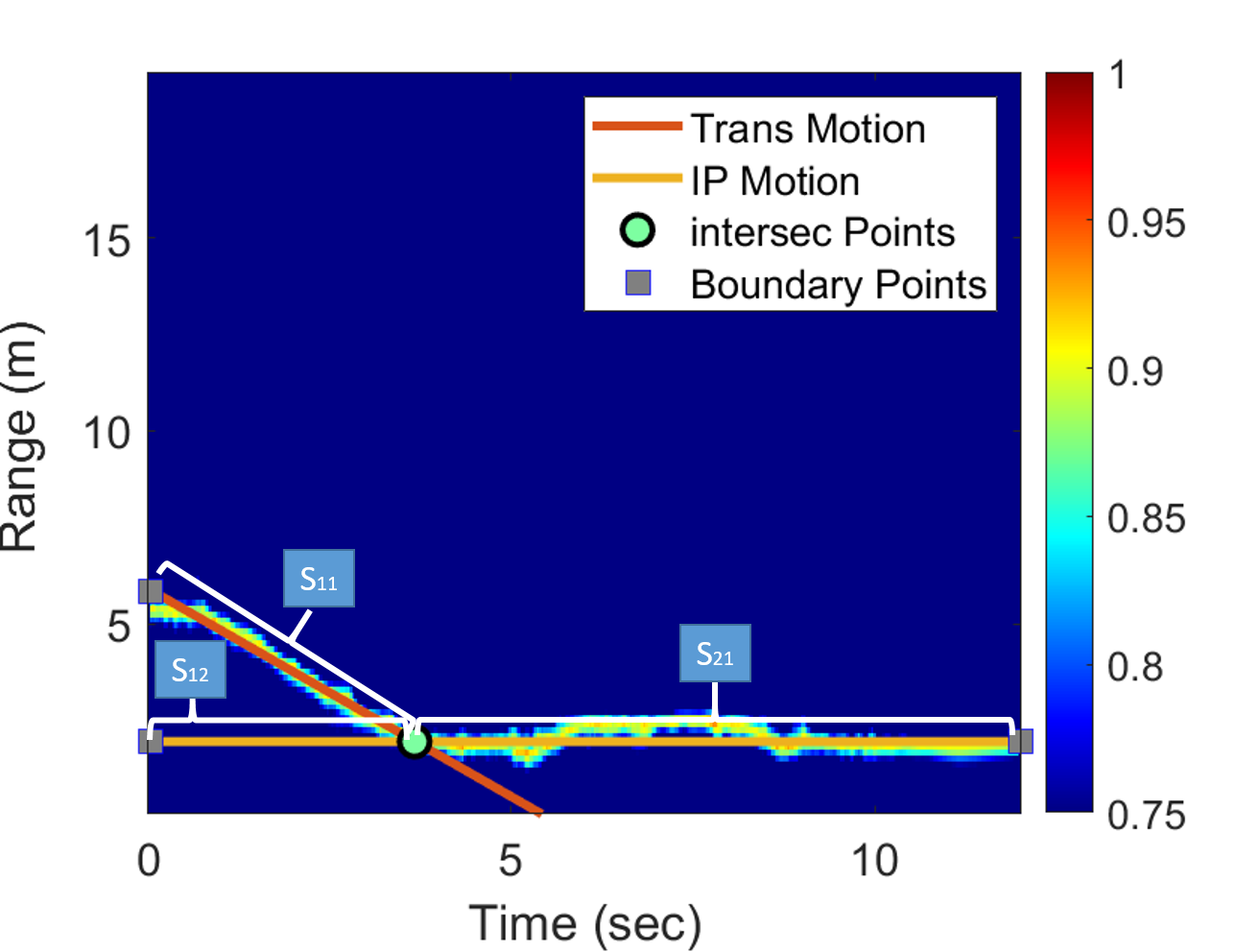}
	\caption{Range lines and intersection point.\label{subfig:RM_Lines}}
\end{figure}

\noindent
Fig.~\ref{subfig:RT1} depicts the Radon transform applied to the range-map of Fig.~\ref{subfig:RM4}, while the angle is incremented in $1^\circ$. One peak at $\theta = 90^\circ$ angle is shown, which refers to an in-place motion. The other peak off $\theta = 90^\circ$ refers to a translation motion. The peak location defines both the angle $\theta$ and a distance $ x'$ from the origin.


The intersection points between the two corresponding lines, indexed by $k$, is given by,
\begin{equation}
\label{eq:intersecEquation}
\left[ {\begin{array}{*{20}{c}}
	{m(k)}&{ - 1}\\
	{m(k + 1)}&{ - 1}
	\end{array}} \right] \cdot \left[ {\begin{array}{*{20}{c}}
	x\\
	{y}
	\end{array}} \right] = \left[ {\begin{array}{*{20}{c}}
	{ - n(k)}\\
	{ - n(k + 1)}
	\end{array}} \right],k = 1,..,K - 1
\end{equation} 
\noindent with, 
\begin{subequations}
	\begin{align} \label{eq:y_m_n_computing}
		y 		&= 	m(k)\cdot x+n(k)\\[.15in]
		\label{eq:m_computing}
		m(k) 	&= 	tan(\frac{\pi}{2} - \theta(k))=cot(\theta(k))\\[.15in]
		\label{eq:n_computing}
		n(k) 	&=  \frac{x'}{sin(\theta(k))}.
	\end{align}
\end{subequations}

\noindent
Fig.~\ref{subfig:RM_Lines} shows the two lines detected by applying  the Radon transform to Fig.~\ref{subfig:RM4} as well as their intersection point obtained by solving the above equations. 
The beginning crossing points with the image boundary at $t_0 = 0s$ are computed by setting $x$ to $0$ in Eq.~\ref{eq:y_m_n_computing}. The only end boundary point at $t_{max} = 12s$ is solved by stetting $x$ to  $t_{max}$. From the two boundary points at $t_0$ to the intersection point at $t = 3.6s$, there are two line segments, shown in Fig.~\ref{subfig:RM_Lines} as $s_{11}$ and $s_{12}$. One of these segments is the ground truth, whereas the other is false. In order to find the ground truth, we compute the normalized signal energy along each segment and use the line of the higher value \cite{Bresenham:1965:ACC:1663347.1663349}. The process is not performed for line segment $s_{21}$ since it is the only line segment which crosses the boundary $t_{max}$ on the right side of the range-map image \cite{aminGuendel:IET}.     

\section{Power Burst Curve (PBC)\label{sec:PBC}}

\begin{figure}[ht]
	\includegraphics[width=0.99\linewidth]{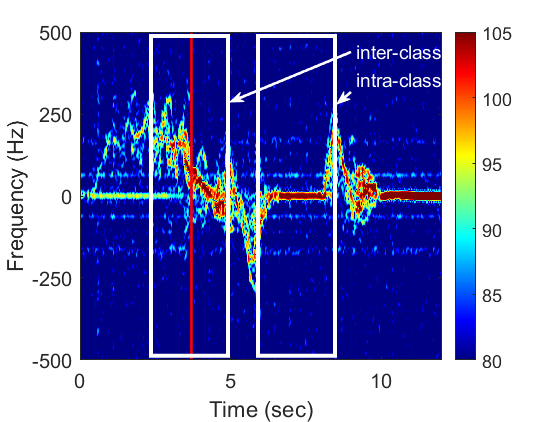}
	\caption{Micro-Doppler image shows the rendered intra- and inter motion separation.\label{subfig:MD_Separation}}
\end{figure}

\begin{figure}[ht]
	\includegraphics[width=0.99\linewidth]{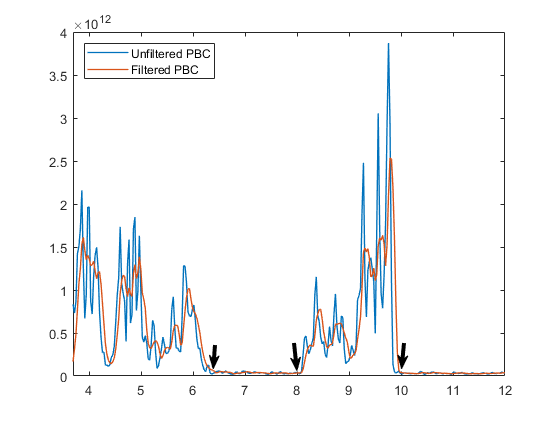}
	\caption{The computed energy of the in-place segment.\label{subfig:PBC3}}
\end{figure}

\noindent To determine whether there is one or a sequence (multiple) of in-place motions, we examine the micro-Doppler signatures over the in-place motion interval. The spectrogram of the motions in the above example is shown in Fig.~\ref{subfig:MD_Separation}. For separating the two consecutive in-place motions, namely sitting down and standing up, we measure the rise and fall of the signal energy in $MD(n,k)$ over slow-time which is known as the Power Burst Curve (PBC) \cite{32_7944316, 33_6889337, amin_Wu:Fall_PBC:7046290}.

The selected frequency bands for power computation are bounded by \(K_{P1} = 20~Hz\) and \(K_{P2} = 270~Hz\)  for  positive-Doppler frequencies and by 
\(K_{N2} =  - 20~Hz\) and \(K_{N1} =  - 270~Hz\) for negative Doppler frequencies. The PBC for the combined frequency bands is given by, 
\begin{equation}
\label{eq:PBC}
\begin{split}
PC(n) = \sum\limits_{k_1 = {K_{P1}}}^{K_{P2}} {{{\left| {MD({k_1},n)} \right|}^2}}  + \\\sum\limits_{{k_2} = {K_{N1}}}^{{K_{N2}}} {{{\left| {MD({k_2},n)} \right|}^2}}, n = 1,2,\dots,N.
\end{split}
\end{equation}
The computation of the above equation results in a fluctuating power curve stemming from intricate micro-Doppler signatures of human motions. Such fluctuation could mistakenly define wrong event boundaries.
To mitigate the above problem, we apply a moving average filter with an extent of $w=5$ samples. 
The filtered PBC is given by,
\begin{equation}
\label{eq:PBC_Filter}
\begin{split}
PC_f(n) = \frac{1}{w}( PC(n) + PC(n - 1) + \dots \\ 
+ PC(n - w)), n = 1,2,\dots,N.
\end{split}
\end{equation}	
The filtered PBC, shown in Fig.~\ref{subfig:PBC3}, is used to determine the onset and offset times of each activity. The threshold has been found empirically as $3\%$ over the minima as, $PC_{f_{min}}+ 0.03 \cdot(PC_{f_{max}}-PC_{f_{min}})$. Based on the this threshold value, the in-place activities are separated according to,       
\begin{equation}
\label{eq:PBC_threshold}
{\color{white}\bar{P}\bar{C}_f(n) =} 
\begin{cases}
Active,~~\text{if } & PC_f(n) \ge threshold\\
Inactive,              &\text{otherwise}.
\end{cases}
\end{equation}

\noindent 
In constructing the state diagram in Sec.~\ref{sec:3hsr}, we included human motions which could easily merge. This represents a challenge to the PBC to separate motion events. So, we rely on the breaking point generated by the Radon transform to indicate in-place motion occurrence and use a window around it to capture the corresponding action \cite{aminGuendel:IET}.


\chapter{\label{chap:State_expResults}State representation and experimental results}

\section{Human motion state representation\label{sec:3hsr}}

\begin{figure}[ht]
\centering
\includegraphics[width=\linewidth]{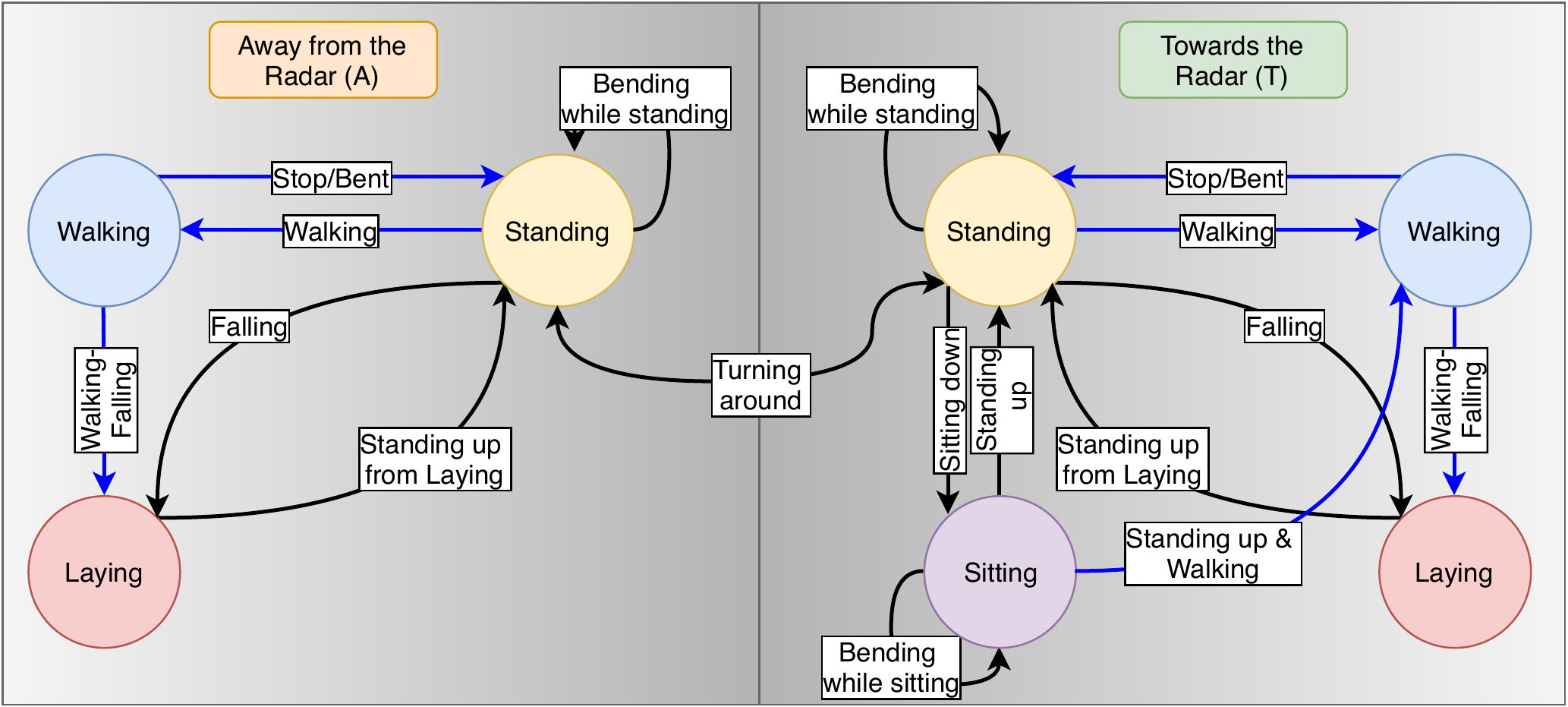}
\caption[State diagram for forward in time motion detection.]{State diagram for forward in time motion detection (black arrows: in-place actions; blue arrows: translation actions). \label{fig:flowgraph_fwd}}
\end{figure}

In this section, we cast human activities as states, namely, walking, standing, sitting, and laying. The latter state commensurates with falling but can be generalized to include intentional actions as laying on floor or bed.  A change, or a transition, from a state to another is performed through an action, or an activity. This is shown in the state diagram in Fig.~\ref{fig:flowgraph_fwd}.
\subsection{Forward in time motion sequence for the "Walking State"\label{subsec:31hsr}}
The person changes from a “walking state” (WS) to a “laying state” (LS) through a falling down action, and transitions to a “standing state” (StS) through a stopping action. The state can also transition to itself through a bending down and up action. The latter case defines contiguous motions in the sense that there is not sufficient time separation between walking followed by bending down and between rebounding from bending and walking in order to be able to declare a (StS) in each case. The classifier must then discriminate between these three actions which are represented by arrows emanating from the WS in Fig.~\ref{fig:flowgraph_fwd}. 

\begin{table}[ht]
\caption{ADL motion according to the human ethogram.\label{tab:ethogram}}
\centering
\scalebox{0.85}{
\vspace{0.15cm}
\begin{tabular}{rl}
\hline
Motion                       & Declaration                                                                         \\ \hline \vspace{-0.25cm}\\ 
T-Walking-Stop/Bent (I)      & Walking towards the radar and stopping/bending contiguously.   \\
T-Walking-Fall (II)          & Walking towards the radar and falling merged contiguously.                \\
A-Walking-Stop/Bent (III)    & Walking away from the radar and stopping/bending contiguously. \\
A-Walking-Fall (IV)          & Walking away from the radar and falling merged contiguously.              \\
Sitting down (V)             & Sitting down motion from the standing, facing the radar.                                        \\
T-Bending w. Standing (VI)   & Bending motion while standing, facing the radar.                                    \\
A-Bending w. Standing (VII)  & Bending motion while standing, away from the radar.                                 \\
T-Falling f. Standing (VIII) & Falling motion from standing, facing the radar.                                     \\
A-Falling f. Standing (IX)   & Falling motion from standing, away from the radar.                                  \\
T-Standing f. Falling (X)    & Standing up motion from falling, facing the radar.                                         \\
A-Standing f. Falling (XI)   & Standing up motion from falling, away from the radar.                                      \\
Standing f. Sitting (XII)    & Standing up motion from sitting, facing the radar.                                         \\
Bending w. Sitting (XIII)    & Bending motion while sitting, facing the radar.                                            \\
Standing up - Walking (XIV)  & Standing up motion merged with walking, towards the radar.                          \\
Start Walking (XV)           & Start walking towards the radar.                                                    \\ \hline
\end{tabular}%
}
\end{table}

\begin{sidewaystable}
	\caption{Fusion classifier for all ADL motion classes.}\vspace{0.15cm}
	\centering
	\label{tab:AllClassesFusion}
	\scalebox{0.9}{
		\begin{tabular}{|l|rrrrrrrrrrrrrrr|} \hline
			\rotatebox[origin=l]{-90}{~\diagbox[dir=SW]{\textbf{Pred. Classes}}{\textbf{True Classes}}~~}  & \rotatebox[origin=l]{-90}{T-Walking-Stop/Bent (I)~} & \rotatebox[origin=l]{-90}{T-Walking-Fall (II)~} & \rotatebox[origin=l]{-90}{A-Walking-Stop/Bent (III)~} & \rotatebox[origin=l]{-90}{A-Walking-Fall (IV)~} & \rotatebox[origin=l]{-90}{Sitting down (V)}  & \rotatebox[origin=l]{-90}{T-Bending w. Standing (VI)~} & \rotatebox[origin=l]{-90}{A-Bending w. Standing (VII)~} & \rotatebox[origin=l]{-90}{T-Falling f. Standing (VIII)~} & \rotatebox[origin=l]{-90}{A-Falling f. Standing (IX)~} & \rotatebox[origin=l]{-90}{T-Standing f. Falling (X)~} & \rotatebox[origin=l]{-90}{A-Standing f. Falling (XI)~} & \rotatebox[origin=l]{-90}{Standing f. Sitting (XII)~} & \rotatebox[origin=l]{-90}{Bending w. Sitting (XIII)~}  & \rotatebox[origin=l]{-90}{Standing up - Walking (XIV)~} & \rotatebox[origin=l]{-90}{Start Walking (XV)~}  \\   \hline \noalign{\vskip 0.05cm} 
			I & 100\% & 0.0\% & 0.0\% & 0.0\% & 0.0\% & 0.0\% & 0.0\% & 0.0\% & 0.0\% & 0.0\% & 0.0\% & 0.0\% & 0.0\% & 0.0\% & 0.0\% \\
			II & 15.9\% & 80.7\% & 0.0\% & 0.0\% & 0.0\% & 0.0\% & 0.0\% & 3.5\% & 0.0\% & 0.0\% & 0.0\% & 0.0\% & 0.0\% & 0.0\% & 0.0\% \\
			III & 0.0\% & 0.0\% & 96.9\% & 2.8\% & 0.0\% & 0.0\% & 0.0\% & 0.0\% & 0.3\% & 0.0\% & 0.0\% & 0.0\% & 0.0\% & 0.0\% & 0.0\% \\
			IV & 0.0\% & 0.0\% & 6.3\% & 90.8\% & 0.0\% & 0.0\% & 0.0\% & 0.0\% & 2.9\% & 0.0\% & 0.0\% & 0.0\% & 0.0\% & 0.0\% & 0.0\% \\
			V & 0.0\% & 0.0\% & 0.0\% & 0.0\% & 98.9\% & 0.0\% & 1.1\% & 0.0\% & 0.0\% & 0.0\% & 0.0\% & 0.0\% & 0.0\% & 0.0\% & 0.0\% \\
			VI & 0.0\% & 0.0\% & 0.0\% & 0.0\% & 0.3\% & 94.0\% & 4.4\% & 0.0\% & 0.0\% & 0.0\% & 0.0\% & 0.8\% & 0.6\% & 0.0\% & 0.0\% \\
			VII & 0.0\% & 0.0\% & 0.0\% & 0.0\% & 0.0\% & 1.4\% & 98.5\% & 0.0\% & 0.0\% & 0.0\% & 0.0\% & 0.0\% & 0.0\% & 0.0\% & 0.0\% \\
			VIII & 0.0\% & 1.2\% & 0.0\% & 0.0\% & 0.0\% & 0.0\% & 0.0\% & 98.5\% & 0.0\% & 0.0\% & 0.0\% & 0.3\% & 0.0\% & 0.0\% & 0.0\% \\
			IX & 0.0\% & 0.0\% & 0.0\% & 0.0\% & 0.1\% & 0.0\% & 0.0\% & 1.4\% & 98.4\% & 0.0\% & 0.0\% & 0.0\% & 0.0\% & 0.0\% & 0.0\% \\
			X & 0.0\% & 0.0\% & 0.0\% & 0.0\% & 2.1\% & 0.0\% & 0.0\% & 0.0\% & 0.0\% & 96.6\% & 1.3\% & 0.0\% & 0.0\% & 0.0\% & 0.0\% \\
			XI & 0.0\% & 0.0\% & 0.0\% & 0.0\% & 0.0\% & 1.4\% & 2.7\% & 1.0\% & 0.0\% & 0.0\% & 89.4\% & 1.0\% & 4.5\% & 0.0\% & 0.0\% \\
			XII & 0.0\% & 0.0\% & 0.0\% & 0.0\% & 0.3\% & 1.5\% & 2.2\% & 0.0\% & 0.0\% & 0.0\% & 0.0\% & 95.3\% & 0.0\% & 0.9\% & 0.0\% \\
			XIII & 0.0\% & 0.0\% & 0.0\% & 0.0\% & 0.0\% & 0.4\% & 0.3\% & 0.0\% & 0.0\% & 0.0\% & 0.2\% & 0.0\% & 99.1\% & 0.0\% & 0.0\% \\
			XIV & 0.0\% & 0.0\% & 0.0\% & 0.0\% & 0.0\% & 0.0\% & 0.0\% & 0.0\% & 0.0\% & 0.0\% & 0.0\% & 1.6\% & 0.0\% & 97.9\% & 0.5\% \\
			XV & 0.0\% & 0.7\% & 0.0\% & 0.0\% & 0.0\% & 0.0\% & 0.0\% & 0.0\% & 0.0\% & 0.0\% & 0.0\% & 0.0\% & 0.0\% & 1.5\% & 97.8\% \\ \hline
		\end{tabular}
	}
\end{sidewaystable} 

\subsection{Backward in time motion sequence for the “Walking State”\label{subsec:32hsr}}
The backward in time sequence of actions considering walking is not entirely reciprocal to the forward sequence. For example walking cannot be preceded by falling but can be followed by it. Also, the WS can be directly reached from the SiS, through standing up action, but cannot be followed by it. This is because a person needs to exhibit short time duration of standing between walking and sitting down for body adjustment which implies that the SiS is preceded by the StS and not by the WS. However, standing up from sitting followed up by walking can be contiguous with no time separation in between to declare a StS. It is noted that such merging is only possible if standing up is in the direction of the follow-on walking motion, otherwise, the person would need to turn around after standing up and walk in the opposite direction which gives rise to a short time interval where the person is in the StS.  Accordingly, a classifier needs to consider only two actions prior to walking towards the radar which are indicated by the arrows entering the walking state in Fig.~\ref{fig:flowgraph_fwd}. 


\subsection{Forward and backward in time motion sequence for the “Sitting State”\label{subsec:33hsr}}
In the forward in time motion sequence, a person can transition from the SiS to itself through a bending action, as shown in Fig.~\ref{fig:flowgraph_fwd}. It can change to the StS by a standing up action or to a WS, as discussed above.  So a classifier applied to the SiS should consider only two in-place classes. For the backward in time motion sequence, a change into the SiS can be performed from the StS only. This is also shown in Fig.~\ref{fig:flowgraph_fwd}. 

\subsection{Forward and backward in time motion sequence for the “Standing State”\label{subsec:34hsr}}
In the forward in time motion sequence, a person can transition from the StS to itself though a bending action, as depicted in Fig.~\ref{fig:flowgraph_fwd}. It can also change to the WS, SiS, or LS. In the backward in time motion sequence, changing into the StS can be from the SiS through standing up, from the WS through stopping, from a LS through standing up from falling, and transition from itself through bending. In this regard, the StS is associated with the highest number of motion classes, or actions, in the forward and backward in time directions.

\subsection{Forward and backward in time motion sequence for the “Laying State”\label{subsec:35hsr}}
In the forward in time motion sequence, a person can change from the LS to the StS through a standing up motion, which is the only possible action. The person can change into the LS from the StS through falling, and from the WS also through falling. This is shown in the state diagram of Fig.~\ref{fig:flowgraph_fwd}, which suggests two possible classifiers, each is applied to the motion actions transitioning in and out of a state. 

Once a state is detected, then the two associated classifiers for the in and out transitions can be applied to infer the previous and follow-on motions, respectively. It should be emphasized that each state can apply a different classifier than the rest.

\subsection{Motions towards and away from the radar\label{subsec:36hsr}}
To account for the possibilities that the motion actions can be performed towards and away from the radar, we make the state diagram in Fig.~\ref{fig:flowgraph_fwd} to consist of two state groups: one group for toward radar motions and the other group for away radar motions. Each group represents one motion direction. A person can transition across the two groups by the means of turning around while standing \cite{aminGuendel:IET}.

\section{Experimental results \label{sec:experimentalResults}}
In this section, we consider three different motion sequences to demonstrate the functionality of the proposed consecutive and contiguous motion detection for ADL. Example~1 includes a critical fall between two walking states. %

 Example~2 demonstrates a motion sequence where a person comes to fall next to a chair and sits down on the chair after standing up from the fall. In this example, the person visits all motion states, (WS, LS, SiS, StS), before turning around and continuing to walk in the opposite direction.  

Example~3 corresponds to a more common sequence of picking an object from the ground from both the StS and the SiS. 

The fused 2-D PCA followed by the NN classifier of Fig.~\ref{fig:fusionGraph} is used in the three examples, except for the motion \circled{1} in Fig.~\ref{fig:SequenceWlkBenSitBenStaWlk}, where we apply micro-Doppler classification. We are able to to account for wrong decisions as in Example-1 and -3. For Example-2, we only account for miss-classification concerning a critical fall.    

We compare the proposed classification approach, which operates on the individual states of Fig.~\ref{fig:flowgraph_fwd}, with the commonly used approach of incorporating all ADL at any given time. The list of actions considered is given in Table~\ref{tab:ethogram}. The results of considering all actions in Table~\ref{tab:ethogram} are shown in Table~\ref{tab:AllClassesFusion}. These results assume isolated and independent separate motions.

\subsection{Example-1: Motion sequence of walking and falling\label{subsec:1_firstEx}}
%

\begin{figure}[ht]
\centering
(a) Micro-Doppler Signature\\
\includegraphics[width=\linewidth]{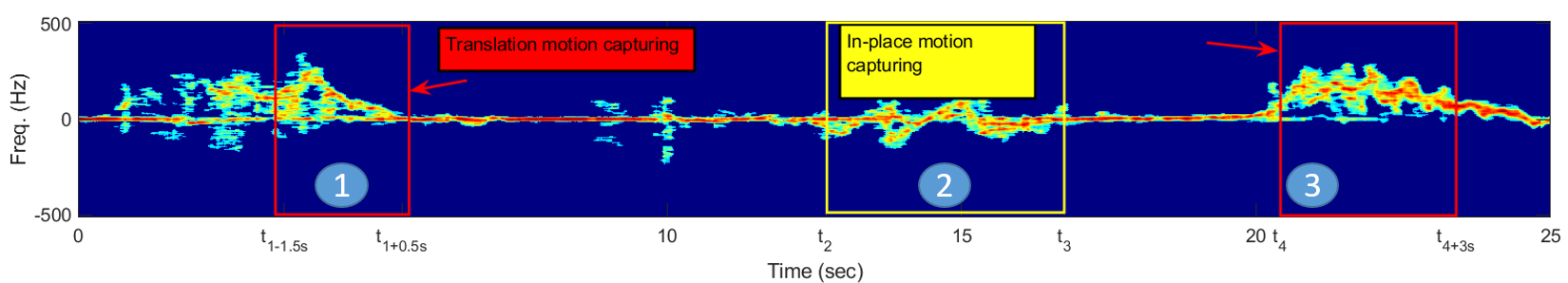}
\\
\vspace{.3cm}
(b) Range-map Profile\\
\includegraphics[width=\linewidth]{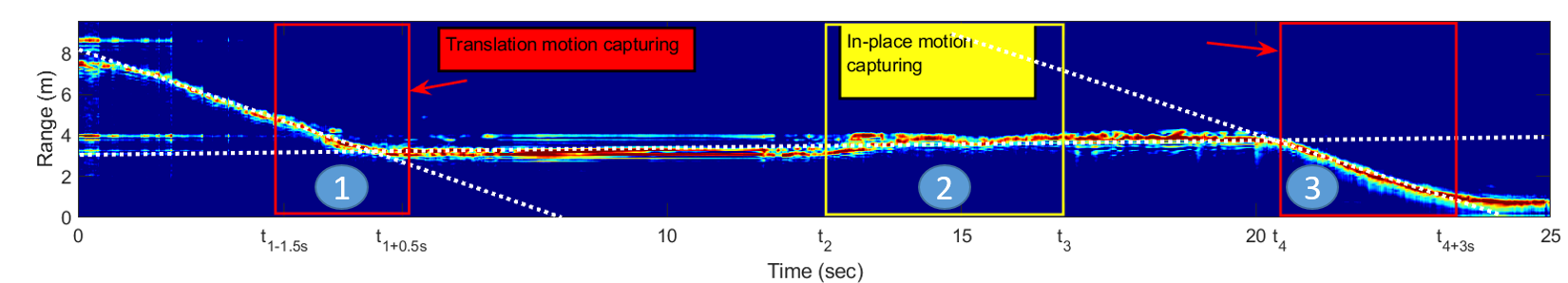}
\caption[Motion sequence of Example-1.]{Motion sequence: walking-falling, standing up from falling, walking. \label{fig:SequenceWlkFalSta}}
\end{figure} 

\begin{figure}[ht]
\begin{center}
    \begin{minipage}[t]{.14\linewidth}
        \centering
        \rotatebox[origin=l]{90}{{\small ~Freq. (Hz)}}~\includegraphics[width=\linewidth]{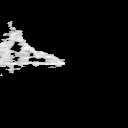}
        \\{\small~~~Time (sec)\\~~~~~~~~(a)}\label{subfig:Ex1_11}
    \end{minipage}
    ~~~~
    \begin{minipage}[t]{.14\linewidth}
        \centering
        \rotatebox[origin=l]{90}{{\small ~Freq. (Hz)}}~\includegraphics[width=\linewidth]{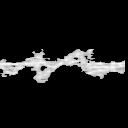}
        \\{\small~~~Time (sec)\\~~~~~~~~(b)}\label{subfig:Ex1_21}
    \end{minipage}
    ~~~~
    \begin{minipage}[t]{.14\linewidth}
    \centering
        \rotatebox[origin=l]{90}{{\small ~Freq. (Hz)}}~\includegraphics[width=\linewidth]{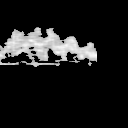}
        \\{\small~~~Time (sec)\\~~~~~~~~(c)}\label{subfig:Ex1_31}
    \end{minipage}  
        \\
\vspace{.3cm}
    \begin{minipage}[t]{.14\linewidth}
        \centering
         \rotatebox[origin=l]{90}{{\small~Range (m)}}~\includegraphics[width=\linewidth]{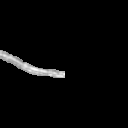}
        \\{\small~~~Time (sec)\\~~~~~~~~(d)}\label{subfig:Ex1_12}
    \end{minipage}
    ~~~~
    \begin{minipage}[t]{.14\linewidth}
        \centering
         \rotatebox[origin=l]{90}{{\small~Range (m)}}~\includegraphics[width=\linewidth]{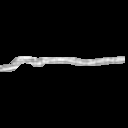}
        \\{\small~~~Time (sec)\\~~~~~~~~(e)}\label{subfig:Ex1_22}
    \end{minipage}
    ~~~~
    \begin{minipage}[t]{.14\linewidth}
    \centering
         \rotatebox[origin=l]{90}{{\small~Range (m)}}~\includegraphics[width=\linewidth]{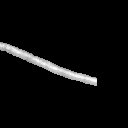}
        \\{\small~~~Time (sec)\\~~~~~~~~(f)}\label{subfig:Ex1_32}
    \end{minipage}
\end{center}  
    \caption[Captured motions for Example-1.]{Captured motions for Example-1 (Micro-Doppler for motions \textcircled{1} to \textcircled{3} are in Fig.~a.-c.; Range-map for motions \textcircled{1} to \textcircled{3} are in Fig.~d.-f.). \label{fig:Ex1_MD_RM}}
\end{figure} 


The example in Fig.~\ref{fig:SequenceWlkFalSta} shows a fall incorporated with a prior walk, followed by a period of laying on the floor, and then standing up from falling to a StS. Finally, the person walks towards the radar.

The captured micro-Doppler signatures and range-map images of Example-1 are shown in Fig.~\ref{fig:Ex1_MD_RM}. The range profiles are shifted to the middle to eliminate biased classification outcome based on the distance to the radar.
%
\subsubsection{Forward in time motion classification} 

\begin{table}[htbp]
\centering 
\caption[Classifier-1.]{Classifier-1: Forward in time classification, {[}$d_{MD}=2$; $d_{RM}=1${]}.}\vspace{0.15cm}
\label{tab:clas_0a}
\scalebox{0.77}{
\begin{tabular}{@{}rrrrrrr@{}}
\toprule
 & \multicolumn{2}{c}{\begin{tabular}[c]{@{}c@{}}Micro-Doppler\\ Predicted\end{tabular}} & \multicolumn{2}{c}{\begin{tabular}[c]{@{}c@{}}Range-map\\ Predicted\end{tabular}} & \multicolumn{2}{c}{\begin{tabular}[c]{@{}c@{}}Fusion\\ Predicted\end{tabular}} \\ \midrule
\multicolumn{1}{r|}{cl.} & (I) & \multicolumn{1}{r|}{(II)} & (I) & \multicolumn{1}{r|}{(II)} & (I) & (II) \\ \midrule
\multicolumn{1}{r|}{(I)} & 98.8\% & \multicolumn{1}{r|}{1.2\%} & 96.6\% & \multicolumn{1}{r|}{3.4\%} & 100\% & 0\% \\
\multicolumn{1}{r|}{(II)} & 11.0\% & \multicolumn{1}{r|}{89.0\%} & 26.8\% & \multicolumn{1}{r|}{73.3\%} & 10.5\% & 89.5\% 
\end{tabular}
} 
\end{table} 
%
\begin{table}[hbt]
\caption[Classifier-2.]{Classifier-2: Forward in time classification, [$d_{MD}=6$; $d_{RM}=2$]. } \vspace{0.15cm}
\label{tab:clas_1a}
\centering
\scalebox{0.77}{
\begin{tabular}{@{}rrrrrrrrr@{}}
\toprule
 & \multicolumn{4}{c}{\begin{tabular}[c]{@{}c@{}}Micro-Doppler\\ Predicted\end{tabular}} & \multicolumn{4}{c}{\begin{tabular}[c]{@{}c@{}}Range-map\\ Predicted\end{tabular}} \\ \midrule
\multicolumn{1}{r|}{cl.} & (I) & (II) & (III) & \multicolumn{1}{r|}{(IV)} & (I) & (II) & (III) & \multicolumn{1}{r|}{(IV)} \\ \midrule
\multicolumn{1}{r|}{(I)} & 99.6\% & 0.2\% & 0.0\% & \multicolumn{1}{r|}{0.2\%} & 98.9\% & 0.9\% & 0.1\% & \multicolumn{1}{r|}{0.1\%} \\
\multicolumn{1}{r|}{(II)} & 0.2\% & 99.8\% & 0.0\% & \multicolumn{1}{r|}{0.0\%} & 1.1\% & 97.5\% & 0.0\% & \multicolumn{1}{r|}{1.4\%} \\
\multicolumn{1}{r|}{(III)} & 0.6\% & 0.0\% & 99.4\% & \multicolumn{1}{r|}{0.0\%} & 1.3\% & 0.0\% & 98.7\% & \multicolumn{1}{r|}{0.0\%} \\
\multicolumn{1}{r|}{(IV)} & 0.3\% & 0.1\% & 0.0\% & \multicolumn{1}{r|}{99.6\%} & 4.6\% & 5.6\% & 0.5\% & \multicolumn{1}{r|}{89.3\%} \\ \cmidrule(r){1-5}
 & \multicolumn{4}{c}{Fusion Predicted} &  &  &  &  \\ \cmidrule(r){1-5}
\multicolumn{1}{r|}{cl.} & (I) & (II) & (III) & \multicolumn{1}{r|}{(IV)} &  &  &  &  \\ \cmidrule(r){1-5}
\multicolumn{1}{r|}{(I)} & 99.9\% & 0.1\% & 0.0\% & \multicolumn{1}{r|}{0.0\%} &  &  &  &  \\
\multicolumn{1}{r|}{(II)} & 0.1\% & 99.9\% & 0.0\% & \multicolumn{1}{r|}{0.0\%} &  &  &  &  \\
\multicolumn{1}{r|}{(III)} & 0.6\% & 0.0\% & 99.4\% & \multicolumn{1}{r|}{0.0\%} &  &  &  &  \\
\multicolumn{1}{r|}{(IV)} & 0.2\% & 0.8\% & 0.3\% & \multicolumn{1}{r|}{98.7\%} &  &  &  & 
\end{tabular}
}
\end{table} 

In Example~1, the Radon transform renders the intersection point between the inclined and horizontal lines in the range-map at $t_1 \approx 5.0s$. This merging between the translation motion and the follow-on in-place motion can indicate a walk that is immediately followed by a fall, a stop, or a bent activity, each is in the direction of walking. The intersection point from the Radon transform typically occurs after the walking ceases. In this case, the PBC cannot time separate the energies of the walking and falling motions. Accordingly, at $t_1$, a window of $2s$ is applied which captures $0.5s$ of the in-place and $1.5s$ of the translation motion (left red rectangulars \circled{1} in Fig.~\ref{fig:SequenceWlkFalSta}). From the state diagram of Fig.~\ref{fig:flowgraph_fwd}, and considering the transitioning out of the WS, the employed classifier deals with only two possible classes, grouped as, (\textsc{I}) walking-stopping \& walking-bending, leading to the StS, and (\textsc{II}) walking-falling, leading to the LS. It is clear from the classification results depicted in Table~\ref{tab:clas_0a} (Classifier-1) that there are no false alarms for falling, but there is a missing probability of $10.5\%$, which is significantly high. If all motion classes are considered including those of the SiS, the missing probability of falling rises to $19.3\%$ (Table~\ref{tab:AllClassesFusion}), which is certainly unacceptable. Since the non-zero missing probability can mistakenly assign a StS instead of a LS for a fall, all in-place motions from the StS and LS, facing the radar, should be considered in the next classifier. According to the state diagram, these motions are: (I) bending while standing, (II) sitting from standing, (III) falling from standing, and (IV) standing up from falling.

The PBC determines that the onset and offset times of the in-place motion \circled{2} are at $t_2 \approx 12.5s$ and $t_3 \approx 16.7s$, respectively. From the corresponding classification Table~\ref{tab:clas_1a} (Classifier-2), the ground truth motion (IV), which is standing up from falling, has a missing probability of $1.3\%$, with no probability of false alarm. In contrast, the missing probability when applying all ADL is shown in Table~\ref{tab:AllClassesFusion} as $3.4\%$.  Comparing Tables~\ref{tab:clas_0a} and \ref{tab:clas_1a}, it is evident that the detection of standing up from falling is more reliable than the detection of falling itself, when it is contiguously merged with walking. This suggests more reliable detection of a fall when it is recovered.

\begin{table}[htbp]
\caption[Classifier-3.]{Classifier-3: Forward- and backward in time classification, \\ $[d_{MD}=14$; $d_{RM}=4]$.}\vspace{0.15cm} 
\label{tab:clas_4}
\centering
\scalebox{0.77}{
\begin{tabular}{@{}rrrrrrr@{}}
\toprule
 & \multicolumn{2}{c}{\begin{tabular}[c]{@{}c@{}}Micro-Doppler\\ Predicted\end{tabular}} & \multicolumn{2}{c}{\begin{tabular}[c]{@{}c@{}}Range-map\\ Predicted\end{tabular}} & \multicolumn{2}{c}{\begin{tabular}[c]{@{}c@{}}Fusion\\ Predicted\end{tabular}} \\ \midrule
\multicolumn{1}{r|}{cl.} & (I) & \multicolumn{1}{r|}{(II)} & (I) & \multicolumn{1}{r|}{(II)} & (I) & (II) \\ \midrule
\multicolumn{1}{r|}{(I)} & 100\% & \multicolumn{1}{r|}{0\%} & 99.2\% & \multicolumn{1}{r|}{0.8\%} & 100\% & 0\% \\
\multicolumn{1}{r|}{(II)} & 0\% & \multicolumn{1}{r|}{100\%} & 0\% & \multicolumn{1}{r|}{100\%} & 0\% & 100\%
\end{tabular}
}
\end{table} 

The above classification result, being in SiS or StS, is maintained until the intersection point at $t_4 \approx 20.5s$ after which, the Radon transform declares a WS. Over the motion window \circled{3}, $t_4$ to  ${t_4+3s}$, the classifier is tasked to discriminate between the only two possible actions that lead to the WS. According to the state diagram, these actions are: (I) starting-walking and (II) standing up from sitting merged with walking. Both motions are classified (Classifier-3 in Table~\ref{tab:clas_4}) with $100\%$ accuracy.
%

\subsubsection{Backward in time motion classification} 
%
%
\begin{table}[htbp]
\caption[Classifier-4.]{Classifier-4: Backward in time classification, [$d_{MD}=7$;~$d_{RM}=2$].\label{tab:clasMat_22}}\vspace{0.15cm}
\centering
\scalebox{0.77}{
\begin{tabular}{@{}rrrrrr@{}}
\toprule
 & \multicolumn{5}{c}{Micro-Doppler predicted} \\ \midrule
\multicolumn{1}{r|}{cl.} & (I) & (II) & (III) & (IV) & (V) \\ \midrule
\multicolumn{1}{r|}{(I)} & 97.8\% & 0.1\% & 1.7\% & 0.0\% & 0.4\% \\
\multicolumn{1}{r|}{(II)} & 0.6\% & 98.0\% & 1.4\% & 0.0\% & 0.0\% \\
\multicolumn{1}{r|}{(III)} & 0.0\% & 1.9\% & 97.9\% & 0.0\% & 0.3\% \\
\multicolumn{1}{r|}{(IV)} & 0.0\% & 0.0\% & 0.0\% & 98.6\% & 1.4\% \\
\multicolumn{1}{r|}{(V)} & 2.2\% & 1.0\% & 1.2\% & 0.1\% & 95.6\% \\ \midrule
\multicolumn{1}{l}{} & \multicolumn{5}{c}{Range-map predicted} \\ \midrule
\multicolumn{1}{r|}{(I)} & 95.5\% & 3.9\% & 0.5\% & 0.0\% & 0.0\% \\
\multicolumn{1}{r|}{(II)} & 2.9\% & 78.1\% & 19.0\% & 0.0\% & 0.0\% \\
\multicolumn{1}{r|}{(III)} & 0.6\% & 7.6\% & 91.7\% & 0.0\% & 0.0\% \\
\multicolumn{1}{r|}{(IV)} & 0.0\% & 1.4\% & 0.0\% & 91.3\% & 7.2\% \\
\multicolumn{1}{r|}{(V)} & 0.9\% & 8.1\% & 0.2\% & 5.8\% & 85.0\% \\ \midrule
\multicolumn{1}{l}{} & \multicolumn{5}{c}{Fusion predicted} \\ \midrule
\multicolumn{1}{r|}{(I)} & 98.7\% & 0.3\% & 0.6\% & 0.2\% & 0.1\% \\
\multicolumn{1}{r|}{(II)} & 0.6\% & 97.9\% & 1.5\% & 0.0\% & 0.0\% \\
\multicolumn{1}{r|}{(III)} & 0.0\% & 1.6\% & 98.4\% & 0.0\% & 0.0\% \\
\multicolumn{1}{r|}{(IV)} & 0.0\% & 0.0\% & 0.0\% & 98.0\% & 2.0\% \\
\multicolumn{1}{r|}{(V)} & 0.0\% & 0.1\% & 0.3\% & 0.8\% & 98.7\%
\end{tabular}
}
\end{table} 
With Table~\ref{tab:clas_4} classification certainty, one can go backward in time and revisit the in-place motions underlying Table~\ref{tab:clas_1a}, but classify previous activities. If Classifier-3 declares a StS, then the actions occurring prior the StS would be (I) standing up from sitting, (II) bending while standing towards the radar, (III) bending while standing away from the radar, (IV) standing up from falling towards the radar, and (V) standing up from falling away from the radar. These actions originate from the SiS, the StS, or the LS. The classification results are shown in Table~\ref{tab:clasMat_22} (Classifier-4). Standing up from falling facing the radar has small confusion with standing up from falling away from the radar. But, it has only $0.2\%$ miss-detection probability and no false alarm probability to other classes aside from standing up from falling. 
It is clear that the results in Tables~\ref{tab:clas_4}+\ref{tab:clasMat_22} of the backward in time Classifiers-3 and -4 are more assertive than those in Tables~\ref{tab:clas_0a}+\ref{tab:clas_1a} of the forward in time motion Classifiers-1 and -2. In this respect, Example~1 underscores the importance of considering both directions in rendering a classification decision \cite{aminGuendel:IET}. 

 \subsection{Example-2: Bidirectional motion sequence of walking, falling and sitting\label{subsec:2_secondEx}}
 %
 \begin{figure}[hbt]
 \centering
 (a) Micro-Doppler Signature\\
 \includegraphics[width=\linewidth]{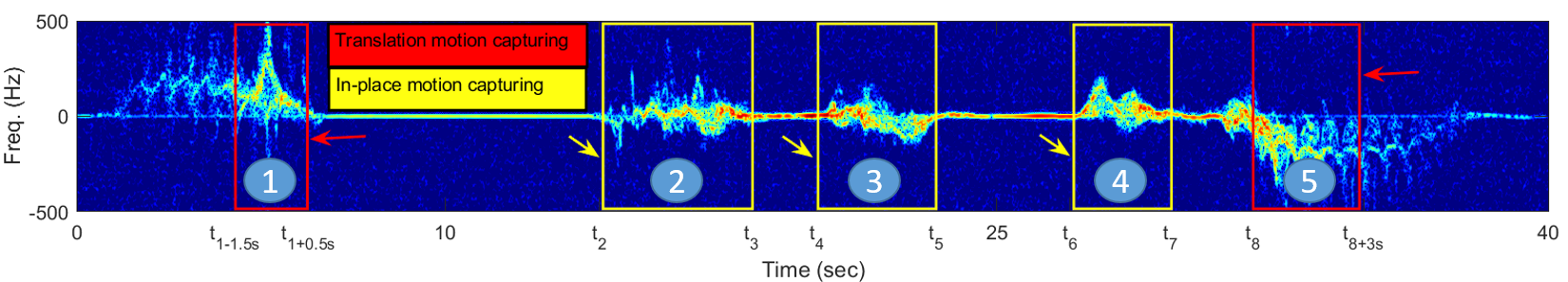}
 (b) Range-map Profile\\
 \includegraphics[width=\linewidth]{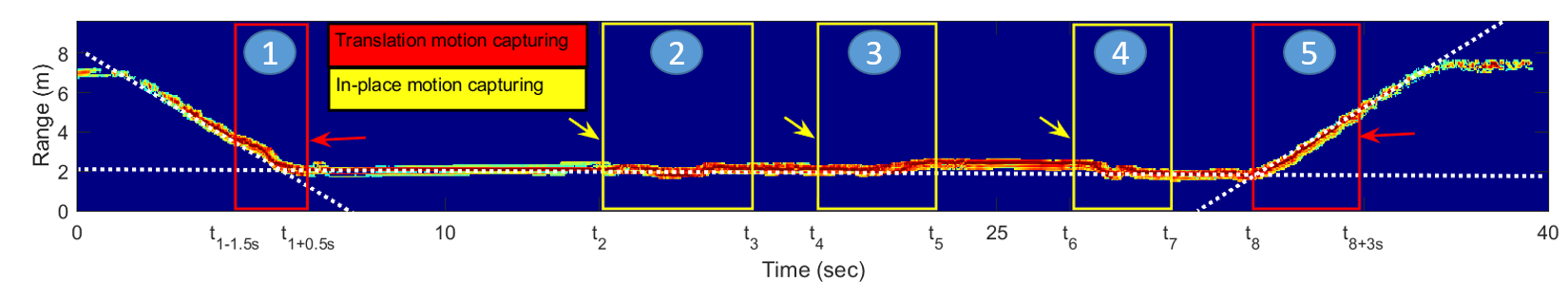}
 \caption[Motion sequence of Example-2.]{Motion sequence: walking-falling, standing up from falling, sitting down, standing up from sitting, turning around and walking. \label{fig:SequenceWlkFalStaSitStaWlk}}
 \end{figure} 
 %
 %
 \begin{figure}[htb]
 	\begin{center}
 		\begin{minipage}[t]{.14\linewidth}
 			\centering
 			\rotatebox[origin=l]{90}{{\small ~Freq. (Hz)}}~\includegraphics[width=\linewidth]{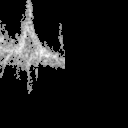}
 			\\{\small~~~Time (sec)\\~~~~~~~~(a)}\label{subfig:Ex2_11}
 		\end{minipage}
 		~~~~
 		\begin{minipage}[t]{.14\linewidth}
 			\centering
 			\rotatebox[origin=l]{90}{{\small ~Freq. (Hz)}}~\includegraphics[width=\linewidth]{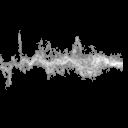}
 			\\{\small~~~Time (sec)\\~~~~~~~~(b)}\label{subfig:Ex2_21}
 		\end{minipage}
 		~~~~
 		\begin{minipage}[t]{.14\linewidth}
 			\centering
 			\rotatebox[origin=l]{90}{{\small ~Freq. (Hz)}}~\includegraphics[width=\linewidth]{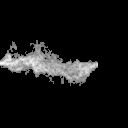}
 			\\{\small~~~Time (sec)\\~~~~~~~~(c)}\label{subfig:Ex2_31}
 		\end{minipage}  
 		~~~~
 		\begin{minipage}[t]{.14\linewidth}
 			\centering
 			\rotatebox[origin=l]{90}{{\small ~Freq. (Hz)}}~\includegraphics[width=\linewidth]{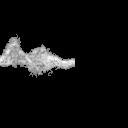}
 			\\{\small~~~Time (sec)\\~~~~~~~~(d)}\label{subfig:Ex2_41}
 		\end{minipage}
 		~~~~
 		\begin{minipage}[t]{.14\linewidth}
 			\centering
 			\rotatebox[origin=l]{90}{{\small ~Freq. (Hz)}}~\includegraphics[width=\linewidth]{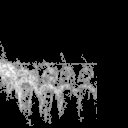}
 			\\{\small~~~Time (sec)\\~~~~~~~~(e)}\label{subfig:Ex2_51}
 		\end{minipage}
 		\\
 		\vspace{.3cm}
 		\begin{minipage}[t]{.14\linewidth}
 			\centering
 			\rotatebox[origin=l]{90}{{\small~Range (m)}}~\includegraphics[width=\linewidth]{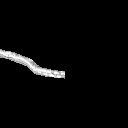}
 			\\{\small~~~Time (sec)\\~~~~~~~~(f)}\label{subfig:Ex2_12}
 		\end{minipage}
 		~~~~
 		\begin{minipage}[t]{.14\linewidth}
 			\centering
 			\rotatebox[origin=l]{90}{{\small~Range (m)}}~\includegraphics[width=\linewidth]{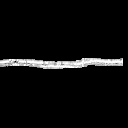}
 			\\{\small~~~Time (sec)\\~~~~~~~~(g)}\label{subfig:Ex2_22}
 		\end{minipage}
 		~~~~
 		\begin{minipage}[t]{.14\linewidth}
 			\centering
 			\rotatebox[origin=l]{90}{{\small~Range (m)}}~\includegraphics[width=\linewidth]{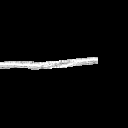}
 			\\{\small~~~Time (sec)\\~~~~~~~~(h)}\label{subfig:Ex2_32}
 		\end{minipage} 
 		~~~~
 		\begin{minipage}[t]{.14\linewidth}
 			\centering
 			\rotatebox[origin=l]{90}{{\small~Range (m)}}~\includegraphics[width=\linewidth]{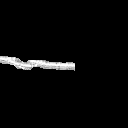}
 			\\{\small~~~Time (sec)\\~~~~~~~~(i)}\label{subfig:Ex2_42}
 		\end{minipage}
 		~~~~
 		\begin{minipage}[t]{.14\linewidth}
 			\centering
 			\rotatebox[origin=l]{90}{{\small~Range (m)}}~\includegraphics[width=\linewidth]{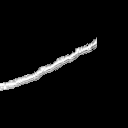}
 			\\{\small~~~Time (sec)\\~~~~~~~~(j)}\label{subfig:Ex2_52}
 		\end{minipage}
 	\end{center}
 	\caption[Captured motions for Example-2.]{Captured motions for Example-2 (Micro-Doppler for motions \textcircled{1} to \textcircled{5} are in Fig.~a.-e.; Range-map for motions \textcircled{1} to \textcircled{5} are in Fig.~f.-j.).\label{fig:Ex2_MD_RM}}
 \end{figure} 
 Example~2 in Fig.~\ref{fig:SequenceWlkFalStaSitStaWlk} demonstrates a fall incorporated with a prior walking, followed by standing up from falling to reach the StS. Afterwards, the person sits down on a chair beside the falling location, where hardly no range swath is noticeable, followed by a standing up activity what leads the person to the StS facing the radar. Finally, the person performs a walking motion away from the radar. Preceding the walking is an integrated turning motion. 
 %
 \subsubsection{Forward in time motion classification} 


 The first two activities \circled{1},~\circled{2} of the Example~2 are the same as for Example~1, and therefore will not be discussed. Table~\ref{tab:clas_0a},~\ref{tab:clas_1a} as well as Classifier-1 and Classifier-2 are applied to declare the person in StS. 

 According to the state diagram in Fig.~\ref{fig:flowgraph_fwd}, the person from the StS has the option of walking in either direction or fulfilling the in-place motions of (I) sitting down, (II) bending from standing, (III) falling from standing. It is noticed that standing up from falling cannot occur twice consecutively; therefore, this motion is not considered.
 %
 \begin{table}[htb]
 \caption[Classifier-5.]{Classifier-5: Forward in time classification, [$d_{MD}=6$; $d_{RM}=2$].} 
 \label{tab:clasMat_13Post}
 \centering
 \scalebox{0.77}{
 \begin{tabular}{rrrrrrrrrr}
 \hline
  & \multicolumn{3}{c}{\begin{tabular}[c]{@{}c@{}}Micro-Doppler\\ Predicted\end{tabular}} & \multicolumn{3}{c}{\begin{tabular}[c]{@{}c@{}}Range-map\\ Predicted\end{tabular}} & \multicolumn{3}{c}{\begin{tabular}[c]{@{}c@{}}Fusion\\ Predicted\end{tabular}} \\ \hline
 \multicolumn{1}{r|}{cl.} & (I) & (II) & \multicolumn{1}{r|}{(III)} & (I) & (II) & \multicolumn{1}{r|}{(III)} & (I) & (II) & (III) \\ \hline
 \multicolumn{1}{r|}{(I)} & 99.9\% & 0.1\% & \multicolumn{1}{r|}{0.0\%} & 98.7\% & 1.3\% & \multicolumn{1}{r|}{0.0\%} & 100\% & 0.0\% & 0.0\% \\
 \multicolumn{1}{r|}{(II)} & 0.1\% & 99.8\% & \multicolumn{1}{r|}{0.1\%} & 1.5\% & 98.5\% & \multicolumn{1}{r|}{0.0\%} & 0.2\% & 99.8\% & 0.0\% \\
 \multicolumn{1}{r|}{(III)} & 0.0\% & 0.9\% & \multicolumn{1}{r|}{99.1\%} & 0.0\% & 1.3\% & \multicolumn{1}{r|}{98.6\%} & 0.0\% & 0.5\% & 99.5\%
 \end{tabular}
 }
 \end{table} 

 The PBC renders a follow-on in-place motion \circled{3} between $t_4 \approx 20.1s$ and $t_5 \approx 23.4s$. According to Classifier-5 (Table~\ref{tab:clasMat_13Post}), the ground truth motion of (I) sitting down has no missing probability, but $0.2\%$ of false alarm probability. In contrast, by considering all motion classes of Table~\ref{tab:AllClassesFusion}, the missing probability becomes $1.1\%$. The classifier correctly puts the person in the SiS. It is noticed that this example does not account for false alarm of $0.2\%$. Therefore, the next assigned in-place motions are: (I) standing up from sitting and (II) bending while sitting.  
 %
 \begin{table}[htb]
 \caption[Classifier-6.]{Classifier-6: Forward in time classification, [$d_{MD}=14$; $d_{RM}=4$].} l
 \label{tab:clasMat_14Post}
 \centering
 \scalebox{0.77}{
 \begin{tabular}{@{}rrrrrrr@{}}
 \toprule
  & \multicolumn{2}{c}{\begin{tabular}[c]{@{}c@{}}Micro-Doppler\\ Predicted\end{tabular}} & \multicolumn{2}{c}{\begin{tabular}[c]{@{}c@{}}Range-map\\ Predicted\end{tabular}} & \multicolumn{2}{c}{\begin{tabular}[c]{@{}c@{}}Fusion\\ Predicted\end{tabular}} \\ \midrule
 \multicolumn{1}{r|}{cl.} & (I) & \multicolumn{1}{r|}{(II)} & (I) & \multicolumn{1}{r|}{(II)} & (I) & (II) \\ \midrule
 \multicolumn{1}{r|}{(I)} & 98.9\% & \multicolumn{1}{r|}{1.1\%} & 99.5\% & \multicolumn{1}{r|}{0.5\%} & 99.6\% & 0.4\% \\
 \multicolumn{1}{r|}{(II)} & 0.1\% & \multicolumn{1}{r|}{99.9\%} & 0.0\% & \multicolumn{1}{r|}{100\%} & 0\% & 100\%
 \end{tabular}
 }
 \end{table} 

 The next in-place motion \circled{4} is detected by the PBC with the onset and offset times at $t_6 \approx 27.0s$ and $t_7 \approx 29.8s$, respectively. The ground truth motion of (I) standing up from sitting has a missing probability of only $0.4\%$ with no false alarm according to Classifier-6 (Table~\ref{tab:clasMat_14Post}). When all motion classes (Table~\ref{tab:AllClassesFusion}) are considered, the missing probability significantly increases to $4.7\%$. 

 The follow-on motion \circled{5} of walking is revealed by the Radon transform. The intersection point has been found at $t_8 \approx 32.1s$. In this case, the person walks away from the radar, and a turning around has occurred. The motion \circled{5} is not classified, since a WS in the opposite direction from the StS facing the radar cannot incorporate any in-place motion (see Fig.~\ref{fig:flowgraph_fwd}).      

 %
 \subsubsection{Backward in time motion classification} 
 With the certainty of the Radon transform, the intersection point  $t_8$ at the end provides the transitioning time from the StS to the WS. Going backwards in time, the former motions can be bidirectional. Therefore, the next applied classifier discriminates between classes of motion performed towards and away from the radar, namely, (I) standing up from sitting, (II) bending while standing towards the radar, (III) bending while standing away from the radar, (IV) standing up from falling towards the radar, and (V) standing up from falling away from the radar. 
 The ground truth motion \circled{4} of (I) of standing up from sitting is classified with a certainty of $98.7\%$ by the Classifier-4 (Table~\ref{tab:clasMat_22}), whereas the accuracy decreases to $95.3\%$ by considering all motion classifier, per  Table~\ref{tab:AllClassesFusion}. 
 
 \begin{table}[htbp]
 	\caption[Classifier-7.]{Classifier-7: Backward in time classification, [$d_{MD}=14$; $d_{RM}=4$].} \vspace{0.15cm}
 	\label{tab:clasMat_23pre}
 	\centering
 	\scalebox{0.77}{
 		\begin{tabular}{@{}rrrrrrr@{}}
 			\toprule
 			& \multicolumn{2}{c}{\begin{tabular}[c]{@{}c@{}}Micro-Doppler\\ Predicted\end{tabular}} & \multicolumn{2}{c}{\begin{tabular}[c]{@{}c@{}}Range-map\\ Predicted\end{tabular}} & \multicolumn{2}{c}{\begin{tabular}[c]{@{}c@{}}Fusion\\ Predicted\end{tabular}} \\ \midrule
 			\multicolumn{1}{r|}{cl.} & (I) & \multicolumn{1}{r|}{(II)} & (I) & \multicolumn{1}{r|}{(II)} & (I) & (II) \\ \midrule
 			\multicolumn{1}{r|}{(I)} & 100\% & \multicolumn{1}{r|}{0.0\%} & 99.4\% & \multicolumn{1}{r|}{0.6\%} & 100\% & 0.0\% \\
 			\multicolumn{1}{r|}{(II)} & 0.0\% & \multicolumn{1}{r|}{99.9\%} & 0.0\% & \multicolumn{1}{r|}{100\%} & 0.0\% & 100\%
 		\end{tabular}
 	}
 \end{table} 

 It is noticed that we do not account for miss-classification in Example-2. Thus, we assign prior the SiS, where the previous motions can be (I) sitting down from standing, or (II) bending from sitting. The classification results in Table~\ref{tab:clasMat_23pre} for the Classifier-7 outperform the results for all motion classes in Table~\ref{tab:AllClassesFusion}. 
 The result of the ground truth motion of sitting down \circled{3} leads back to the StS. It is noted that sitting down can be merged with a prior turning motion, hence, the persons orientation is uncertain. Therefore, bidirectional classes leading to the StS will be considered for motion \circled{2}, as by detecting motion \circled{4}. The true motion of (IV) standing up from falling facing the radar is detected with $98.0\%$ with Classifier-4, shown in Table~\ref{tab:clasMat_22}, whereas most of the confusion occurs again the motion (V) -- standing up from falling away from the radar. The motions (IV) and (V) have the same outcome of the LS, while the persons orientation is secondary. Indeed prior to the LS, a falling has most likely occurred, either from the StS, or the WS. Since a WS has been priorly detected by the Radon transform, a walking-falling motion \circled{1} has appeared. Thus, a fall from the StS can be excluded.   

\subsection{Example-3: Bidirectional motion sequence of picking an object\label{subsec:3_thirdEx}}
%

\begin{figure}[!ht]
\centering
(a) Micro-Doppler Signature\\
\includegraphics[width=\linewidth]{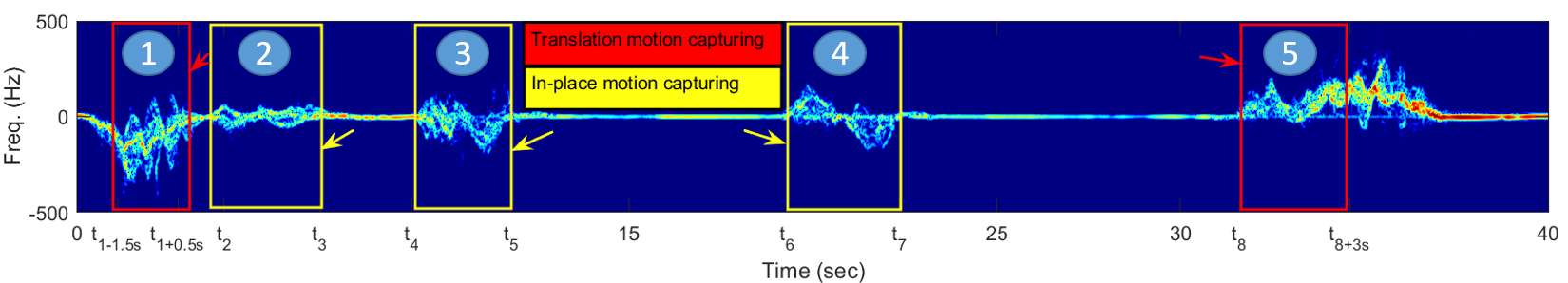}
\\
\vspace{.3cm}
(b) Range-map Profile\\
\includegraphics[width=\linewidth]{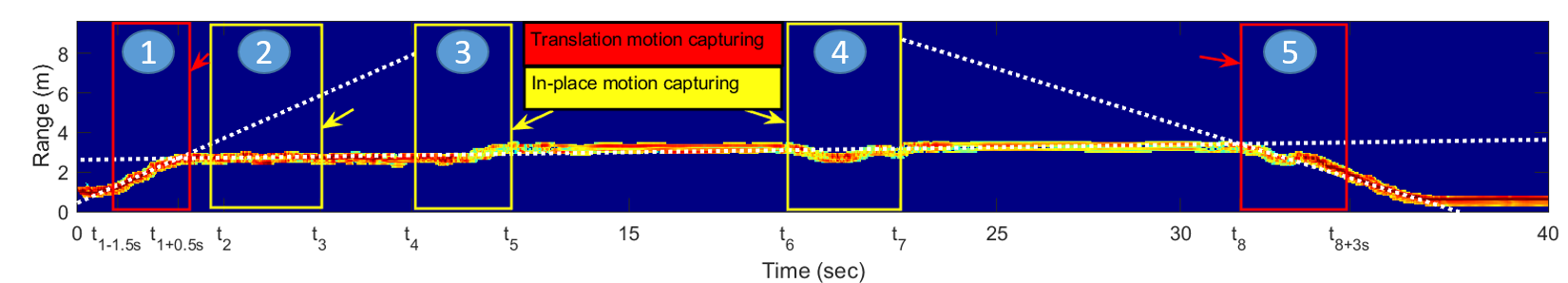}
\caption[Motion sequence of Example-3]{Motion sequence: walking-stopping, bending from standing, turning around and sitting down, bending from sitting, standing up from sitting merged with walking. \label{fig:SequenceWlkBenSitBenStaWlk}}
\end{figure} 

\begin{figure}[htb]
\begin{center}
    \begin{minipage}[t]{.14\linewidth}
        \centering
        \rotatebox[origin=l]{90}{{\small ~Freq. (Hz)}}~\includegraphics[width=\linewidth]{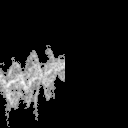}
        \\{\small~~~Time (sec)\\~~~~~~~~(a)}\label{subfig:Ex3_11}
    \end{minipage}
    ~~~~
    \begin{minipage}[t]{.14\linewidth}
        \centering
        \rotatebox[origin=l]{90}{{\small ~Freq. (Hz)}}~\includegraphics[width=\linewidth]{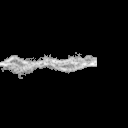}
        \\{\small~~~Time (sec)\\~~~~~~~~(b)}\label{subfig:Ex3_21}
    \end{minipage}
    ~~~~
    \begin{minipage}[t]{.14\linewidth}
    \centering
        \rotatebox[origin=l]{90}{{\small ~Freq. (Hz)}}~\includegraphics[width=\linewidth]{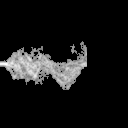}
        \\{\small~~~Time (sec)\\~~~~~~~~(c)}\label{subfig:Ex3_31}
    \end{minipage}  
     ~~~~
    \begin{minipage}[t]{.14\linewidth}
        \centering
        \rotatebox[origin=l]{90}{{\small ~Freq. (Hz)}}~\includegraphics[width=\linewidth]{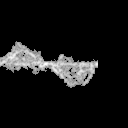}
        \\{\small~~~Time (sec)\\~~~~~~~~(d)}\label{subfig:Ex3_41}
    \end{minipage}
    ~~~~
    \begin{minipage}[t]{.14\linewidth}
    \centering
        \rotatebox[origin=l]{90}{{\small ~Freq. (Hz)}}~\includegraphics[width=\linewidth]{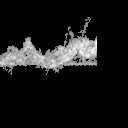}
        \\{\small~~~Time (sec)\\~~~~~~~~(e)}\label{subfig:Ex3_51}
    \end{minipage}
    \\
\vspace{.3cm}
    \begin{minipage}[t]{.14\linewidth}
        \centering
        \rotatebox[origin=l]{90}{{\small~Range (m)}}~\includegraphics[width=\linewidth]{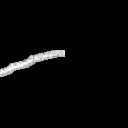}
        \\{\small~~~Time (sec)\\~~~~~~~~(f)}\label{subfig:Ex3_12}
    \end{minipage}
    ~~~~
    \begin{minipage}[t]{.14\linewidth}
        \centering
        \rotatebox[origin=l]{90}{{\small~Range (m)}}~\includegraphics[width=\linewidth]{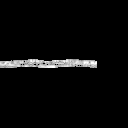}
        \\{\small~~~Time (sec)\\~~~~~~~~(g)}\label{subfig:Ex3_22}
    \end{minipage}
    ~~~~
    \begin{minipage}[t]{.14\linewidth}
    \centering
        \rotatebox[origin=l]{90}{{\small~Range (m)}}~\includegraphics[width=\linewidth]{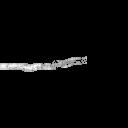}
        \\{\small~~~Time (sec)\\~~~~~~~~(h)}\label{subfig:Ex3_32}
    \end{minipage} 
     ~~~~
    \begin{minipage}[t]{.14\linewidth}
        \centering
        \rotatebox[origin=l]{90}{{\small~Range (m)}}~\includegraphics[width=\linewidth]{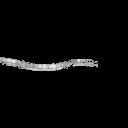}
        \\{\small~~~Time (sec)\\~~~~~~~~(i)}\label{subfig:Ex3_42}
    \end{minipage}
    ~~~~
    \begin{minipage}[t]{.14\linewidth}
    \centering
        \rotatebox[origin=l]{90}{{\small~Range (m)}}~\includegraphics[width=\linewidth]{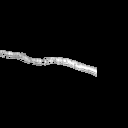}
        \\{\small~~~Time (sec)\\~~~~~~~~(j)}\label{subfig:Ex3_52}
    \end{minipage}
\end{center}
\caption[Captured motions for Example-3.]{Captured motions for Example-3 (Micro-Doppler for motions \textcircled{1} to \textcircled{5} are in Fig.~a.-e.; Range-map for motions \textcircled{1} to \textcircled{5} are in Fig.~f.-j.).\label{fig:Ex3_MD_RM}}
\end{figure} 

Example~3 in Fig.~\ref{fig:SequenceWlkBenSitBenStaWlk} shows a common ADL motion sequence of walking away from the radar, stopping and bending in the walking direction, and picking an object close to a chair location. Then, the person turns and sits down. While sitting, the person picks an object from the floor. After remaining in the SiS for a period of time, the person stands up and walks towards the radar. For the latter activity, the in-place motion and walking are merged, and are time inseparable by the Radon transform or by the energy detector.

The captured micro-Doppler signatures and range-map images of Example-3 (Fig.~\ref{fig:SequenceWlkBenSitBenStaWlk}) are shown in Fig.\ref{fig:Ex3_MD_RM}. The range profiles are shifted, as in Example-1, to compensate biased classification based on the distance to the radar. 
The fused 2-D PCA followed by the NN classifier of Fig.~\ref{fig:fusionGraph} is used for the motions \circled{2} to \circled{5}. Motion \circled{1} is classified based only on spectrogram with the micro-Doppler image (Fig~\ref{fig:Ex3_MD_RM}.a). 
%
\subsubsection{Forward in time motion classification}
The first intersection point is determined by the Radon transform at $t_1 \approx 2.8s$, and a $2s$ window is used for capturing the motion \circled{1} in Fig.~\ref{fig:SequenceWlkBenSitBenStaWlk}. The binary Classifier-5 is applied for the motions (I) walking-stopping \& walking-bending and (II) walking-falling for the direction facing away from the radar. It is noticed that the classifier differs from those in the previous examples, since the direction of translation is reversed. 
%
\begin{table}[htbp]
\centering 
\caption[Classifier-8.]{Classifier-8: Forward in time classification, {[}$d_{MD}=10$; $d_{RM}=2${]}.}\vspace{0.15cm}
\label{tab:clasMat_0b}
\scalebox{0.77}{
\begin{tabular}{@{}rrrrrrr@{}}
\toprule
 & \multicolumn{2}{c}{\begin{tabular}[c]{@{}c@{}}Micro-Doppler\\ Predicted\end{tabular}} & \multicolumn{2}{c}{\begin{tabular}[c]{@{}c@{}}Range-map\\ Predicted\end{tabular}} & \multicolumn{2}{c}{\begin{tabular}[c]{@{}c@{}}Fusion\\ Predicted\end{tabular}} \\ \midrule
\multicolumn{1}{r|}{cl.} & (I) & \multicolumn{1}{r|}{(II)} & (I) & \multicolumn{1}{r|}{(II)} & (I) & (II) \\ \midrule
\multicolumn{1}{r|}{(I)} & 99.3\% & \multicolumn{1}{r|}{0.7\%} & 94.8\% & \multicolumn{1}{r|}{5.2\%} & 99.3\% & 0.7\% \\
\multicolumn{1}{r|}{(II)} & 1.1\% & \multicolumn{1}{r|}{98.9\%} & 8.1\% & \multicolumn{1}{r|}{91.9\%} & 3.3\% & 96.7\%
\end{tabular}
}
\end{table} 
In this case, as shown in Table~\ref{tab:clasMat_0b} (Classifier-8), the micro-Doppler classification outperforms fusion classification and, therefore, will preferably be used. It gives $99.3\%$ correct classification rate. 
Although, there is only $0.7\%$ of miss-detection, the LS can still be assigned instead of the ground truth, the StS. 
Accordingly, the next Classifier-9 should consider classes for both states in walking direction, namely, (I) turning \& sitting down,  (II) bending while standing away from the radar, (III) falling from standing away from the radar, and (IV) standing up from falling away from the radar.  Conforming to the state diagram in Fig.~\ref{fig:flowgraph_fwd}, class (I) leads to the SiS, class (III) to the LS and class (II) and class (IV) lead to the StS. 

The PBC determines the next in-place motion \circled{2} to have the onset and offset times at $t_2 \approx 3.7s$ and $t_3 \approx 6.6s$, respectively. 
%
\begin{table}[hbt]
\caption[Classifier-9.]{Classifier-9: Forward in time classification, [$d_{MD}=10$; $d_{RM}=2$]. } \vspace{0.15cm}
\label{tab:clasMat_1b_4classes} 
\centering
\scalebox{0.77}{
\begin{tabular}{@{}rrrrrrrrr@{}}
\toprule
 & \multicolumn{4}{c}{\begin{tabular}[c]{@{}c@{}}Micro-Doppler\\ Predicted\end{tabular}} & \multicolumn{4}{c}{\begin{tabular}[c]{@{}c@{}}Range-map\\ Predicted\end{tabular}} \\ \midrule
\multicolumn{1}{r|}{cl.} & (I) & (II) & (III) & \multicolumn{1}{r|}{(IV)} & (I) & (II) & (III) & \multicolumn{1}{r|}{(IV)} \\ \midrule
\multicolumn{1}{r|}{(I)} & 99.9\% & 0.1\% & 0.0\% & \multicolumn{1}{r|}{0.0\%} & 98.0\% & 0.3\% & 0.2\% & \multicolumn{1}{r|}{1.5\%} \\
\multicolumn{1}{r|}{(II)} & 0.0\% & 99.6\% & 0.0\% & \multicolumn{1}{r|}{0.4\%} & 0.2\% & 99.8\% & 0.0\% & \multicolumn{1}{r|}{0.0\%} \\
\multicolumn{1}{r|}{(III)} & 1.5\% & 0.0\% & 97.0\% & \multicolumn{1}{r|}{1.5\%} & 3.0\% & 0.1\% & 95.2\% & \multicolumn{1}{r|}{1.7\%} \\
\multicolumn{1}{r|}{(IV)} & 0.0\% & 4.4\% & 0.0\% & \multicolumn{1}{r|}{95.6\%} & 4.7\% & 2.2\% & 0.9\% & \multicolumn{1}{r|}{92.3\%} \\ \cmidrule(r){1-5}
\multicolumn{1}{r|}{~} & \multicolumn{4}{c|}{Fusion Predicted} &  &  &  &  \\ \cmidrule(r){1-5}
\multicolumn{1}{r|}{cl.} & (I) & (II) & (III) & \multicolumn{1}{r|}{(IV)} &  &  &  &  \\ \cmidrule(r){1-5}
\multicolumn{1}{r|}{(I)} & 99.9\% & 0.1\% & 0.0\% & \multicolumn{1}{r|}{0.0\%} &  &  &  &  \\
\multicolumn{1}{r|}{(II)} & 0.0\% & 99.9\% & 0.0\% & \multicolumn{1}{r|}{0.1\%} &  &  &  &  \\
\multicolumn{1}{r|}{(III)} & 0.5\% & 0.9\% & 98.0\% & \multicolumn{1}{r|}{0.6\%} &  &  &  &  \\
\multicolumn{1}{r|}{(IV)} & 0.0\% & 1.2\% & 0.0\% & \multicolumn{1}{r|}{98.8\%} &  &  &  & 
\end{tabular}
}
\end{table} 
The corresponding classification results of Classifier-9 are shown in Table~\ref{tab:clasMat_1b_4classes}. The true motion \circled{2} of (II) bending while standing has $0.1\%$ missing probability. This number rises to $1.5\%$ when all motion classes are considered, per Table~\ref{tab:AllClassesFusion}. More critical is the probability of false alarm associated with class (I) and class (III), which would assign a wrong state. Therefore, the next classifier should consider motions from the LS and the StS away from the radar, as used for motion \circled{2} plus the motions stemming from the SiS facing the radar, i.e., (V) standing up from sitting and (VI) bending while sitting. 
%
%
\begin{table}[htbp]
\caption[Classifier-10.]{Classifier-10: Forward in time classification, [$d_{MD}=10$;~$d_{RM}=2$].\label{tab:clasMat_1b_6classes}} \vspace{0.15cm}
\centering
\scalebox{0.77}{
\begin{tabular}{@{}rrrrrrr@{}}
\toprule
 & \multicolumn{6}{c}{Micro-Doppler predicted} \\ \midrule
\multicolumn{1}{r|}{cl.} & (I) & (II) & (III) & (IV) & (V) & (VI) \\ \midrule
\multicolumn{1}{r|}{(I)} & 99.8\% & 0.2\% & 0.0\% & 0.0\% & 0.0\% & 0.0\% \\
\multicolumn{1}{r|}{(II)} & 0.0\% & 99.5\% & 0.0\% & 0.5\% & 0.0\% & 0.0\% \\
\multicolumn{1}{r|}{(III)} & 2.4\% & 0.0\% & 95.8\% & 1.9\% & 0.0\% & 0.0\% \\
\multicolumn{1}{r|}{(IV)} & 0.0\% & 6.2\% & 0.2\% & 90.8\% & 0.7\% & 2.1\% \\
\multicolumn{1}{r|}{(V)} & 0.3\% & 1.6\% & 0.0\% & 0.3\% & 97.1\% & 0.6\% \\
\multicolumn{1}{r|}{(VI)} & 0.0\% & 0.4\% & 0.0\% & 0.4\% & 0.0\% & 99.2\% \\ \midrule
\multicolumn{1}{l}{} & \multicolumn{6}{c}{Range-map predicted} \\ \midrule
\multicolumn{1}{r|}{(I)} & 98.1\% & 0.1\% & 0.1\% & 1.8\% & 0.0\% & 0.0\% \\
\multicolumn{1}{r|}{(II)} & 0.1\% & 91.6\% & 0.0\% & 0.0\% & 0.2\% & 8.1\% \\
\multicolumn{1}{r|}{(III)} & 6.1\% & 0.6\% & 91.6\% & 0.7\% & 0.9\% & 0.0\% \\
\multicolumn{1}{r|}{(IV)} & 7.8\% & 1.7\% & 0.1\% & 76.7\% & 2.2\% & 11.5\% \\
\multicolumn{1}{r|}{(V)} & 0.3\% & 0.6\% & 0.0\% & 0.1\% & 96.9\% & 2.0\% \\
\multicolumn{1}{r|}{(VI)} & 0.0\% & 5.5\% & 0.0\% & 0.0\% & 0.0\% & 94.5\% \\ \midrule
\multicolumn{1}{l}{} & \multicolumn{6}{c}{Fusion predicted} \\ \midrule
\multicolumn{1}{r|}{(I)} & 99.8\% & 0.2\% & 0.0\% & 0.0\% & 0.0\% & 0.0\% \\
\multicolumn{1}{r|}{(II)} & 0.0\% & 99.9\% & 0.0\% & 0.1\% & 0.0\% & 0.1\% \\
\multicolumn{1}{r|}{(III)} & 0.7\% & 0.8\% & 97.4\% & 1.1\% & 0.0\% & 0.0\% \\
\multicolumn{1}{r|}{(IV)} & 0.0\% & 1.0\% & 0.0\% & 95.1\% & 0.0\% & 3.9\% \\
\multicolumn{1}{r|}{(V)} & 0.4\% & 0.7\% & 0.0\% & 0.7\% & 98.1\% & 0.1\% \\
\multicolumn{1}{r|}{(VI)} & 0.0\% & 0.4\% & 0.0\% & 0.1\% & 0.0\% & 99.5\%
\end{tabular}
}
\end{table} 
The classification results for Classifier-10 are shown in Table~\ref{tab:clasMat_1b_6classes}.
The next in-place motion begins at $t_4 \approx 9.1s$ and ends at $t_5 \approx 11.8s$. The ground truth motion \circled{3} of (I) turning \& sitting down is classified with a certainty of $99.8\%$ ($0.2\%$ miss-detection). Similarly as before, prescence of false alarm can lend to a wrong state, which broadens the possible next motions.  

The next considered in-place motions, shown in Classifier-11 (Table~\ref{tab:clasMat_1b_8classes}), are (I) turning \& sitting down,  (II) bending while standing away from the radar, (III) falling from standing away from the radar, (IV) standing up from falling away from the radar,  (V) standing up from sitting, (VI) bending while sitting, (VII) bending from standing towards the radar, and (VIII) falling from standing towards the radar. 
\begin{table}[htbp]
\caption[Classifier-11.]{Classifier-11: Forward in time classification, [$d_{MD}=10$;~$d_{RM}=2$].\label{tab:clasMat_1b_8classes}}\vspace{0.15cm} 
\centering
\scalebox{0.77}{
\begin{tabular}{@{}rrrrrrrrr@{}}
\toprule
 & \multicolumn{8}{c}{Micro-Doppler predicted} \\ \midrule
\multicolumn{1}{r|}{cl.} & (I) & (II) & (III) & (IV) & (V) & (VI) & \multicolumn{1}{l}{(VII)} & \multicolumn{1}{l}{(VIII)} \\ \midrule
\multicolumn{1}{r|}{(I)} & 99.9\% & 0.0\% & 0.0\% & 0.0\% & 0.0\% & 0.0\% & 0.1\% & 0.0\% \\
\multicolumn{1}{r|}{(II)} & 0.0\% & 99.3\% & 0.0\% & 0.3\% & 0.0\% & 0.0\% & 0.4\% & 0.0\% \\
\multicolumn{1}{r|}{(III)} & 1.5\% & 0.0\% & 96.9\% & 1.6\% & 0.0\% & 0.0\% & 0.0\% & 0.0\% \\
\multicolumn{1}{r|}{(IV)} & 0.0\% & 5.6\% & 0.2\% & 90.3\% & 0.5\% & 2.5\% & 0.9\% & 0.0\% \\
\multicolumn{1}{r|}{(V)} & 0.3\% & 1.3\% & 0.0\% & 0.3\% & 97.4\% & 0.5\% & 0.0\% & 0.1\% \\
\multicolumn{1}{r|}{(VI)} & 0.0\% & 0.2\% & 0.0\% & 0.5\% & 0.0\% & 99.1\% & 0.1\% & 0.0\% \\
\multicolumn{1}{r|}{(VII)} & 0.1\% & 2.0\% & 0.0\% & 0.0\% & 0.1\% & 0.3\% & 97.4\% & 0.0\% \\
\multicolumn{1}{r|}{(VIII)} & 0.0\% & 0.0\% & 0.0\% & 0.0\% & 1.2\% & 0.5\% & 0.0\% & 98.3\% \\ \midrule
\multicolumn{1}{l}{} & \multicolumn{8}{c}{Range-map predicted} \\ \midrule
\multicolumn{1}{r|}{(I)} & 98.1\% & 0.0\% & 0.2\% & 0.8\% & 0.0\% & 0.0\% & 0.8\% & 0.0\% \\
\multicolumn{1}{r|}{(II)} & 0.2\% & 88.3\% & 0.0\% & 0.0\% & 0.0\% & 6.7\% & 4.8\% & 0.0\% \\
\multicolumn{1}{r|}{(III)} & 2.3\% & 0.0\% & 94.6\% & 0.1\% & 0.8\% & 0.0\% & 0.1\% & 2.1\% \\
\multicolumn{1}{r|}{(IV)} & 7.7\% & 3.9\% & 0.0\% & 75.0\% & 0.5\% & 10.5\% & 1.0\% & 1.3\% \\
\multicolumn{1}{r|}{(V)} & 0.6\% & 0.8\% & 0.0\% & 0.0\% & 94.1\% & 2.7\% & 1.5\% & 0.3\% \\
\multicolumn{1}{r|}{(VI)} & 0.0\% & 3.2\% & 0.0\% & 0.0\% & 0.7\% & 93.6\% & 2.4\% & 0.0\% \\
\multicolumn{1}{r|}{(VII)} & 1.2\% & 16.2\% & 0.0\% & 0.0\% & 2.4\% & 14.9\% & 65.3\% & 0.0\% \\
\multicolumn{1}{r|}{(VIII)} & 0.0\% & 0.0\% & 0.0\% & 0.0\% & 1.1\% & 0.0\% & 0.4\% & 98.5\% \\ \midrule
\multicolumn{1}{l}{} & \multicolumn{8}{c}{Fusion predicted} \\ \midrule
\multicolumn{1}{r|}{(I)} & 99.7\% & 0.0\% & 0.0\% & 0.0\% & 0.0\% & 0.0\% & 0.3\% & 0.0\% \\
\multicolumn{1}{r|}{(II)} & 0.0\% & 99.3\% & 0.0\% & 0.0\% & 0.0\% & 0.0\% & 0.7\% & 0.0\% \\
\multicolumn{1}{r|}{(III)} & 0.3\% & 0.1\% & 98.1\% & 0.5\% & 0.0\% & 0.0\% & 0.0\% & 1.0\% \\
\multicolumn{1}{r|}{(IV)} & 0.0\% & 1.4\% & 0.0\% & 95.8\% & 0.1\% & 2.6\% & 0.1\% & 0.1\% \\
\multicolumn{1}{r|}{(V)} & 0.4\% & 0.6\% & 0.0\% & 0.4\% & 98.0\% & 0.3\% & 0.2\% & 0.0\% \\
\multicolumn{1}{r|}{(VI)} & 0.0\% & 0.2\% & 0.0\% & 0.1\% & 0.0\% & 99.5\% & 0.2\% & 0.0\% \\
\multicolumn{1}{r|}{(VII)} & 0.2\% & 1.9\% & 0.0\% & 0.0\% & 0.1\% & 0.3\% & 97.4\% & 0.0\% \\
\multicolumn{1}{r|}{(VIII)} & 0.0\% & 0.0\% & 0.0\% & 0.0\% & 0.8\% & 0.0\% & 0.0\% & 99.2\%
\end{tabular}
}
\end{table} 

The PBC finds the next in-place motion \circled{4} between $t_6 \approx 19.3s$ and $t_7 \approx 22.4s$. The ground truth motion of (VI) bending while sitting is detected with a certainty of $99.5\%$. Then, the person keeps sitting until the Radon transform detects a WS towards the radar at $t_8 \approx 31.7s$. From that point, the person must have been in a SiS or StS until time $t_8$. According to the state diagram in Fig.~\ref{fig:flowgraph_fwd}, there are only two arrows coming into the WS facing the radar. 
Therefore, the captured motion \circled{5} of (II) standing up merged with walking is classified versus (I) starting-walking from standing by using Classifier-3 (Table~\ref{tab:clas_4}). 
Since both classes are classified with an accuracy of $100\%$, we declare the leading state before the standing up merged with walking to be the SiS. In contrast, the ground truth motion of standing up merged with walking is classified with $97.9\%$ when applying all considered ADL motion classes.

\subsubsection{Backward in time motion classification} 
As in Example~1, backward in time classification describing pre-walking activities. Table~\ref{tab:clas_4} (Classifier-3) provides $100\%$ classification accuracy for (I) starting-walking and (II) standing up from sitting merged with walking.
This time, however, motion (II) is classified and leads to the SiS. Going back in time, the Classifier-9 (Table~\ref{tab:clasMat_23pre}) can be used to discriminate between previous activities, namely, (I) sitting down from standing, and (II) bending from sitting. With the ground truth motion \circled{4} of (II) bending from sitting correctly declared, the SiS remains in effect with a certainty of $100\%$ classification rate. Going further back in time, the same classifier is applied for motion \circled{3} where the ground truth motion of (I) sitting down form standing puts the person in the StS. At this point, the orientation of the person is uncertain. Thus, bidirectional classes leading to the StS must be considered. The five classes of Classifier-4 are shown in Table~\ref{tab:clasMat_22}, dealing with (I) standing up from sitting, (II) bending while standing towards the radar, (III) bending while standing away from the radar, (IV) standing up from falling towards the radar, and (V) standing up from falling away from the radar. The true motion \circled{2} of (III) bending away from the radar while standing has an accuracy of $98.4\%$. Further, the result ascertains the turning around motion between bending from standing and sitting down. The person remains in the StS for a short period before the Radon transform detects a previous WS. The walking does not incorporate falling, since it cannot occur before the StS \cite{aminGuendel:IET}. 
%
\subsection{Comparison of forward- and backward in time classification to all ADL classes}
\begin{figure}[ht]
\centering
\includegraphics[clip, trim=35.13mm 18.9mm 35.13mm 18.9mm, width=0.7\linewidth]{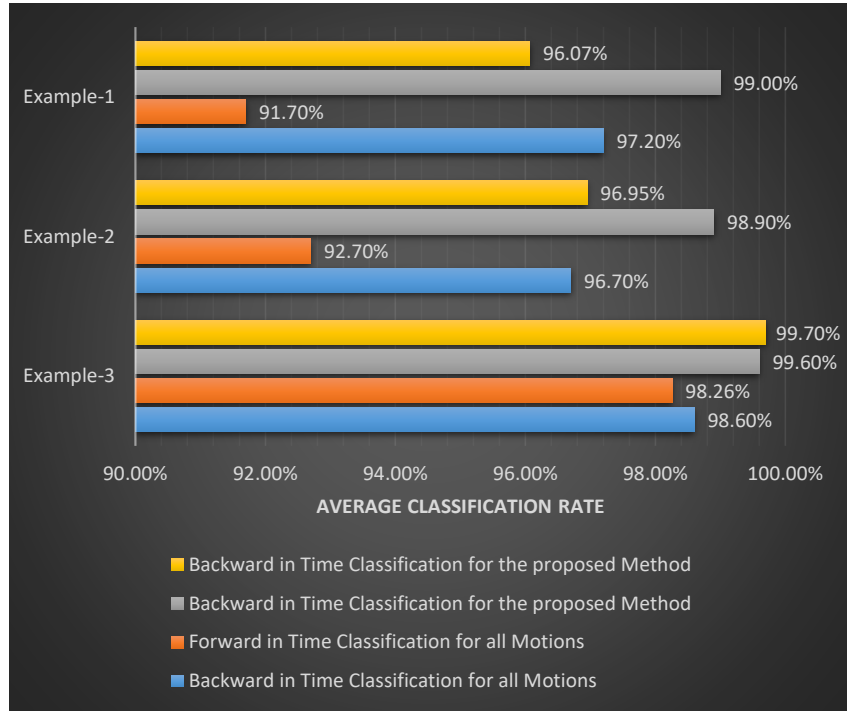}
\caption[The average classification rate diagram for the three examples.]{Shows the average classification rate for the three examples by applying forward and backward in time motion classification for the proposed method and by considering all motion classes. \label{fig:averageDiagram}}
\end{figure} 
In this section, we show the average classification rate in Fig.~\ref{fig:averageDiagram} for the proposed method by classifying ADL motions forward and backward in time. Further, we compare the results of limiting the number of motion classes per the state diagram to the more general case by considering all ADL motion classes.

Beginning with Example-1 (Fig.~\ref{fig:SequenceWlkFalSta}), the average classification rate is $96.07\%$ for the forward in time classification for the motion sequence of walking-falling, standing up from falling and start walking. The low classification rate of the merged motion walking-falling of $89.5\%$ is in impotent factor. Considering the backward in time motion classification, the average classification rate increases to $99.00\%$. By using one classifier with all motion classes, the average classification rate for forward in time classification decreased to $91.70\%$, since the certainty of classifying walking-falling is even worse with $80.7\%$. By going backward in time, the average classification rate of $97.20\%$ which is lower than that for the proposed method.       

 The average classification rate of Example-2 (Fig.~\ref{fig:SequenceWlkFalStaSitStaWlk}) behaves similar as Example-1, since the motion of walking-falling has a negative impact for the forward in time classification. Accordingly, the average classification rate is $96.95\%$ for the proposed method and $92.70\%$ by using all motion classes. For going backwards in time, the average classification rate is $98.90\%$ for the proposed method and decreases to $96.70\%$ by using all classes. In that example, the forward in time classification rate for the proposed method outperforms both, the forward and backward classification method by using an all motion classifier.

Different from the previous examples, Example-3 (Fig.~\ref{fig:SequenceWlkBenSitBenStaWlk}) shows a forward in time motion classification outperforming marginally the backward in time classification for the proposed method with $99.70\%$ compared to $99.60\%$. The average classification rate of $98.26\%$ and $98.60\%$ for forward- and backward in time motion classification, respectively, are achieved all motion classifier \cite{aminGuendel:IET}. 

\chapter{Summary\label{chap:Conclusion}} 

\section{Conclusion}
In this thesis, we considered classification of human ADL accounting for possible adjacent and sometimes inseparable contiguous motions. These motions, like falling and bending, may not have clear individual event time boundaries, especially when are merged with walking. Since walking defines the translation motion and can be easily discerned from pronounced changes in the range, we use it as the building block that guides the classifiers and resolve ambiguities which may occur in classifying in-place motions. The latter include sitting down, standing up, falling, and bending from standing or sitting.  The Radon transform is applied to the range-map to detect the translation motion, whereas an energy detector, referred to as Power Burst Curve (PBC), is used to provide the onset or/and offset times of in-place motions.

The human ethogram, which is the catalogue of human behavior and possible motion sequences, was used to reduce the number of motion classes classified at any given time. This was made possible by considering human motions as states that consist of standing, sitting, walking, and laying. The actions of standing up, sitting down, falling and bending are cast as transitional motions that link the different states. At a given state, and according to the ethogram, there are known and limited transitioning actions in and out of it. Reducing the number of classes and the use of state diagram improve classification compared to using all ADL. It also allows changing the features to best commensurate with the given state. We used 2-D PCA with NN classifier and changed the number of principal components according to the state and transitioning actions.

The thesis proposed performing classifications of ADL in both forward and backward in time and showed that the results are often different. This suggests performing classifications in both time directions to obtain different levels of assertions regarding the incurred activities.

\section{Further work}
\begin{figure}[ht]
	\centering
	\includegraphics[width=.7\linewidth]{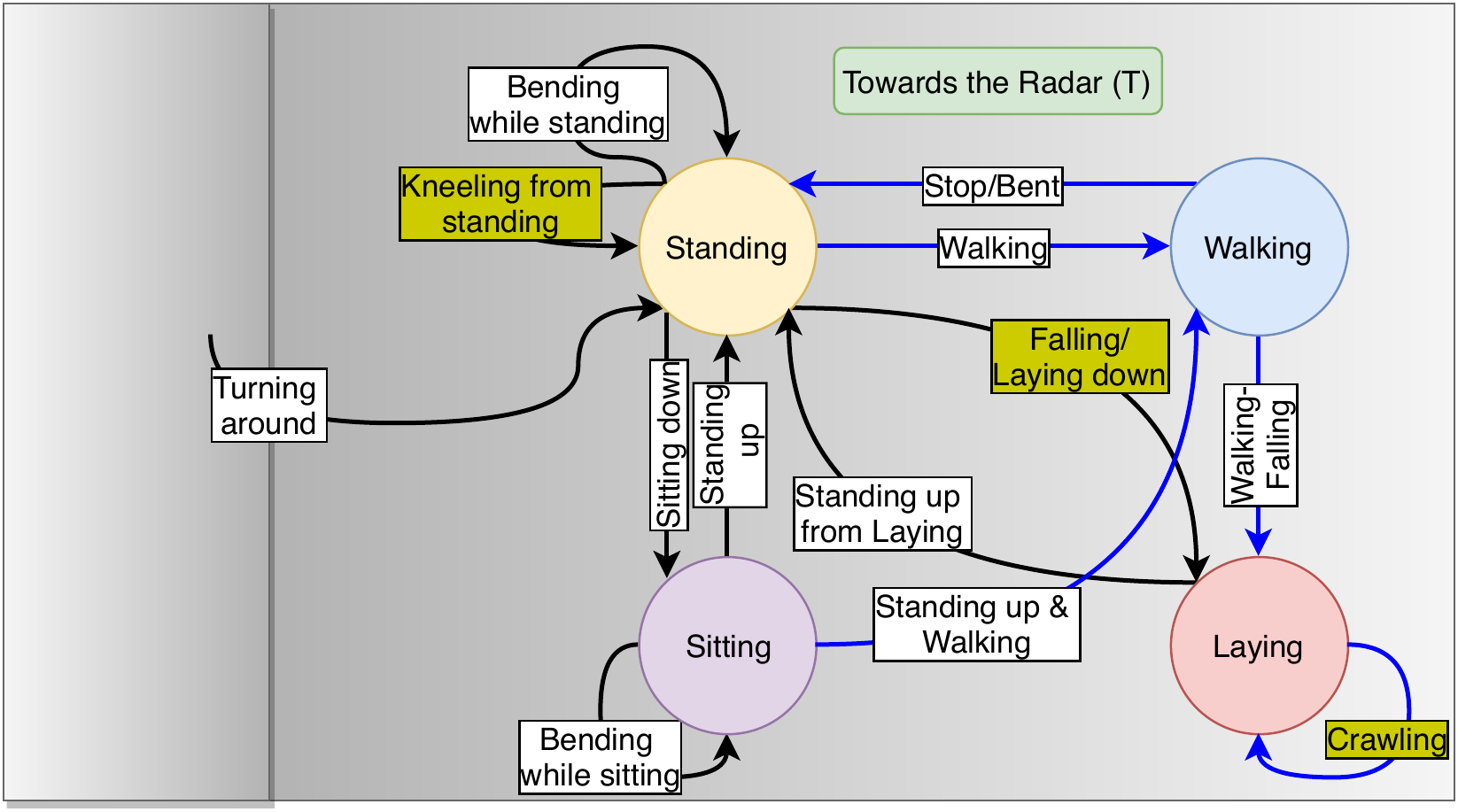}
	\caption[Extended state diagram for forward in time classification for motions towards the radar.]{Extended state diagram for forward in time classification for motions towards the radar (black arrows: in-place actions; blue arrows: translation actions). \label{fig:1radarconfmc9}}
\end{figure}
The work in this thesis considers, from our point of view, the most critical and valuable motions. However,  the state diagram in Fig.~\ref{fig:flowgraph_fwd} can be extended as desired. For example, Fig.~\ref{fig:1radarconfmc9} shows a possible generalization with additional motions of \textit{Laying down}, \textit{Crawling}, and \textit{Kneeling from standing} for towards the radar.

Further research can be investigated in separating multiple targets based on the proposed method by using the Radon transform. Also people counting is certainly a topic of interest in ADL environments. By using more advanced radar technology, e.g., by applying MIMO functionality, more research can be investigated to determine the person intricate motions, including sideways.          

The work in this thesis uses 2D-PCA with the Nearest Neighbor classification. However, the proposed approach can be applied with other methods, all can benefit from the state diagram. The selected feature extraction method and classifier in this thesis have given the best reliable results. Other methods, however, can be computational simpler and move sustainable for real-time applications.


\begin{spacing}{0.9}

\cleardoublepage
\bibliography{References/references} 

\begin{thebibliography}{10}

\bibitem{1_aminSP}
{M. G. Amin}, {Y. D. Zhang}, {F. Ahmad}, and {K. C. Ho}, ``Radar signal
  processing for elderly fall detection: The future for in-home monitoring,''
  {\em IEEE Signal Processing Magazine}, vol.~33, no.~2, pp.~71--80, 2016.

\bibitem{amin2017radar}
M.~G. Amin, {\em Radar for Indoor Monitoring: Detection, Classification, and
  Assessment}.
\newblock CRC Press, 2017.

\bibitem{8_4801689}
Y.~{Kim} and H.~{Ling}, ``Human activity classification based on
  micro-{Doppler} signatures using a support vector machine,'' {\em IEEE
  Transactions on Geoscience and Remote Sensing}, vol.~47, pp.~1328--1337, May
  2009.

\bibitem{mobasseri2009time}
B.~G. Mobasseri and M.~G. Amin, ``A time-frequency classifier for human gait
  recognition,'' in {\em Optics and Photonics in Global Homeland Security V and
  Biometric Technology for Human Identification VI} (B.~V. Kumar, S.~Prabhakar,
  A.~A. Ross, C.~S. Halvorson, and Šárka O.~Southern, eds.), vol.~7306,
  pp.~434 -- 442, International Society for Optics and Photonics, SPIE, 2009.

\bibitem{10_5404249}
S.~S. {Ram}, C.~{Christianson}, Y.~{Kim}, and H.~{Ling}, ``Simulation and
  analysis of human micro-{Dopplers} in through-wall environments,'' {\em IEEE
  Transactions on Geoscience and Remote Sensing}, vol.~48, pp.~2015--2023,
  April 2010.

\bibitem{9_7348882}
R.~M. {Narayanan} and M.~{Zenaldin}, ``Radar micro-{Doppler} signatures of
  various human activities,'' {\em IET Radar, Sonar Navigation}, vol.~9, no.~9,
  pp.~1205--1215, 2015.

\bibitem{12_Hong2011}
{J. Hong}, {S. Tomii}, and {T. Ohtsuki}, ``Cooperative fall detection using
  {Doppler} radar and array sensor,'' {\em in 2013 IEEE 24th International
  Symposium on Personal Indoor and Mobile Radio Communications},
  pp.~3492--3496, 2013.

\bibitem{13_Su2015}
{B. Y. Su}, {K. C. Ho}, {M. J. Rantz}, and {M. Skubic}, ``{Doppler} radar fall
  activity detection using the wavelet transform,'' {\em IEEE Transactions on
  Biomedical Engineering}, vol.~62, no.~3, pp.~865--875, 2015.

\bibitem{14_Wu2015}
{Q. Wu}, {Y. D. Zhang}, {W. Tao}, and {M. G. Amin}, ``Radar-based fall
  detection based on {Doppler} time-frequency signatures for assisted living,''
  {\em Sonar Navigation IET Radar}, vol.~9, no.~2, pp.~173--183, 2015.

\bibitem{LeKernec8746868}
J.~{Le Kernec}, F.~{Fioranelli}, C.~{Ding}, H.~{Zhao}, L.~{Sun}, H.~{Hong},
  J.~{Lorandel}, and O.~{Romain}, ``Radar signal processing for sensing in
  assisted living: The challenges associated with real-time implementation of
  emerging algorithms,'' {\em IEEE Signal Processing Magazine}, vol.~36,
  pp.~29--41, July 2019.

\bibitem{kim2016hand}
Y.~Kim and B.~Toomajian, ``Hand gesture recognition using {micro-Doppler}
  signatures with convolutional neural network,'' {\em IEEE Access}, vol.~4,
  pp.~7125--7130, 2016.

\bibitem{wang2016interacting}
S.~Wang, J.~Song, J.~Lien, I.~Poupyrev, and O.~Hilliges, ``{Interacting with
  Soli: Exploring fine-grained dynamic gesture recognition in the
  radio-frequency spectrum},'' in {\em Proceedings of the 29th Annual Symposium
  on User Interface Software and Technology}, (Tokyo, Japan), Oct. 2016.

\bibitem{skaria2019hand}
S.~Skaria, A.~Al-Hourani, M.~Lech, and R.~J. Evans, ``{Hand-gesture recognition
  using two-Antenna {Doppler} radar with deep convolutional neural networks},''
  {\em IEEE Sensors Journal}, vol.~19, no.~8, pp.~3041--3048, 2019.

\bibitem{maminzz}
M.~G. Amin, Z.~Zeng, and T.~Shan, ``{Hand gesture recognition based on radar
  micro-{Doppler} signature envelopes},'' in {\em Proceedings of the 2019 IEEE
  Radar Conference (RadarConf)}, (Boston, MA), Apr. 2019.

\bibitem{zhang2016dynamic}
S.~Zhang, G.~Li, M.~Ritchie, F.~Fioranelli, and H.~Griffiths, ``Dynamic hand
  gesture classification based on radar micro-{Doppler} signatures,'' in {\em
  Proceedings of the 2016 CIE International Conference on Radar (RADAR)},
  (Guangzhou, China), Oct. 2016.

\bibitem{GurbuzAmin:DLindoor}
S.~Z. {Gurbuz} and M.~G. {Amin}, ``Radar-based human-motion recognition with
  deep learning: Promising applications for indoor monitoring,'' {\em IEEE
  Signal Processing Magazine}, vol.~36, pp.~16--28, July 2019.

\bibitem{3_7842648}
{S. Z. Gurbuz}, {C. Clemente}, {A. Balleri}, and {J. Soraghan},
  ``Micro-{Doppler}-based in-home aided and unaided walking recognition with
  multiple radar and sonar systems,'' {\em IET Radar, Sonar Navigation},
  vol.~11, no.~1, pp.~107--115, 2017.

\bibitem{Seifert8613848}
A.~{Seifert}, M.~G. {Amin}, and A.~M. {Zoubir}, ``Toward unobtrusive in-home
  gait analysis based on radar micro-{Doppler} signatures,'' {\em IEEE
  Transactions on Biomedical Engineering}, vol.~66, pp.~2629--2640, Sep. 2019.

\bibitem{5_7944373}
{M. S. Seyfioğlu}, {S. Z. Gürbüz}, {A. M.Özbayoğlu}, and {M. Yüksel},
  ``Deep learning of micro-{Doppler} features for aided and unaided gait
  recognition,'' in {\em Proceedings of the 2017 IEEE Radar Conference
  (RadarConf)}, pp.~1125--1130, May 2017.

\bibitem{6_seifert2019RadarConf}
{A.-K. Seifert}, {A. M. Zoubir}, and {M. G. Amin}, ``Detection of gait
  asymmetry using indoor {D}oppler radar,'' in {\em Proceedings of the IEEE
  Radar Conference (RadarConf), Boston, MA}, April 2019.

\bibitem{shahFioranelli:ADL:8894733}
S.~A. {Shah} and F.~{Fioranelli}, ``{RF} sensing technologies for assisted
  daily living in healthcare: A comprehensive review,'' {\em IEEE Aerospace and
  Electronic Systems Magazine}, vol.~34, pp.~26--44, Nov 2019.

\bibitem{amin2019rf}
M.~G. Amin, A.~Ravisankar, and R.~G. Guendel, ``{RF} sensing for continuous
  monitoring of human activities for home consumer applications,'' in {\em Big
  Data: Learning, Analytics, and Applications}, vol.~10989, pp.~33 -- 44,
  International Society for Optics and Photonics, SPIE, May 2019.

\bibitem{aminGuendel2019radarConf}
M.~G. Amin and R.~G. Guendel, ``Radar human motion recognition using motion
  states and two-way classifications,'' in {\em Proceedings of the 2020 IEEE
  Radar Conference (RadarConf), (submitted)}, 2020.

\bibitem{jokanovic2018fall}
B.~Jokanovi{\'c} and M.~Amin, ``Fall detection using deep learning in
  range-{Doppler} radars,'' {\em IEEE Transactions on Aerospace and Electronic
  Systems}, vol.~54, no.~1, pp.~180--189, 2018.

\bibitem{amin1992time}
M.~G. Amin, ``Time-frequency spectrum analysis and estimation for nonstationary
  random processes. {Time}-frequency signal analysis methods and
  applications,'' {\em Ed: B. Boashash, Longman Chesire 1992; Melbourne,
  Australia}, pp.~208--232, 1992.

\bibitem{SetlurAmin:SPIE}
P.~Setlur, M.~G. Amin, and F.~Ahmad, ``Analysis of micro-doppler signals using
  linear fm basis decomposition,'' in {\em Radar Sensor Technology} (R.~N.
  Trebits and J.~L. Kurtz, eds.), vol.~6210, pp.~201 -- 211, International
  Society for Optics and Photonics, SPIE, 2006.

\bibitem{baris:DataCubeProcessing8691492}
B.~{Erol} and M.~G. {Amin}, ``Radar data cube processing for human activity
  recognition using multi subspace learning,'' {\em IEEE Transactions on
  Aerospace and Electronic Systems}, pp.~1--1, 2019.

\bibitem{park2016microdoppler_bc}
J.~Park, R.~J. Javier, T.~Moon, and Y.~Kim, ``Micro-{Doppler} based
  classification of human aquatic activities via transfer learning of
  convolutional neural networks,'' in {\em Sensors}, 2016.

\bibitem{humanEthogram}
B.~Jesness, ``A human ethogram: Its scientific acceptability and importance
  (now new, because new technology allows investigation of the hypotheses) (an
  early must read),'' 02 1985.

\bibitem{baris:GANbased:8835589}
B.~{Erol}, S.~Z. {Gurbuz}, and M.~G. {Amin}, ``{GAN-based} synthetic radar
  micro-{Doppler} augmentations for improved human activity recognition,'' in
  {\em Proceedings of the 2019 IEEE Radar Conference (RadarConf)}, pp.~1--5,
  April 2019.

\bibitem{amin2016radar}
M.~G. Amin, Y.~D. Zhang, F.~Ahmad, and K.~D. Ho, ``Radar signal processing for
  elderly fall detection: The future for in-home monitoring,'' {\em IEEE Signal
  Processing Magazine}, vol.~33, no.~2, pp.~71--80, 2016.

\bibitem{amin_Wu:Fall_PBC:7046290}
Q.~{Wu}, Y.~D. {Zhang}, W.~{Tao}, and M.~G. {Amin}, ``Radar-based fall
  detection based on {Doppler} time–frequency signatures for assisted
  living,'' {\em IET Radar, Sonar Navigation}, vol.~9, no.~2, pp.~164--172,
  2015.

\bibitem{baris:ASILOMAR:WidebandFall7869686}
B.~{Erol}, M.~G. {Amin}, B.~{Boashash}, F.~{Ahmad}, and Y.~D. {Zhang},
  ``Wideband radar based fall motion detection for a generic elderly,'' in {\em
  Proceedings of the 2016 50th Asilomar Conference on Signals, Systems and
  Computers}, pp.~1768--1772, Nov 2016.

\bibitem{pcaKNN:8252120}
M.~{Punjabi} and G.~L. {Prajapati}, ``Lazy learner and pca: An evolutionary
  approach,'' in {\em Proceedings of the 2017 Computing Conference},
  pp.~312--316, July 2017.

\bibitem{rajaguru2017knn}
H.~Rajaguru and S.~Prabhakar, {\em KNN Classifier and K-Means Clustering for
  Robust Classification of Epilepsy from EEG Signals. A Detailed Analysis}.
\newblock Anchor Academic Publishing, 2017.

\bibitem{website2500B}
{Ancortek Inc.}, ``{SDR-KIT 2500B}.''
\newblock (retrieved: 09/20/2019 ).

\bibitem{art:fmcwRadarModel}
J.-J. Lin, Y.-P. Li, W.-C. Hsu, and T.-S. Lee, ``Design of an {FMCW} radar
  baseband signal processing system for automotive application,'' {\em
  SpringerPlus}, vol.~5, 12 2016.

\bibitem{fmcwRadar_8443507}
Y.~{Sun}, T.~{Fei}, F.~{Schliep}, and N.~{Pohl}, ``Gesture classification with
  handcrafted micro-{Doppler} features using a {FMCW} radar,'' in {\em
  Proceedings of the 2018 IEEE MTT-S International Conference on Microwaves for
  Intelligent Mobility (ICMIM)}, pp.~1--4, April 2018.

\bibitem{art:hanningWindow467238}
{W. {Chen}} and N.~C. {Griswold}, ``An efficient recursive time-varying
  {Fourier} transform by using a half-sine wave window,'' in {\em Proceedings
  of IEEE-SP International Symposium on Time- Frequency and Time-Scale
  Analysis}, pp.~284--286, Oct 1994.

\bibitem{aminGuendel:IET}
M.~G. Amin and R.~G. Guendel, ``Radar classifications of consecutive and
  contiguous human motions,'' {\em IET Radar, Sonar Navigation, (under
  review)}, 2019.

\bibitem{eclean:Kulpa}
K.~{Kulpa}, ``The clean type algorithms for radar signal processing,'' in {\em
  2008 Microwaves, Radar and Remote Sensing Symposium}, pp.~152--157, Sep.
  2008.

\bibitem{baris:FusionPCA8835840}
B.~{Erol} and M.~{Amin}, ``Generalized pca fusion for improved radar human
  motion recognition,'' in {\em Proceedings of the 2019 IEEE Radar Conference
  (RadarConf)}, pp.~1--5, April 2019.

\bibitem{Hoorfar:4815956}
K.~M. {Yemelyanov}, N.~{Engheta}, A.~{Hoorfar}, and J.~A. {McVay}, ``Adaptive
  polarization contrast techniques for through-wall microwave imaging
  applications,'' {\em IEEE Transactions on Geoscience and Remote Sensing},
  vol.~47, pp.~1362--1374, May 2009.

\bibitem{chihaoui2016survey}
M.~Chihaoui, A.~Elkefi, W.~Bellil, and C.~Ben~Amar, ``A survey of {2D} face
  recognition techniques,'' {\em Computers}, vol.~5, no.~4, p.~21, 2016.

\bibitem{29_wininger2013basis}
{K. L. Wininger}, ``Basis of {CT}: the {R}adon transform,'' {\em Radiologic
  technology}, vol.~84, no.~4, pp.~413--418, 2013.

\bibitem{book:principleOfModernRadar}
W.~L. Melvin and J.~A. Scheer, {\em Principles of Modern Radar: Advanced
  Techniques}.
\newblock Edison NJ: Scitech Publishing, 2013.

\bibitem{30_4599159}
L.~{Cirillo}, A.~{Zoubir}, and M.~{Amin}, ``Parameter estimation for locally
  linear fm signals using a time-frequency {Hough} transform,'' {\em IEEE
  Transactions on Signal Processing}, vol.~56, pp.~4162--4175, Sep. 2008.

\bibitem{Bresenham:1965:ACC:1663347.1663349}
J.~E. Bresenham, ``Algorithm for computer control of a digital plotter,'' {\em
  IBM Syst. J.}, vol.~4, pp.~25--30, Mar. 1965.

\bibitem{32_7944316}
B.~{Erol}, M.~G. {Amin}, and B.~{Boashash}, ``Range-{Doppler} radar sensor
  fusion for fall detection,'' in {\em Proceedings of the 2017 IEEE Radar
  Conference (RadarConf)}, pp.~0819--0824, May 2017.

\bibitem{33_6889337}
{L. R. Rivera}, {E. Ulmer}, {Y. D. Zhang}, {W. Tao}, and {M. G. Amin},
  ``Radar-based fall detection exploiting time-frequency features,'' in {\em
  Proceedings of the 2014 IEEE China Summit International Conference on Signal
  and Information Processing (ChinaSIP)}, pp.~713--717, July 2014.

\end{thebibliography}
\bibliographystyle{ieeetr}



\end{spacing}


%
%

\printthesisindex 

\end{document}